\documentclass[aps,prd,onecolumn,preprintnumbers,showpacs,nofootinbib,amssymb]{
revtex4}
\usepackage{graphicx}
\usepackage{amsmath}
\usepackage{amssymb}
\usepackage{amsfonts}
\usepackage{bm}
\usepackage{color}

\def\be{\begin{equation}}
\def\ee{\end{equation}}
\def\bea{\begin{eqnarray}}
\def\eea{\end{eqnarray}}

\begin{document}

\title{Statistical mechanics of self-gravitating systems in general
relativity:\\
I. The quantum Fermi gas}
\author{Pierre-Henri Chavanis}
\affiliation{Laboratoire de Physique Th\'eorique, Universit\'e de Toulouse,
CNRS, UPS, France}

\begin{abstract} 

We develop a general
formalism to determine the statistical equilibrium states of
self-gravitating systems in general relativity and complete previous works on
the subject. Our results are valid for an
arbitrary form of entropy but, for illustration, we explicitly consider the
Fermi-Dirac entropy for fermions. The maximization of entropy
at fixed mass-energy and particle number determines  the distribution function
of the system and its equation of state. It also implies the
Tolman-Oppenheimer-Volkoff equations of hydrostatic equilibrium and the
Tolman-Klein relations. Our paper provides all the necessary equations that are
needed to construct the caloric curves of self-gravitating fermions
in general relativity as done in recent works. We consider
the nonrelativistic
limit $c\rightarrow +\infty$ and recover the equations obtained within the
framework of Newtonian gravity. We also discuss the inequivalence of
statistical ensembles as well as the relation
between the dynamical and thermodynamical stability of self-gravitating systems
in Newtonian gravity and
general relativity.

\end{abstract}


\maketitle

\section{Introduction}
\label{sec_iso}

Self-gravitating fermions play an important role in different areas of
astrophysics. They appeared in the context of white dwarfs, neutron
stars and dark matter halos where the fermions are electrons, neutrons and
massive neutrinos respectively.
We start by a brief history of the subject giving an exhaustive list of
references.\footnote{A detailed historic of the statistical
mechanics of
self-gravitating systems (classical and quantum) in Newtonian gravity and
general relativity is given in Refs. \cite{acf,acb}.}

Soon after the discovery of the quantum statistics by Fermi 
\cite{fermi,fermibis} and Dirac  \cite{dirac} in 1926, Fowler \cite{fowler}
used this ``new thermodynamics'' to solve the
puzzle of the extreme high density of white dwarfs, which
could not be explained by classical physics \cite{eddingtonbook}. He understood
that white dwarfs owe their stability to
the quantum pressure of the degenerate electron gas resulting  from the Pauli
exclusion principle \cite{paulipple}. He considered a completely degenerate
electron gas at $T=0$
based on the fact that the temperature in white dwarfs is much smaller
than
the Fermi temperature ($T\ll T_F$). He also used
Newtonian gravity which is a very good approximation to describe white dwarfs
in general.
The first models
\cite{stoner29,milne,chandra31nr} of
white dwarfs  were based on  the
nonrelativistic equation of state of a Fermi
gas and provided the corresponding mass-radius relation. Stoner
\cite{stoner29} developed an
analytical approach based on a uniform density approximation for the star.
Chandrasekhar \cite{chandra31nr}  derived the exact mass-radius relation of
nonrelativistic white dwarfs by
applying the theory of polytropes of index $n=3/2$ \cite{emden}. It was
then realized that
 special relativity must be taken into account at high densities. When the
relativistic equation of state is employed it was found that white dwarfs can
exist only below a maximum mass $M_{\rm max}=1.42\, M_{\odot}$ 
\cite{frenkel,anderson,stoner30,chandra31,chandra31composite,stonertyler,
landau32,stoner32,stoner32eddington,chandra32ZA,chandra34autre,chandra34,
chandra35},
now known as the Chandrasekhar limiting
mass. Frenkel \cite{frenkel} was
the first to mention
that relativistic effects become important when the mass of white dwarfs
becomes larger than the solar mass but he did not
envision the existence of an upper mass limit. The
maximum mass of white dwarfs was
successively derived by Anderson \cite{anderson}, Stoner
\cite{stoner30}, Chandrasekhar \cite{chandra31} and Landau
\cite{landau32} using different methods.
Anderson \cite{anderson} first obtained an estimate of the maximum mass but his
relativistic treatment of the problem was erroneous. Stoner
\cite{stoner30} corrected the
mistakes of Anderson and derived the complete mass-radius relation
of white dwarfs and their maximum mass. Anderson \cite{anderson} and Stoner
\cite{stoner30} both used an
analytical approach based
on the uniform density approximation previously introduced by
Stoner \cite{stoner29} in the nonrelativistic limit.
Chandrasekhar \cite{chandra31} and Landau \cite{landau32} 
obtained the exact value
of the maximum mass by considering an ultrarelativistic electron gas and
applying the theory of polytropes of index $n=3$ \cite{emden}. Finally,
Chandrasekhar
\cite{chandra35} 
obtained the complete mass-radius relation of white
dwarfs and the maximum mass by numerically solving the equation of hydrostatic
equilibrium with the relativistic equation of state. At the maximum mass, the
radius of the star vanishes. Close to the maximum
mass,
general
relativity must be taken into account as first considered by Kaplan 
\cite{kaplan} and Chandrasekhar and Tooper \cite{ct}. In that case, the radius
of the star at $M_{\rm max}$ is finite, being equal to $R_{*}=1.03\times 10^3\,
{\rm km}$ ($246$ times the corresponding Schwarzschild radius), instead of
vanishing as in the
Newtonian treatment.

Contrary to the case of white dwarfs, neutron stars were
predicted theoretically
before being observed. The neutron was predicted by
Rutherford \cite{rutherford} as early as 1920 and finally discovered by Chadwick
in 1932 \cite{chadwick}. The first explicit
prediction of
neutron stars with extremely high density and very small
radius  was made by Baade and Zwicky  in December 1933
\cite{baadez34a,baadez34aa,baadez34b}. Remarkably,
they anticipated that neutron stars could result from supernova explosion. The
first calculation of neutron star model was performed
by Oppenheimer and Volkoff \cite{ov} who assumed matter to be composed of
an ideal gas of free neutrons at high density. They worked at zero temperature
at which the neutrons are completely degenerate and employed the 
relativistic equation of state of an ideal fermion gas previously derived in
the case of white dwarfs. They used general relativity
because of the high mass and density of neutron stars.  They found that
equilibrium configurations exist only below a maximum mass $M_{\rm max}=0.710\,
M_{\odot}$, now
known as the Oppenheimer-Volkoff limiting mass.\footnote{It is not
well-known that, at the same period, Zwicky \cite{z1,z2} also attempted to
determine
the maximum mass of neutron stars that he interpreted as Schwarzschild's mass.
Using
heuristic arguments, he obtained an expression of the form $M_{\rm
max}^{\rm Zwicky}=k\, m_n(e^2/Gm_pm_e)^{3/2}=91\, k\,  M_{\odot}$, where $k$ is
a
dimensionless number assumed to be of order unity. He
mentioned the need to take into account the equation of state of the neutrons
but did not quote   Oppenheimer and Volkoff \cite{ov} (actually, Oppenheimer and
Zwicky
refused to acknowledge each other's papers, see \cite{opp}). The maximum mass
obtained by Oppenheimer and Volkoff \cite{ov} is $M_{\rm max}^{\rm OV}=0.384\,
(\hbar
c/G)^{3/2}m_n^{-2}=0.710\, M_{\odot}$. Matching the
two
expressions of the maximum mass obtained by Zwicky and Oppenheimer and Volkoff,
we get $k=7.80\times 10^{-3}$ and $\alpha\equiv e^2/\hbar c=13.4
m_pm_e/m_n^2\simeq 13.4 {m_e}/{m_p}$. Interestingly, this equation
provides a
relation between the fine-structure
constant
$\alpha\simeq 1/137$ and the
ratio $m_e/m_p$ between the electron mass and the proton mass (we have
used $m_n\simeq m_p$). This type of relationships has been proposed in the past
by several authors \cite{kragh} using heuristic arguments or pure
numerology (see \cite{ouf} for a recent discussion).}  Their radius
at $M_{\rm max}$ is $R_*=9.16\, {\rm km}$ ($4.37$ times the
corresponding Schwarzschild radius).
Detailed
studies of
neutron stars with
more realistic equations of state taking into account
the repulsive effect of nuclear forces were made between 1958 and 1967
\cite{hww,cameron,zeldovichHS,hsa,zeldovichstiff,as1,as2,zeldovich,zeldovichNM,
wheeler,saakyan,dk,chandra64,tooper1,chiu,mz,sv,inman,tooper2,harrison,
harrisonstiff,as3,thorne,tc,mt,btm,wheeler66,cazzola} (see a review
in Ref. \cite{htww65}), but, apart from these works, neutron stars were not
much
studied because it was estimated that their residual thermal radiation would be
too faint to be observed. The situation changed when pulsars were discovered by
Hewish {\it et al.} \cite{hewish}
in 1968. The same year, Gold \cite{gold1} proposed that pulsars were rotating
neutron stars, and this is generally accepted today. Too massive stars cannot
reach a quiescent equilibrium state and  rather collapse into a
black hole which is a Schwarzschild
\cite{schwarzschild1,schwarzschild2} singularity of spacetime in the
nonrotating case or a Kerr \cite{kerr} singularity of
spacetime in the rotating case. This continued gravitational contraction was
originally described by Oppenheimer and Snyder \cite{os} in 1939 but the concept
of
black holes as real physical objects took a long time to be accepted. It is now
believed that supermassive black holes, surrounded by an accretion disk, lie in
the core of galaxies and power
active galactic nuclei and quasars.\footnote{The name ``black hole'' was
popularized by Wheeler
\cite{wheelerBH,wheelerBH2} but it appeared earlier \cite{ewing,rosenfeld},
being probably introduced by Dicke in analogy with the
Black Hole prison of Calcutta (see Ref. \cite{bhhl}). The term ``quasar'' was
invented by Chiu \cite{chiuquasar} to name ``quasi-stellar radio sources''.}

The fermionic models of white dwarfs and neutron stars were exported to the
case of dark matter halos, assuming
that
dark matter
is made of massive neutrinos as originally proposed by Markov \cite{markov} and
Cowsik and McClelland \cite{cmc1,cmc2}. A lower bound on the fermion mass was
obtained by Tremaine and Gunn \cite{tg} using constraints arising from the
Vlasov equation. The first
models decribed dark matter
halos at
$T=0$ using the  equation of state of a completely degenerate fermion gas either
in the nonrelativistic limit \cite{cmc2,r,cgr,fjr,vlt,vtt,viollierseul,tv,du}
or in 
general relativity
\cite{markov,gao,gao2,kallman,zcls,bgr,bmv,nm,narain,kupi}. Subsequent
models considered dark matter halos at finite temperature showing that they
have a ``core-halo'' structure consisting in a dense core (fermion ball)
surrounded by a dilute isothermal atmosphere
leading to flat rotation curves. Most models were based on the
ordinary Fermi-Dirac distribution in Newtonian gravity
\cite{ir,bvn,csmnras,robert,pt,dark,ispolatov,rieutord,ptdimd,ijmpb,mb,
ddvs1,ddvs2,vss,vsedd,vs2,rsu,vsbh} or
general relativity 
\cite{cls,cl,gmr,merafina,bvr,ar,bmtv,btv,arf,sar,rar,acf,rc}.\footnote{The
self-gravitating Fermi gas at finite temperature was also studied in Refs.
\cite{juttnerQ,chandrabook,wares,margrave,ht,em,edwards,
boshkayev1,boshkayev2,roupasNS} in
the context of
stellar
structure (white dwarfs and neutron stars).} Other models
were based on the more realistic 
fermionic King model (describing tidally truncated fermionic dark matter halos)
in Newtonian
gravity
\cite{stella,imrs,chavmnras,paolis,clm2} or general
relativity \cite{krut}. Some authors \cite{bvn,bvr,bmtv,btv,rar,krut}
have proposed that a
fermion ball could mimic a supermassive black hole that is purported to exist at
the center of a galaxy but some difficulties with this scenario were pointed out
in \cite{nature,reid,genzel}. The
status of the fermion ball scenario is still not clearly settled. Recently,
it has been proposed that the fermion ball may represent a large bulge instead
of mimicking a black hole
\cite{modeldm}. The self-gravitating Fermi gas was also studied  in
relation to the violent relaxation of
collisionless
self-gravitating systems described by the Lynden-Bell \cite{lb} distribution
which is formally similar to the Fermi-Dirac
distribution
\cite{ktb1,csr,ktb2,csmnras,robert,chavmnras,proceedingsDubrovnik}.

The study of phase transitions in the self-gravitating Fermi gas is an
important problem in itself. To make the mathematical study well-posed, we need
to confine the system in a finite region of space in order to have a finite
mass. Indeed, the density of an unbounded isothermal self-gravitating system
decreases as $r^{-2}$ at large distances leading to an infinite mass. Phase
transitions in the self-gravitating Fermi
gas  were first analyzed for box-confined models both in the
nonrelativistic limit
\cite{ht,bvn,csmnras,robert,pt,dark,ispolatov,rieutord,ptdimd,ijmpb} (see a
review in
\cite{ijmpb}) and in general relativity \cite{bvr,acf,rc}. The study of phase
transitions in the more realistic case of tidally truncated self-gravitating
fermions described by the fermionic King model \cite{stella,chavmnras} has been
performed in Ref. \cite{clm2} in  the nonrelativistic limit. These
studies describe the transition between a gaseous phase and a condensed
phase corresponding to a compact object (white dwarf, neutron star, fermion
ball), or the collapse of the system towards a black hole when its mass is too
large. These phase transitions may be related to the
onset of the red-giant structure (leading
to white dwarfs)  in a late phase of stellar evolution and
to the supernova phenomenon (leading to neutron stars)
\cite{pomeau1,pomeau2,calettre}. General
phase diagrams have been obtained in Refs. \cite{ijmpb,acf,clm2}.

The basic equations describing a Newtonian self-gravitating gas of
fermions at arbitrary temperature consist in the condition
of hydrostatic equilibrium combined with the ideal 
equation of
state of the Fermi gas \cite{chandrabook}.\footnote{These equations are
equivalent to the Fermi-Dirac-Poisson equations. They constitute the
so-called finite temperature Thomas-Fermi (TF) model. At $T=0$,
the
Fermi-Dirac distribution reduces to a step function and we recover
the TF model. We recall that the original TF
model \cite{thomas,fermiatom1,fermiatom2,fermiatom3} describes the distribution
of electrons in an atom which results from the balance between the
quantum pressure (Pauli's exclusion principle), the electrostatic repulsion of
the electrons, and the electrostatic attraction of the nucleus. For
self-gravitating systems, the TF model describes the distribution of fermions in
a star or in a
dark matter halo which results from the balance between
the quantum pressure (Pauli's exclusion principle) and the gravitational
attraction of the fermions. The
rigorous mathematical justification of
the thermodynamic limit for
self-gravitating fermions in a box leading to the finite 
temperature TF model is given in Refs.
\cite{levyleblond,htf,hnt,messer,baumgartner,hseul,ns1,ns2,bh,messerpt1,
messerpt2}.} These equations can be obtained all at
once from the maximum  entropy principle of
statistical mechanics. It
determines
the most probable state of the system at statistical equilibrium. This
variational principle was introduced in astrophysics by Ogorodnikov
\cite{ogo1,ogo2}, Antonov \cite{antonov} and
Lynden-Bell and Wood \cite{lbw} in the context of
stellar systems and was further studied by several authors
\cite{ipser74,tvh,hkpart1,nakada,hs,hk3,khd,katzpoincare1,
katz2,ih,inagaki,katzking,lecarkatz,ms,sflp,lp,kiessling,paddyapj,dvsc,
katzokamoto,dvs1,dvs2,aaiso,crs,sc,grand,lifetime,cn,clm1} (see the reviews
\cite{paddy,katzrevue,ijmpb}). The
statistical equilibrium state of
a classical self-gravitating system is obtained by maximizing the Boltzmann
entropy at fixed energy and particle number in the microcanonical ensemble or by
minimizing the Boltzmann free energy at fixed particle number in the canonical
ensemble. Statistical ensembles are inequivalent
for self-gravitating systems \cite{paddy,katzrevue,ijmpb} and for other
systems with long-range
interactions \cite{cdr,campabook}. The maximum
entropy principle was
then extended to the case of self-gravitating fermions by replacing the
Boltzmann entropy by the Fermi-Dirac entropy
\cite{lb,ht,bvn,csr,csmnras,robert,proceedingsDubrovnik,pt,dark,ispolatov,
rieutord,ptdimd,ijmpb,clm2}. The
maximum entropy principle has also been applied to generalized forms of entropy
in Refs.
\cite{antonovthesis,ipser74,ih,thlb,ano,gen,physa,lang,cstsallis,
aaantonov,super,assise,nfp,cc,entropy}.\footnote{A functional of the form
$S=-\int
C(\overline{f})\, d{\bf r}d{\bf v}$, where $C(\overline{f})$ is any convex
function of the coarse-grained distribution function $\overline{f}({\bf r},{\bf
v},t)$, was introduced by Antonov \cite{antonovthesis} for collisionless
stellar systems, and called ``quasi-entropy'' (see Ref. \cite{ano}). The same
functional was reintroduced independently by Tremaine {\it et al.} \cite{thlb}
in relation to the theory of violent relaxation \cite{lb} and called
``$H$-function''. A functional of the form $S=-\int
C({f})\, d{\bf r}d{\bf v}$, without the bar on $f$, was
introduced by Ipser \cite{ipser74,ih} in his study on the dynamical stability
of collisionless stellar systems with respect to the Vlasov-Poisson
equations. An effective thermodynamical
formalism involving generalized entropic functionals  of the form $S=-\int
C({f})\, d{\bf r}d{\bf v}$ was developed by Chavanis
\cite{gen,physa,lang,cstsallis,aaantonov,super,assise,nfp,cc,entropy} who gave 
various interpretations of these functionals.} For
example, the Tsallis entropy \cite{tsallis} (which is related
to one of the
functionals considered by Ipser \cite{ipser74} -- see \cite{cstsallis}) leads to
power-law
distribution
functions
\cite{ipser74,pp,gen,physa,aapoly,ts1,ts2,tsprl,grand,lang,cstsallis,aaantonov}.
They correspond to the stellar polytropes introduced
by
Eddington \cite{eddington16} in 1916 as particular steady states of the
Vlasov-Poisson equations.

The basic equations describing a spherically symmetric self-gravitating
gas
of fermions at arbitrary temperature in general relativity consist in the
Tolman-Oppenheimer-Volkov \cite{tolman,ov}
equations of hydrostatic equilibrium\footnote{There is debate
\cite{debate1,debate2,debate3} on the name that should be given to these
equations: OV or TOV? As far as we can judge, Eq. (\ref{n57}) was
first written by
Tolman \cite{tolman} 
while Eq. (\ref{n58b}) was first written by Oppenheimer and Volkoff \cite{ov}.
Therefore, Eq. (\ref{n57}) should be called the Tolman equation and Eq.
(\ref{n58b}) should be
called the OV equation. If we consider that Eq. (\ref{n58b}) is a
rather direct consequence of Eq. (\ref{n57}) then it may be called the TOV
equation as well
\cite{debate2}. However, Ref. \cite{debate3} stresses some
fundamental differences between
Eqs. (\ref{n57}) and (\ref{n58b}). To
complement this discussion, we note that Chandrasekhar
\cite{chandra64,chandra72}
rederived Eq. (\ref{n58b}) without referring to Oppenheimer and
Volkoff \cite{ov}, maybe considering that
this equation results almost immediately from the Einstein field equations
(\ref{n56})-(\ref{n58})
with the metric (\ref{n50}). In his review on the
first thirty years of
general relativity \cite{cp} he writes: ``Equations (60) and (61) are
often referred to as the Oppenheimer-Volkoff equations though they are contained
in Schwarzschild's paper in very much these forms.''} combined with the ideal
 equation of
state   of the relativistic Fermi gas \cite{chandrabook},\footnote{The equation
of state
of a completely degenerate ($T=0$) Fermi gas at arbitrary densities (i.e. for
any degree of relativistic motion) was first derived by Frenkel  \cite{frenkel}
in a not well-known paper. It was rederived  independently by Stoner
\cite{stoner30,stoner32} and Chandrasekhar \cite{chandra35}. The
equation of state of a gas of fermions at arbitrary temperature was first
derived by Juttner \cite{juttnerQ} extending his earlier work on the
relativistic theory of an ideal classical gas \cite{juttner1,juttner2}. These
different results are exposed in the
classical monograph of Chandrasekhar \cite{chandrabook} on stellar
structure.}
and the
Tolman-Klein \cite{tolman,klein} relations stating that in general relativity
the temperature and
the chemical potential are not uniform at statistical equilibrium. 
These equations can be obtained all at once from the maximum entropy
principle of statistical mechanics. This
variational principle was introduced in relativistic
astrophysics by
Tolman \cite{tolman} (see Appendix \ref{sec_tolman})
and further developed by several
authors \cite{cocke,hk0,kh,khk,kam,kh2,hk,ipser80,sorkin,sy,sy2,bvrelat,
aarelat1,aarelat2,gaobis,gaoE,roupas1,gsw,fg,roupas1E,
schiffrin,psw,fhj}. The maximum entropy principle is valid for a general
form of entropy.  It was specifically
applied to the self-gravitating black-body radiation in
\cite{tolman,sorkin,aarelat2}, to self-gravitating fermions described
by the Fermi-Dirac entropy in \cite{bvrelat} and to classical particles
described by the Boltzmann entropy in \cite{kh,khk,kh2,ipser80,roupas}. The
statistical
equilibrium state of a general relativistic self-gravitating system is obtained
by maximizing the entropy at fixed mass-energy and particle
number in the
microcanonical ensemble or by
minimizing the free energy at fixed particle number in the canonical
ensemble. Again,
the statistical ensembles are inequivalent in general
relativity.

In this paper, we synthesize previous works on the subject and develop
a general formalism to determine the statistical equilibrium states of
self-gravitating systems in general relativity. Our results are valid for an
arbitrary form of entropy but, for illustration, we explicitly consider the
case of fermions described by the Fermi-Dirac entropy.\footnote{The case of
classical
particles described by the Boltzmann entropy is specifically considered in
our companion paper \cite{rgb} (Paper II).} Our paper provides all
the
necessary
equations that are needed to construct the caloric curves of self-gravitating
fermions in general relativity as done in recent works \cite{bvr,acf,rc}.
The present paper is organized as follows. In Sec. \ref{sec_smnrf} we develop
the statistical
mechanics of nonrelativistic self-gravitating fermions.
In Sec. \ref{sec_grf} we develop
the statistical mechanics of relativistic self-gravitating fermions within
the framework of general relativity. In Sec. \ref{sec_nr} we consider the
nonrelativistic
limit $c\rightarrow +\infty$ and recover from the general relativistic
formalism the equations obtained within the framework of Newtonian gravity.
Throughout the paper, we discuss the relation between 
dynamical and thermodynamical stability of self-gravitating systems in Newtonian
gravity and general relativity.

\section{Statistical mechanics of nonrelativistic self-gravitating fermions}
\label{sec_smnrf}

In this section, we consider the statistical mechanics of nonrelativistic
self-gravitating fermions. We use a
presentation that can be extended in 
general relativity (see Sec. \ref{sec_grf}). In particular, we assume since
the start that
the system is spherically symmetric. It can be shown that the maximum entropy
state of a nonrotating self-gravitating system is necessarily spherically
symmetric so this assumption is not restrictive.

\subsection{Hydrostatic equilibrium of gaseous spheres in Newtonian gravity}
\label{sec_heng}

\subsubsection{Newton's law}
\label{sec_nl}

The gravitational
potential $\Phi(r)$ is determined from the mass density $\rho(r)$ by the Poisson
equation 
\begin{equation}
\label{n2}
\frac{1}{r^2}\frac{d}{dr}\left (r^2\frac{d\Phi}{dr}\right )=4\pi G
\rho.
\end{equation}
The mass contained within a sphere of radius $r$ is
\begin{equation}
\label{n3}
M(r)=\int_0^r \rho(r') 4\pi {r'}^2\, dr'\quad \Rightarrow \quad
\frac{dM}{dr}=4\pi \rho r^2.
\end{equation}
Multiplying the Poisson equation by $r^2$ and integrating between $0$
and $r$, we obtain Newton's law
\begin{equation}
\label{n4}
\frac{d\Phi}{dr}=\frac{GM(r)}{r^2}.
\end{equation}
We note that Newton's law is valid for any spherically symmetic
distribution of matter, steady or unsteady.

Let us assume that the system occupies a region of radius $R$ and contains a
mass
$M(R)=M$. In the empty space outside the system, Newton's law becomes 
\begin{equation}
\label{n4c}
\frac{d\Phi}{dr}=\frac{GM}{r^2}\quad \Rightarrow \quad
\Phi(r)=-\frac{GM}{r} \qquad (r\ge R).
\end{equation}
This leads to the boundary condition
\begin{equation}
\label{n4b}
\Phi(R)=-\frac{GM}{R}.
\end{equation}

\subsubsection{Condition of hydrostatic equilibrium}
\label{sec_hng}

We now consider a self-gravitating gas at equilibrium.
The condition of hydrostatic
equilibrium 
\begin{equation}
\label{n1}
\frac{dP}{dr}=-\rho\frac{d\Phi
} {d r }
\end{equation}
expresses the balance between the pressure gradient and the
gravitational force. Using Newton's law (\ref{n4}), we can
rewrite the condition of hydrostatic equilibrium  (\ref{n1}) as
\begin{equation}
\label{n5}
\frac{dP}{dr}=-\rho\frac{GM(r)}{r^2}.
\end{equation}
Multiplying Eq. (\ref{n5}) by $r^2/\rho$, taking the derivative of this relation
with
respect to $r$, and using Eq. (\ref{n3}), we obtain the fundamental differential
equation of hydrostatic equilibrium:
\begin{equation}
\label{n6}
\frac{1}{r^2}\frac{d}{dr}\left (\frac{r^2}{\rho}\frac{dP}{dr}\right
)=-4\pi G \rho.
\end{equation}
This equation can also be obtained by dividing Eq. (\ref{n1})  by
$\rho$, forming the Laplacian, and using the Poisson equation (\ref{n2}).

\subsubsection{Barotropic equation of state}
\label{sec_beos}

If the gas is described by a barotropic equation of state, $P=P(\rho)$, Eq.
(\ref{n6}) determines a differential equation for $\rho(r)$ of the form
\begin{equation}
\label{n6a}
\frac{1}{r^2}\frac{d}{dr}\left (\frac{r^2}{\rho}P'(\rho)\frac{d\rho}{dr}\right
)=-4\pi G \rho.
\end{equation}
On the other hand, integrating Eq. (\ref{n1}), we find that 
\begin{equation}
\label{n6b}
d\Phi=-\frac{dP}{\rho} \qquad \Rightarrow\qquad 
\Phi(r)=-\int^{\rho(r)}\frac{P'(\rho)}{\rho}\, d\rho,
\end{equation}
implying that the gravitational potential is a function
$\Phi=\Phi(\rho)$ of the density or, inversely, that the density is a function
$\rho=\rho(\Phi)$ of
the gravitational potential. Substituting this relation into the Poisson
equation (\ref{n2}) we obtain a differential equation for $\Phi(r)$ of the form
\begin{equation}
\label{n6c}
\frac{1}{r^2}\frac{d}{dr}\left (r^2\frac{d\Phi}{dr}\right
)=4\pi G \rho(\Phi).
\end{equation}
We can also introduce the enthalpy $h$ through the relation
\begin{equation}
\label{ent1}
dh=\frac{dP}{\rho}\quad \Rightarrow \quad
h(r)=\int^{\rho(r)}\frac{P'(\rho)}{\rho}\, d\rho.
\end{equation}
For a barotropic gas,  the enthalpy is a function $h=h(\rho)$ of the
density or, inversely, the density is a  function $\rho=\rho(h)$ of the
enthalpy. From Eq. (\ref{n1}), we obtain
\begin{equation}
\label{ent2}
h(r)+\Phi(r)={\rm cst}.
\end{equation}
Taking the Laplacian of this relation and using the Poisson equation (\ref{n2})
we
obtain a differential equation for $h(r)$ of the form
\begin{equation}
\label{ent3}
\frac{1}{r^2}\frac{d}{dr}\left (r^2\frac{dh}{dr}\right
)=-4\pi G\rho(h).
\end{equation}
Eqs. (\ref{n6a}), (\ref{n6c}) and (\ref{ent3}) are
clearly
equivalent. They are also equivalent to the two first order
differential equations for $M(r)$ and $\Phi(r)$:
\begin{equation}
\label{comb}
\frac{dM}{dr}=4\pi \rho r^2,\qquad \frac{d\Phi}{dr}=\frac{GM(r)}{r^2},
\end{equation}
with $\rho=\rho(\Phi)$. These equations can be combined into the
single equation (\ref{n6c}). Using Eq. (\ref{ent2}) they can also be expressed
in terms of $h$. Finally, using Eqs. (\ref{n3}) and (\ref{n5}),
we can easily obtain the following differential equation
\begin{equation}
\label{n6d}
P'\left (\frac{M'}{4\pi r^2}\right )(r^2 M''-2rM')+GM(r)M'=0
\end{equation}
for the mass profile $M(r)$. It has to be supplemented by the boundary
conditions $M(0)=0$ and $M(R)=M$.

\subsection{Local variables}
\label{sec_lnr}

We consider a system of self-gravitating
fermions described by the 
distribution function $f({\bf r},{\bf p})$ such that $f({\bf r},{\bf p})\,
d{\bf r}d{\bf p}$
gives the number density of fermions at position ${\bf r}$ with impulse ${\bf
p}$. We
introduce the particle number density
\begin{equation}
\label{n7}
n=\int f\, d{\bf p}
\end{equation}
and the kinetic energy density
\begin{equation}
\label{n8}
\epsilon_{\rm kin}=\int f E_{\rm kin}(p) \, d{\bf p},
\end{equation}
where 
\begin{equation}
\label{n9}
E_{\rm kin}(p)=\frac{p^2}{2m} 
\end{equation}
is the kinetic energy  of a particle. The local pressure is given by
\cite{chandrabook}
\begin{equation}
\label{n10}
P=\frac{1}{3}\int f p \frac{dE_{\rm kin}}{dp}\, d{\bf p}=\frac{1}{3}\int f
\frac{{p}^2}{m}\, d{\bf p}.
\end{equation}
We have the relation
\begin{equation}
\label{n11}
P=\frac{2}{3}\epsilon_{\rm kin}
\end{equation}
between the pressure and the kinetic energy density. It is valid for an
arbitrary distribution function (see Appendix \ref{sec_b}).
Finally, we introduce the Fermi-Dirac entropy density
\begin{eqnarray}
\label{n12}
s=-k_B\frac{g}{h^3}\int\Biggl\lbrace \frac{f}{f_{\rm max}}\ln \frac{f}{f_{\rm
max}}
+\left (1-\frac{f}{f_{\rm max}}\right )\ln\left
(1-\frac{f}{f_{\rm max}}\right )\Biggr\rbrace\, d{\bf p},
\end{eqnarray}
where $f_{\rm max}=g/h^3$ is the maximum possible value of the distribution
function fixed by the Pauli exclusion principle and $g$ is the spin
multiplicity of quantum states ($g=2$ for particles of spin $1/2$). The
Fermi-Dirac entropy
can be obtained from a combinatorial analysis. It
is equal to the logarithm of the number of {\it microstates}
(complexions) -- characterized by the specification of the position and the
impulse
of all the fermions $\lbrace {\bf r}_i,{\bf p}_i\rbrace$ -- corresponding to a
given
{\it macrostate} -- characterized by the (smooth) distribution function $f({\bf
r},{\bf
p})$ giving the density of fermions in a macrocell $({\bf r}, {\bf r}+d{\bf
r}; {\bf p}, {\bf p}+d{\bf p})$,
irrespectively of their precise position in the cell. The microstates must
respect
the Pauli exclusion principle, i.e., there cannot be more than $g$ particles
in the same microcell of volume $h^3$. A counting analysis taking into account
the Pauli
exclusion principle leads to the expression (\ref{n12}) of the
entropy (see, e.g., Refs. \cite{ptdimd,ijmpb} for details). 

{\it Remark:} In this paper, to be specific, we consider a  system of
fermions associated with
the Fermi-Dirac entropy (\ref{n12}). However, as shown in Appendices
\ref{sec_b}-\ref{sec_alt}, our approach is more
general. It is actually valid for any kind of particles described by a
(generalized) entropy of the
form (\ref{qgd9}).

\subsection{Global variables}
\label{sec_gnr}

The particle number is 
\begin{equation}
\label{n13}
N=\int n \,  4\pi r^2\, dr.
\end{equation}
The mass is
\begin{equation}
\label{n14}
M=Nm=\int \rho\,  4\pi r^2\, dr,
\end{equation}
where $\rho=nm$ is the mass density. The energy is  $E=E_{\rm kin}+W$, 
where\footnote{We use the same
symbol for the kinetic energy of one particle [see Eq. (\ref{n9})] and for the
total kinetic energy [see Eq. (\ref{n15})]. In
general, there is no ambiguity.}
\begin{equation}
\label{n15}
E_{\rm kin}=\int \epsilon_{\rm kin}\,  4\pi r^2\,
dr=\frac{3}{2}\int P\, 4\pi r^2\,
dr
\end{equation}
is the  kinetic energy and
\begin{equation}
\label{n16}
W=\frac{1}{2}\int
\rho \Phi\,  4\pi r^2\, dr =-\int\rho \frac{GM(r)}{r}\,  4\pi
r^2\,
dr 
\end{equation}
is the gravitational potential energy (the second expression -- only valid for
spherical systems -- is derived in Appendix \ref{sec_v}). 
The entropy is
\begin{eqnarray}
\label{n17}
S=\int s \,  4\pi r^2\, dr.
\end{eqnarray}

A statistical equilibrium state exists
only if the
system is confined within a box of radius $R$ otherwise it would
evaporate \cite{ptdimd,ijmpb}. In the microcanonical ensemble, the particle
number $N$ and the energy
$E$ are
conserved.  The statistical equilibrium state
of the system
is obtained by maximizing the Fermi-Dirac entropy $S$ at fixed energy $E$ and
particle number $N$:
\begin{eqnarray}
\max\ \lbrace {S}\, |\,  E, N \,\, {\rm fixed} \rbrace.
\end{eqnarray}
This determines the ``most
probable'' state of an isolated system. To solve this maximization problem, we
proceed
in two steps.\footnote{We use this ``two-steps'' procedure because (i) it can be
easily extended to general relativity and (ii) it is useful for studying
the sign of the second variations of entropy determining the thermodynamical
stability of the system  (see Appendix
\ref{sec_alt}). }
We first maximize the entropy density $s(r)$ at fixed kinetic energy density
$\epsilon_{\rm kin}(r)$ and particle number density $n(r)$ with respect to
variations
on $f({\bf r},{\bf p})$. This gives us the Fermi-Dirac distribution (\ref{n20})
which corresponds to the condition of
local thermodynamic equilibrium.
Then, we
substitute this distribution in the entropy density (\ref{n12}) to express it as
a function of $\epsilon_{\rm kin}(r)$ and $n(r)$. Finally, we maximize the 
entropy $S$ at fixed energy $E$ and particle number $N$ with respect
to variations on $\epsilon_{\rm kin}(r)$ and $n(r)$. This gives us  the mean
field Fermi-Dirac distribution (\ref{n37}) which is the statistical equilibrium
state of the system. In Appendix \ref{sec_os},
we maximize the entropy $S$ at fixed energy $E$ and particle number
$N$ with
respect to variations on $f({\bf r},{\bf p})$ (one-step process) and directly
obtain  the mean
field
Fermi-Dirac distribution (\ref{n37}).

\subsection{Maximization of the entropy density at fixed kinetic energy
density and particle number density}
\label{sec_lmnr}

\subsubsection{Local thermodynamic equilibrium}

We first maximize the entropy density (\ref{n12}) at fixed  kinetic energy
density
(\ref{n8})
and
particle number density (\ref{n7}). We write the variational problem for the
first
variations (extremization) under the form
\begin{equation}
\label{n19}
\frac{\delta s}{k_B}-\beta(r)\delta\epsilon_{\rm kin}+\alpha(r)\delta n=0,
\end{equation}
where $\beta(r)$ and $\alpha(r)$ are local (position dependent) Lagrange
multipliers.
This leads to the Fermi-Dirac distribution function
\begin{equation}
\label{n20}
f({\bf r},{\bf p})=\frac{g}{h^3}\frac{1}{1+e^{\beta(r)p^2/2m-\alpha(r)}},
\end{equation}
where $\alpha(r)$ and $\beta(r)$ are determined in terms of $n(r)$
and $\epsilon_{\rm
kin}(r)$ by substituting Eq. (\ref{n20}) into Eqs. (\ref{n7}) and
(\ref{n8}) (see Eqs. (\ref{n24}) and (\ref{n25}) below). By computing the
second variations of $s$, we can easily show (see Appendix \ref{sec_ts}) that 
Eq.
(\ref{n20}) is the
global maximum of
$s(r)$ at fixed $\epsilon_{\rm
kin}(r)$ and $n(r)$. Therefore, Eq. (\ref{n20}) corresponds to the condition of
local thermodynamic equilibrium. Introducing the local temperature $T(r)$ and
the local
chemical potential $\mu(r)$ by the relations 
\begin{equation}
\label{n21}
\beta(r)=\frac{1}{k_B T(r)}\qquad {\rm and}\qquad \alpha(r)=\frac{\mu(r)}{k_B
T(r)},
\end{equation}
the Fermi-Dirac
distribution (\ref{n20}) can be rewritten as
\begin{equation}
\label{n22}
f({\bf r},{\bf p})=\frac{g}{h^3}\frac{1}{1+e^{\lbrack p^2/2m-\mu(r)\rbrack/k_B
T(r)}}.
\end{equation}
On the other hand,
the variational principle (\ref{n19}) reduces to
\begin{equation}
\label{n23}
d s=\frac{d\epsilon_{\rm kin}}{T}-\frac{\mu}{T} dn,
\end{equation}
which corresponds to the first
law of thermodynamics. This law is valid for an arbitrary form of entropy
(see Appendix \ref{sec_osts}).

\subsubsection{Local variables}

Substituting the Fermi-Dirac distribution
(\ref{n22}) into Eqs. (\ref{n7}),
(\ref{n8}) and (\ref{n10}), we get
\begin{equation}
\label{n24}
n(r)=\frac{g}{h^3}\int \frac{1}{1+e^{\lbrack
p^2/2m-\mu(r)\rbrack/k_B T(r)}}  \, d{\bf p},
\end{equation}
\begin{equation}
\label{n25}
\epsilon_{\rm kin}(r)=\frac{g}{h^3}\int \frac{p^2/2m}{1+e^{\lbrack
 p^2/2m -\mu(r)\rbrack/k_B T(r)}}  \, d{\bf p},
\end{equation}
\begin{equation}
\label{n26}
P(r)=\frac{g}{3h^3}\int  \frac{p^2/m}{1+e^{\lbrack
p^2/2m-\mu(r)\rbrack/k_B T(r)}} \, d{\bf p}=\frac{g}{h^3}k_B
T(r)\int \ln\left (1+e^{-\lbrack
 p^2/2m-\mu(r)\rbrack/k_B T(r)}\right )  \, d{\bf p},
\end{equation}
where the second equality in Eq. (\ref{n26}) has been obtained from an
integration by parts. Eqs.
(\ref{n24}) and (\ref{n25}) determine $T(r)$ and $\mu(r)$ as a function of
$n(r)$ and $\epsilon_{\rm kin}(r)$.  They also determine the equation of
state $P=P\lbrack n(r),T(r)\rbrack$ in implicit form. On the other hand,
substituting the
Fermi-Dirac distribution function
(\ref{n22}) into Eq. (\ref{n12}), and using Eqs. (\ref{n24})-(\ref{n26}), we
obtain  after some calculations the relation
\begin{equation}
\label{n28}
s(r)=\frac{\epsilon_{\rm kin}(r)+P(r)-\mu(r) n(r)}{T(r)}.
\end{equation}
Actually, this relation, which is called the integrated Gibbs-Duhem relation, is
a
general
relation valid for an arbitrary form of entropy (see Appendix
\ref{sec_gd}). Using Eq.
(\ref{n11}), it reduces to the form
\begin{equation}
\label{n28b}
s(r)=\frac{\frac{5}{3}\epsilon_{\rm kin}(r)-\mu(r) n(r)}{T(r)}.
\end{equation}
Finally, combining the first law of
thermodynamics (\ref{n23}) with the integrated Gibbs-Duhem relation (\ref{n28})
we obtain
the identity
\begin{equation}
\label{n29}
d\left (\frac{P}{T}\right )=n\, d\left (\frac{\mu}{T}\right
)-\epsilon_{\rm kin} \, d\left (\frac{1}{T}\right ).
\end{equation}
We also have the identities
\begin{equation}
\label{dent2}
d\left (\frac{\epsilon_{\rm kin}}{n}\right )=-Pd\left
(\frac{1}{n}\right )+Td\left (\frac{s}{n}\right )\qquad {\rm and}\qquad s
dT-dP+nd\mu=0
\end{equation}
which correspond to the standard form of the first law of
thermodynamics and the local Gibbs-Duhem relation. These results are valid for
an arbitrary form of entropy (see Appendix \ref{sec_gd}).

\subsection{Maximization of the entropy at fixed
energy and particle
number}
\label{sec_gmnr}

\subsubsection{Variational principle}

Using the integrated Gibbs-Duhem relation (\ref{n28}), the entropy can be
written as
\begin{equation}
\label{n30}
S=\int_0^R \frac{\epsilon_{\rm kin}(r)+P(r)-\mu(r) n(r)}{T(r)}\, 
4\pi
r^2\, dr.
\end{equation}
The functionals $S$, $E$ and $N$ depend on  $\epsilon_{\rm
kin}(r)$ and $n(r)$. We now maximize $S$ at fixed $E$ and
$N$. We write the variational problem for the first variations
(extremization) under the form
\begin{equation}
\label{n31}
\frac{\delta S}{k_B}-\beta_0 \delta E+\alpha_0\delta N=0,
\end{equation}
where $\beta_0$ and $\alpha_0$ are global (uniform) Lagrange multipliers. Taking
the first
variations of  $E$ and $N$ from Eqs. (\ref{n13})-(\ref{n16}), and taking
the first
variations of   $S$ from Eq. (\ref{n17}) by using the
first law of thermodynamics (\ref{n23}), we obtain
\begin{eqnarray}
\label{n32}
\int \frac{\delta \epsilon_{\rm kin}}{k_BT}\, dV-\int
\frac{\mu}{k_BT}\delta n \, dV
-\beta_0 \int \delta\epsilon_{\rm kin} \, dV-\beta_0 \int
m\Phi \delta n \, dV+\alpha_0 \int
\delta n \, dV=0,
\end{eqnarray}
where we have introduced the abbreviation $dV=4\pi r^2\, dr$ for
the volume
element. In Eq. (\ref{n32}) the variations on $\delta \epsilon_{\rm kin}$ and
$\delta n$ must
vanish
individually.

\subsubsection{Variations on $\delta \epsilon_{\rm kin}$}

The vanishing of Eq. (\ref{n32}) with respect to variations on $\delta
\epsilon_{\rm kin}$ gives
\begin{eqnarray}
\label{n33}
\frac{1}{k_B T(r)}=\beta_0,
\end{eqnarray}
implying that the temperature is uniform at statistical equilibrium. In the
following, we will denote
the temperature by $T$
and we will replace $\beta_0$ by $\beta$. Therefore, the first relation of Eq.
(\ref{n21}) becomes 
\begin{eqnarray}
\label{n33b}
\beta=\frac{1}{k_B T}.
\end{eqnarray}
Since the temperature is uniform, the Fermi-Dirac distribution 
(\ref{n22}) and the local variables
(\ref{n24})-(\ref{n26}) can be rewritten as
\begin{equation}
\label{n22u}
f({\bf r},{\bf p})=\frac{g}{h^3}\frac{1}{1+e^{\lbrack p^2/2m-\mu(r)\rbrack/k_B
T}},
\end{equation}
\begin{equation}
\label{n24u}
n(r)=\frac{g}{h^3}\int \frac{1}{1+e^{\lbrack
p^2/2m-\mu(r)\rbrack/k_B T}}  \, d{\bf p},
\end{equation}
\begin{equation}
\label{n25u}
\epsilon_{\rm kin}(r)=\frac{g}{h^3}\int \frac{p^2/2m}{1+e^{\lbrack
 p^2/2m -\mu(r)\rbrack/k_B T}}  \, d{\bf p},
\end{equation}
\begin{equation}
\label{n26u}
P(r)=\frac{g}{3h^3}\int  \frac{p^2/m}{1+e^{\lbrack
p^2/2m-\mu(r)\rbrack/k_B T}}  \, d{\bf p}=\frac{g}{h^3}k_B
T\int \ln\left (1+e^{-\lbrack
 p^2/2m-\mu(r)\rbrack/k_B T}\right )  \, d{\bf p}.
\end{equation}
On the other hand, Eq. (\ref{n29}) reduces to
\begin{equation}
\label{n29red}
dP=n \, d\mu.
\end{equation}
On the other
hand,
eliminating formally $\mu(r)$ between Eqs.
(\ref{n24u}) and (\ref{n26u}), we see that the equation of state of the Fermi
gas at statistical equilibrium is
barotropic: $P(r)=P[n(r),T]$ (see Appendix \ref{sec_scaling} for a more explicit
expression). Therefore,
according to Eq. (\ref{n29red}) we
have
$\mu(r)=\mu[n(r),T]$ with
\begin{equation}
\label{alt9t}
\mu'(n,T)=\frac{P'(n,T)}{n} \quad \Rightarrow
\quad \mu(n,T)=\int^{n}\frac{P'(n',T)}{n'}\, dn',
\end{equation}
and
\begin{equation}
\label{n29redb}
\frac{dP}{dr}=n \frac{d\mu}{dr}.
\end{equation}
In Eq. (\ref{alt9t}), the derivative is with respect to the
variable $n$.

\subsubsection{Variations on $\delta n$}

The vanishing of Eq. (\ref{n32}) with respect to variations on $\delta n$ gives
\begin{eqnarray}
\label{n34}
\frac{\mu(r)}{k_B T}=\alpha_0-\beta m\Phi(r).
\end{eqnarray}
Using the second relation of Eq. (\ref{n21}), becoming 
\begin{eqnarray}
\label{n34b}
\alpha(r)=\frac{\mu(r)}{k_B
T},
\end{eqnarray}
we can rewrite Eq. (\ref{n34}) as
\begin{eqnarray}
\label{n35}
\alpha(r)=\alpha_0-\beta m\Phi(r), \qquad {\rm or}\qquad \mu(r)=\mu_0-
m\Phi(r)
\end{eqnarray}
with
$\mu_0=\alpha_0 k_B T$. The chemical potential $\mu(r)$ is not uniform at
statistical equilibrium when a
gravitational potential is present. However, the {\it total} chemical
potential $\mu_{\rm tot}(r)\equiv \mu(r)+m\Phi(r)$ is uniform  at statistical
equilibrium ($\mu_{\rm tot}(r)=\mu_0$). This is the
Gibbs law.
Taking the derivative of Eq. (\ref{n34}) with respect to $r$ and using Eq.
(\ref{n4}) we get
\begin{eqnarray}
\label{n34r}
\frac{d\mu}{dr}=- m\frac{d\Phi}{dr}=-\frac{GM(r)m}{r^2}.
\end{eqnarray}
On the other hand, from Eqs. (\ref{n4b}) and (\ref{n35}), we obtain
\begin{eqnarray}
\label{n35b}
\mu(R)=\mu_0+\frac{GMm}{R}.
\end{eqnarray}

\subsubsection{Condition of hydrostatic equilibrium}
\label{sec_chd}

Combining Eqs. (\ref{n29redb}) and (\ref{n34r}), we obtain the condition of
hydrostatic equilibrium
\begin{equation}
\label{henq}
\frac{dP}{dr}=- \rho\frac{d\Phi}{dr}=-\rho(r)\frac{GM(r)}{r^2}.
\end{equation}
Therefore, the condition of statistical equilibrium, obtained by
extremizing the 
entropy at fixed energy and particle number, implies the condition of
hydrostatic equilibrium. This condition was not assumed in the preceding
calculations. It results from the thermodynamical variational principle
(maximization of entropy at fixed energy and particle number). The intrinsic
reason of this result will be given in Sec. \ref{sec_dsnv}.

\subsubsection{Entropy}

Using Eqs. (\ref{n33b}) and (\ref{n35}), the entropy density 
(\ref{n28b}) can be rewritten as
\begin{equation}
\label{n41}
s(r)=\frac{\frac{5}{3}\epsilon_{\rm
kin}(r)-\mu_0 n(r)+\rho(r)\Phi(r)}{T}.
\end{equation}
Integrating Eq. (\ref{n41}) over the whole configuration, we find that the
entropy is given at statistical equilibrium  by 
\begin{equation}
\label{n46}
S=-\frac{\mu_0}{T}N+\frac{5E_{\rm kin}}{3T}+\frac{2W}{T}.
\end{equation}
We emphasize that the results derived in this section are valid
for an
arbitrary form
of entropy.

\subsection{Canonical ensemble: Minimization of the free energy at
fixed particle number}

In the previous sections, we worked in the microcanonical ensemble in which the
particle number and the energy are fixed. We now consider the canonical ensemble
where the system is in contact with a heat
bath fixing the temperature $T$. In that case, the relevant thermodynamical
potential is the free energy
\begin{equation}
\label{n18}
F=E-T S.
\end{equation}
In the canonical
ensemble, the statistical
equilibrium state of the system
is obtained by minimizing the Fermi-Dirac free energy $F$ at fixed particle
number $N$:
\begin{eqnarray}
\min\ \lbrace F \, |\,  N \,\, {\rm fixed} \rbrace.
\end{eqnarray}
This determines the ``most probable'' state of a system in contact with a
thermal bath.

Minimizing the free energy $F=E-TS$ at fixed $N$ is equivalent to
maximizing the Massieu function $J=S/k_B-\beta E$ at fixed $N$ (the Massieu
function is the Legendre transform of the entropy with respect to the energy).
To solve this maximization problem we proceed in two steps. We first
maximize the Massieu function  $J$  at fixed kinetic energy density
$\epsilon_{\rm kin}(r)$ and particle number density $n(r)$ under
variations of $f({\bf r},{\bf p})$. Since the kinetic energy
density $\epsilon_{\rm kin}(r)$ determines the kinetic energy $E_{\rm kin}$
and since the  particle number
density $n(r)$ determines the potential energy $W$,
this is equivalent to maximizing the entropy $S$  at fixed $\epsilon_{\rm
kin}(r)$ and $n(r)$. This returns the results of Sec. \ref{sec_lmnr}. Using
these results, we can express the Massieu function $J=S/k_B-\beta E$ in terms of
$\epsilon_{\rm kin}(r)$ and $n(r)$. We now
maximize $J$ at
fixed particle number $N$ under
variations of $\epsilon_{\rm kin}(r)$ and $n(r)$. The first variations
(extremization) can be
written
as
\begin{equation}
\label{jn}
\delta \left (\frac{S}{k_B}-\beta E\right )+\alpha_0\delta N=0.
\end{equation}
Since $\beta$ is a constant, this variational principle is equivalent to Eq.
(\ref{n31}) so we get the
same results as in Sec. \ref{sec_gmnr} (for the first variations). Finally,
using Eqs. (\ref{n46}) and (\ref{n18}), we find that the
free energy is given at statistical equilibrium 
by
\begin{equation}
\label{n47}
F=\mu_0 N-W-\frac{2}{3}E_{\rm kin}.
\end{equation}

\subsection{Equations determining the statistical equilibrium state}

In this section, we provide the equations that determine the statistical
equilibrium state of a gas of self-gravitating fermions. These equations can be
easily extended to a distribution function arising from a generalized
form of entropy (see Appendix \ref{sec_osts}).

\subsubsection{Local variables in terms of $\Phi(r)$}

Substituting Eqs. (\ref{n33}) and (\ref{n34}) into Eqs.
(\ref{n22}) and (\ref{n24})-(\ref{n26}), we obtain
\begin{equation}
\label{n37}
f({\bf r},{\bf p})=\frac{g}{h^3}\frac{1}{1+e^{-\alpha_0}e^{\beta
(p^2/2m+m\Phi(r))}},
\end{equation}
\begin{equation}
\label{n38}
n(r)=\frac{g}{h^3}\int \frac{1}{1+e^{-\alpha_0}e^{\beta ({p}^2/2m+m\Phi(r))}}\,
{\rm
d}{\bf p},
\end{equation}
\begin{equation}
\label{n39}
\epsilon_{\rm kin}(r)=\frac{g}{h^3}\int
\frac{{p^2}/{2m}}{1+e^{-\alpha_0}e^{\beta
({p}^2/2m+m\Phi(r))}}\, {\rm
d}{\bf p},
\end{equation}
\begin{equation}
\label{n40}
P(r)=\frac{g}{3h^3}\int \frac{{p^2}/{m}}{1+e^{-\alpha_0}e^{\beta
({p}^2/2m+m\Phi(r))}}\, {\rm
d}{\bf p}=\frac{g}{h^3}k_B T\int \ln\left
\lbrack 1+e^{\alpha_0}e^{-\beta(p^2/2m+m\Phi(r))}\right \rbrack \, d{\bf
p}.
\end{equation}
These equations determine the statistical equilibrium state of
the self-gravitating Fermi gas. Taking the derivative of the
pressure from Eq. (\ref{n40}) with
respect to $r$, and using Eq. (\ref{n38}), we recover the condition of
hydrostatic equilibrium (\ref{henq}).  We show in Appendix \ref{sec_heang}
that this
result remains valid for an arbitrary form of entropy.

\subsubsection{Equation for $\rho(r)$}

The density of particles (\ref{n38})  and the pressure (\ref{n40})  are related
to the gravitational potential by
\begin{equation}
\rho(r)=\frac{4\pi g\sqrt{2}m^{5/2}}{h^3\beta^{3/2}}I_{1/2}\left
\lbrack e^{-\alpha_{0}+\beta m\Phi(r)}\right \rbrack,
\label{cp1}
\end{equation}
\begin{equation}
P(r)=\frac{8\pi g\sqrt{2}m^{3/2}}{3h^3\beta^{5/2}}I_{3/2}\left
\lbrack e^{-\alpha_{0}+\beta m\Phi(r)}\right \rbrack,
\label{cp2}
\end{equation}
where $I_{n}$ denotes the Fermi integral
\begin{equation}
I_{n}(t)=\int_{0}^{+\infty}\frac{x^{n}}{1+t e^{x}}\, dx.
\label{cp3}
\end{equation}
We recall the identity
\begin{equation}
I'_{n}(t)=-\frac{n}{t}I_{n-1}(t), \qquad (n>0),
\label{cp4}
\end{equation}
which can be established from Eq. (\ref{cp3}) by an integration by parts. 
Eqs. (\ref{cp1}) and
(\ref{cp2}) determine
the equation of state $P(r)=P[\rho(r),T]$ of the nonrelativistic Fermi gas at
finite temperature in parametric form with parameter $\alpha(r)=\alpha_0-\beta
m\Phi(r)$. Substituting this equation of state into the fundamental
differential equation of hydrostatic equilibrium (\ref{n6}), we obtain a
differential equation for the density profile of the form 
\begin{equation}
\label{dir}
\frac{1}{r^2}\frac{d}{dr}\left (\frac{r^2}{\rho}\frac{dP(\rho)}{dr}\right
)=-4\pi G \rho,
\end{equation}
where $P(\rho)$ is the Fermi-Dirac equation of state.

\subsubsection{Equation for $\Phi(r)$}

The distribution function of a system of
self-gravitating fermions at statistical equilibrium is given by Eq.
(\ref{n37}). Substituting this distribution function into the Poisson equation
(\ref{n2}), using Eq. (\ref{n7}), we
obtain a differential equation for the gravitational potential of the form
\begin{equation}
\label{dim1}
\frac{1}{r^2}\frac{d}{dr}\left (r^2\frac{d\Phi}{dr}\right )=4\pi G
\frac{gm}{h^3}\int \frac{1}{1+e^{-\alpha_0}e^{\beta
(p^2/2m+m\Phi(r))}}\, d{\bf p}.
\end{equation}
It is called the Fermi-Dirac-Poisson equation or the Thomas-Fermi
equation at finite temperature. Using Eq. (\ref{cp1}), it 
can be rewritten as 
\begin{equation}
\label{cp5}
\frac{1}{r^2}\frac{d}{dr}\left (r^2\frac{d\Phi}{dr}\right )=\frac{16\pi^2
g\sqrt{2}Gm^{5/2}}{h^3\beta^{3/2}}I_{1/2}\left
\lbrack e^{-\alpha_{0}+\beta m\Phi(r)}\right \rbrack.
\end{equation}
Once the gravitational potential $\Phi(r)$ has been determined by solving Eq.
(\ref{cp5}), the density
$\rho(r)$ is given by  Eq. (\ref{cp1}). The general procedure to obtain the
density profiles and the caloric curves of the self-gravitating Fermi gas at
finite temperature from the differential equation (\ref{cp5}) is explained in
detail in \cite{ptdimd}. The study of phase transitions in the nonrelativistic
self-gravitating Fermi gas has been performed
in
\cite{ht,bvn,csmnras,robert,pt,dark,ispolatov,rieutord,ptdimd,ijmpb}.\footnote{
Refs.
\cite{ht,bvn} work in the canonical ensemble while
Refs. \cite{csmnras,robert} work in
the microcanonical ensemble. Refs.
\cite{pt,dark,ispolatov,rieutord,ptdimd,ijmpb} consider both
canonical and microcanonical ensembles.} The complete
canonical and microcanonical phase diagrams are given in \cite{ijmpb}. Phase
transitions in the fermionic King model have been studied in \cite{clm2}.

\subsubsection{Equation for fugacity $z(r)$}

Introducing the fugacity 
\begin{equation}
\label{cp6}
z(r)=e^{\alpha(r)}=e^{\mu(r)/k_B T}
\end{equation}
and using Eq. (\ref{n34}) giving 
\begin{equation}
\label{cp6}
z(r)=e^{\alpha_0}e^{-\beta m\Phi(r)},
\end{equation}
we can rewrite the Fermi-Dirac distribution function (\ref{n37}) as
\begin{equation}
\label{cp7}
f({\bf r},{\bf p})=\frac{g}{h^3}\frac{1}{1+\frac{1}{z(r)}e^{\beta
p^2/2m}}.
\end{equation}
On the other hand, the density and the pressure from Eqs. (\ref{cp1}) and
(\ref{cp2}) can be expressed in terms of $z$ as
\begin{equation}
\rho(r)=\frac{4\pi g\sqrt{2}m^{5/2}}{h^3\beta^{3/2}}I_{1/2}\left
\lbrack\frac{1}{z(r)}\right \rbrack,
\label{cp8}
\end{equation}
\begin{equation}
P(r)=\frac{8\pi g\sqrt{2}m^{3/2}}{3h^3\beta^{5/2}}I_{3/2}\left
\lbrack\frac{1}{z(r)}\right \rbrack.
\label{cp9}
\end{equation}
Taking the derivative of Eq. (\ref{cp9}) with respect to $r$ and using Eq.
(\ref{cp4}), we obtain
\begin{equation}
P'(r)=\rho(r)\frac{k_B T}{m}\frac{z'(r)}{z(r)}.
\label{cp10}
\end{equation}
Starting from the condition of hydrostatic
equilibrium (\ref{n5}), using Eq. (\ref{cp10}), multiplying the resulting
expression by $r^2$, taking the derivative with respect to $r$ and using Eqs.
(\ref{n3}) and (\ref{cp8}), we finally obtain the following differential
equation for $z(r)$:
\begin{equation}
\frac{z''}{z}+\frac{2}{r}\frac{z'}{z}-\left (\frac{z'}{z}\right
)^2+\frac{16\pi^2g\sqrt{2}Gm^{7/2}}{h^3\beta^{1/2}}I_{1/2}\left
(\frac{1}{z}\right)=0.
\label{p11}
\end{equation}
We can obtain the density profiles and the caloric curves of
the self-gravitating Fermi gas at finite temperature from the differential
equation (\ref{p11}) as done in \cite{muller}.

\subsection{Thermodynamical stability and ensembles inequivalence}
\label{sec_tsei}

We have seen that the statistical equilibrium states in the
microcanonical and 
canonical ensembles are the same. Indeed, the extrema of entropy at fixed
energy and particle number coincide with the extrema of free energy at fixed
particle number. This is a general result of statistical mechanics which is
due to the fact that the first order variational problems (\ref{n31}) and
(\ref{jn}) coincide \cite{cc}. 
However, the thermodynamical {\it stability} of these equilibrium states, which
is
related to the sign of the
second order variations of the appropriate thermodynamical potential (entropy or
free energy),
may be
different in the microcanonical and canonical ensembles. In this sense, the
statistical ensembles are inequivalent for self-gravitating systems
\cite{paddy,katzrevue,ijmpb}. This
is a specificity
of systems
with long-range interactions whose energy is nonadditive \cite{cdr,campabook}. 
It can be shown that canonical stability implies
microcanonical stability while the reciprocal is wrong \cite{cc}. Therefore,
there are more stable equilibrium states in the microcanonical ensemble than in
the canonical ensemble.\footnote{For example, equilibrium states
with a negative
specific heat may be stable in the microcanonical ensemble while they are always
unstable in the canonical ensemble \cite{lbw,thirring} (see Appendix B of
\cite{acb}).} Basically, this is
because the microcanonical ensemble is more constrained (because of the
conservation of energy) than the canonical ensemble.

The thermodynamical stability of an equilibrium state can be investigated by
studying the sign of the second order variations of the appropriate
thermodynamical potential (entropy or free energy) and reducing this study to an
eigenvalue
problem. This method allows us to determine the form of the perturbation
that triggers the thermodynamical instability. We refer to
\cite{antonov,paddyapj,paddy,sc,aaiso,grand} for a detailed discussion of
this stability problem in the case of classical self-gravitating systems.

On the other hand, the thermodynamical stability of the system can be directly
settled from the topology of the series of equilibria $\beta(-E)$ by using the
Poincar\'e
criterion \cite{poincare} (see \cite{lbw,katzpoincare1,ijmpb} for some
applications of this
method in relation to the statistical mechanics of self-gravitating systems):

(i) In the microcanonical ensemble, if we plot $\beta(-E)$ at fixed $N$, a 
change
of stability can occur only at a turning point of energy. A mode of stability is
lost if the curve  $\beta(-E)$ rotates clockwise and gained if it rotates
anticlockwise. Since $S$ and $E$ reach their extrema at the
same points (in view of the fact that $\delta S/k_B=\beta\delta E$), the curve
$S(E)$ displays spikes at
its extremal points.

(ii) In the canonical ensemble, if we plot  $\beta(-E)$ at fixed $N$, a
change of stability can occur only at a turning point of temperature. A mode of
stability is lost if the curve  $\beta(-E)$  rotates clockwise and gained if it
rotates anticlockwise. Since $J$ and $\beta$ reach their
extrema at the same points (in view of the fact that $\delta J=-E\delta \beta$),
the curve $J(\beta)$ displays spikes at its extremal points. We can also
consider the curve $\alpha_0(N)$ at fixed $T$. If we plot $\alpha_0(N)$, a 
change of stability
can  occur only at a turing point of particle number $N$. A mode of stability is
lost if
the
curve $\alpha_0(N)$  rotates clockwise and gained if it rotates anticlockwise.
Since $J$ and $N$ reach their extrema at the same
points (in view of the fact that $\delta J=-\alpha_0\delta N$),
the curve $J(N)$ displays spikes at its extremal points.

\subsection{Dynamical stability}
\label{sec_dsn}

\subsubsection{Vlasov-Poisson equations}
\label{sec_dsnv}

The distribution function $f({\bf r},{\bf v})$ of a system of
self-gravitating fermions at statistical equilibrium is given by Eq.
(\ref{n37}). It is of the form
$f=f(\epsilon)$ with $f'(\epsilon)<0$, where $\epsilon=v^2/2+\Phi(r)$ is the
energy of a particle by unit of mass. 
Therefore, at statistical equilibrium, the
distribution function depends only on the individual energy $\epsilon$
of the particles  and is monotonically decreasing. It is
shown in Appendices
\ref{sec_osts} and \ref{sec_hea} that these properties remain valid for a
general form of 
entropy. Since $f({\bf r},{\bf v})$ is a function of
the energy
$\epsilon$,  which is a constant of the motion, it is a particular steady state
of the Vlasov-Poisson equations.
This is a special case of the Jeans theorem \cite{jeansth}.\footnote{According
to the Jeans theorem \cite{jeansth}, a spherical stellar system in collisionless
equilibrium has a distribution function of the form $f=f(\epsilon,L)$ where
$\epsilon$ is the energy and $L$ is the angular momentum. We note that an
extremum of entropy $S$ at fixed energy $E$ and particle number $N$ leads to a
distribution function that
depends only on $\epsilon$, not on $L$. Therefore, an extremum of
entropy at fixed energy and particle number is necessarily isotropic.}
Therefore, a
statistical equilibrium state (extremum of entropy at fixed energy and particle
number) is a steady state of the Vlasov-Poisson equations.  Furthermore, it
can be
shown  that thermodynamical stability implies
dynamical stability with respect to the Vlasov-Poisson equations \cite{ih}.
Therefore, a stable thermodynamical equilibrium state (maximum of entropy at
fixed
energy and particle
number) is always dynamically
stable.\footnote{This result is very general  (being valid for an arbitrary
entropic functional and for any
long-range potential of interaction) and stems from the fact
that the entropy (which is a particular Casimir), the energy and the particle
number are
conserved by the
Vlasov-Poisson equations (see \cite{cc} and Appendix \ref{sec_dsv}).} In
general, the reciprocal is wrong.
This is the case  in Newtonian gravity.
Indeed, it can be
shown \cite{doremus71,doremus73,gillon76,sflp,ks,kandrup91} that {\it all} the
distribution functions of
the form $f=f(\epsilon)$ with $f'(\epsilon)<0$ are dynamically stable with
respect to the Vlasov-Poisson equations. As
a result, in Newtonian gravity, all
the statistical 
equilibrium states (i.e. all the extrema of entropy at fixed energy and particle
number) are
dynamically stable, even those that are thermodynamically
unstable.

\subsubsection{Euler-Poisson equations}
\label{sec_dsne}

We have seen that the statistical equilibrium state of a system of
self-gravitating fermions is described by a barotropic equation of state
of the form $P(r)=P[\rho(r)]$ and that it 
satisfies the condition of hydrostatic equilibrium (\ref{n1}). It is shown in
Appendices \ref{sec_osts} and \ref{sec_hea} that these properties remain
valid for an arbitrary form of entropy. As a result, the system is in a
steady state of the
Euler-Poisson equations.  Furthermore, it can be
shown (see \cite{aaantonov} and Appendices \ref{sec_altce} and \ref{sec_dseNG})
that
the
thermodynamical stability of a self-gravitating system  in the canonical
ensemble is equivalent to its dynamical stability with
respect to the Euler-Poisson equations.

\subsubsection{Kinetic  equations}

It can also be shown that the thermodynamical stability of a self-gravitating
system in the microcanonical ensemble is equivalent to its dynamical stability
with respect to the Landau-Poisson equations \cite{gen2}
and
that the
thermodynamical stability of a self-gravitating system in the canonical ensemble
is equivalent to its dynamical stability with respect to the 
Kramers-Poisson, damped
Euler-Poisson, and  Smoluchowski-Poisson equations
\cite{gen,gen2,sc,nfp}. These results are natural since these kinetic equations
describe
 the thermodynamical (secular) evolution of the system in the microcanonical and
canonical ensembles respectively. In particular, they satisfy an
$H$-theorem for the entropy in the microcanonical ensemble and for the free
energy in the canonical ensemble.\footnote{This is at variance with the Vlasov
and Euler equations which do not
satisfy an $H$-theorem.} Using Lyapunov's direct method
it can be shown that they relax towards a stable thermodynamical equilibrium
state.

\subsection{Particular limits}

The Fermi-Dirac distribution (\ref{n37}) can be written as
\begin{equation}
\label{zw1}
f({\bf r},{\bf p})=\frac{g}{h^3}\frac{1}{1+e^{\lbrack E_{\rm
kin}(p)-\mu(r)\rbrack/k_B
T}},
\end{equation}
where $E_{\rm kin}=p^2/2m$ and $\mu(r)=\mu_0-m\Phi(r)$. Let us consider
particular limits of this distribution function.

\subsubsection{The completely degenerate Fermi gas (ground state)}

The completely degenerate limit corresponds to   $T\rightarrow
0$, $\mu(r)>0$
finite, and $\alpha(r)=\mu(r)/k_B T\rightarrow +\infty$. In
that case, the chemical potential is positive and large compared to the
temperature. This yields the Fermi distribution (or Heaviside
function):
\begin{eqnarray}
\label{zw3}
f({\bf r},{\bf
p})=\frac{g}{h^3}\quad {\rm if}\quad E_{\rm kin}(p)< E_{F}(r) \quad (p<
p_F(r)),
\end{eqnarray}
\begin{eqnarray}
\label{zw4}
f({\bf r},{\bf
p})=0 \quad {\rm if}\quad E_{\rm kin}(p)> E_{F}(r) \quad (p> p_F(r)),
\end{eqnarray}
where 
\begin{equation}
\label{zw2b}
E_F(r)=\mu(r)=\mu_0-m\Phi(r),
\end{equation}
and
\begin{equation}
\label{zw2}
p_F(r)=\sqrt{2m\mu(r)}=\sqrt{2m(\mu_0-m\Phi(r))}
\end{equation}
are the Fermi energy and the Fermi impulse.  The density and the pressure are
given by
\begin{equation}
\rho=\int
fm\, d{\bf
p}=\int_{0}^{p_{F}}\frac{g}{h^3}m 4\pi p^{2}\, dp=\frac{4\pi g
m}{3h^3}p_{F}^3({r}),
\label{zw5}
\end{equation}
\begin{equation}
P=\frac{1}{3}\int
f \frac{p^2}{m}\, d{\bf
p}=\frac{1}{3}\int_{0}^{p_{F}}\frac{g}{h^3}\frac{1}{m}p^2 4\pi p^{2}\,
dp=\frac{4\pi g}{15 m h^3}p_{F}^5({r}).
\label{zw6}
\end{equation}
Eliminating the Fermi impulse between these two expressions, we find that  the
equation of state of the nonrelativistic Fermi gas at $T=0$ is
\begin{equation}
P=K\rho^{5/3}, \qquad K=\frac{1}{5}\left (\frac{3h^3}{4\pi g m^4}\right )^{2/3}.
\label{zw7}
\end{equation}
This is the equation of state of a polytrope of index $\gamma=5/3$ ({\it i.e.}
$n=3/2$). Substituting Eq. (\ref{zw7}) into Eq. (\ref{dir}), the fundamental
equation of hydrostatic equilibrium can be written as
\begin{equation}
\label{dirb}
\frac{1}{r^2}\frac{d}{dr}\left (r^2\frac{d\rho^{2/3}}{dr}\right
)=-\frac{8\pi G}{5K} \rho.
\end{equation}
Alternatively, substituting Eq. (\ref{zw5}) with the relation from Eq.
(\ref{zw2}) into the
Poisson equation (\ref{n2}), we obtain the Thomas-Fermi
equation
\begin{equation}
\label{n2b}
\frac{1}{r^2}\frac{d}{dr}\left (r^2\frac{d\Phi}{dr}\right )=\frac{16\pi^2
g G m}{3h^3}\left\lbrack 2m(\mu_0-m\Phi(r))\right\rbrack^{3/2}.
\end{equation}
These equations can both be reduced to the Lane-Emden equation
\cite{chandrabook}. For a given value of $\rho_0=\rho(0)$ or $\Phi_0=\Phi(0)$,
we can solve these
equations until the point where the density vanishes:
$\rho(R)=0$. This
determines the radius $R$ of the configuration (when
$T=0$ we do not need a box to confine the system). We can then compute the
corresponding mass
$M$.\footnote{Since $\rho(R)=0$ we find that
$\mu_0=m\Phi(R)=-GMm/R$. Therefore, $\mu_0<0$.} By
varying $\rho_0$ or $\Phi_0$, we get the mass-radius relation $M(R)$. The 
mass-radius relation and the corresponding density profiles of the
self-gravitating Fermi gas at $T=0$ are given in 
\cite{chandrabook}.

{\it Remark:} For the self-gravitating Fermi gas at $T=0$, the free energy
$F$ reduces to the energy $E=E_{\rm kin}+W$. Therefore, a stable equilibrium
state at $T=0$ is a minimum of $E$ at fixed $N$ (ground state). We have seen
that the
nonrelativistic Fermi gas at $T=0$ is described by the polytropic equation of
state (\ref{zw7}). Using Eqs. (\ref{n11}) and (\ref{zw7}) the energy can be
written as $E=E_{\rm
kin}+W$ where $E_{\rm
kin}=\frac{3}{2}\int P\, d{\bf r}=\frac{3}{2}K\int\rho^{5/3}\, d{\bf r}$ is
the kinetic energy. This is the same as the energy ${\cal W}=U+W$ of a
polytropic gas of index $n=3/2$, where $U=\int \rho\int^{\rho}
\frac{P(\rho')}{{\rho'}^2}\,
d\rho'\,
d{\bf
r}=\frac{3}{2}K\int\rho^{5/3}\, d{\bf r}$ is the internal energy (see
Appendix \ref{sec_dseNG}).\footnote{See Appendix C of \cite{wddimd} for a more
general discussion.} Therefore, for
the 
self-gravitating Fermi gas at $T=0$, we explicitly recover the fact that the
condition of thermodynamical stability 
coincides with the condition of dynamical stability with respect to the
Euler-Poisson equations (see Sec. \ref{sec_dsne}).

\subsubsection{The nondegenerate Fermi gas (classical limit)}

The nondegenerate (classical) limit corresponds to 
\begin{equation}
\label{mb2}
\frac{E_{\rm kin}(p)-\mu(r)}{k_B T}\gg 1 \quad {\it i.e.} \quad \beta
E_{\rm kin}(p)-\alpha(r)\gg
1.
\end{equation}
This yields the Maxwell-Boltzmann distribution
\begin{equation}
\label{mb3}
f({\bf r},{\bf p})=\frac{g}{h^3} e^{-\lbrack
E_{\rm kin}(p)-\mu(r)\rbrack/k_B
T}.
\end{equation}
The condition from Eq. (\ref{mb2}) is always fulfilled when
$\alpha(r)\rightarrow
-\infty$,
whatever the value of $\beta E_{\rm kin}(p)$. Therefore, the
condition
$\mu(r)\rightarrow -\infty$, $T$ finite and $\alpha(r)=\mu(r)/k_B
T\rightarrow -\infty$ implies the nondegenerate (classical)
limit. In that case, the chemical
potential is negative and large (in absolute value) as compared to the
temperature. However, this is not the only case where the Maxwell-Boltzmann
distribution is valid. We can be in the classical limit for arbitrary
values of $\alpha(r)$ (positive or negative) provided that $\beta
E_{\rm kin}(p)-\alpha(r)\gg 1$.
The classical limit is specifically studied in Paper II.

\section{Statistical mechanics of general relativistic fermions}
\label{sec_grf}

In this section, we consider the statistical mechanics of self-gravitating 
fermions within the framework of  general relativity. We use a presentation
similar to the one developed in Sec. \ref{sec_smnrf} to treat  the
statistical mechanics of self-gravitating 
fermions within the framework of Newtonian gravity.

\subsection{Hydrostatic equilibrium of gaseous spheres
in general relativity}
\label{sec_hegr}

\subsubsection{Einstein equations}
\label{sec_ee}

The Einstein field equations of general
relativity are expressed as  
\begin{equation}
R_{\mu\nu}-{1\over 2}g_{\mu\nu}R=-{8\pi G\over c^{4}} T_{\mu\nu},
\label{n48}
\end{equation}
where $R_{\mu\nu}$ is the Ricci tensor, $T_{\mu\nu}$ is the energy-momentum
tensor
and $g_{\mu\nu}$ is the metric tensor defined by
\begin{equation}
ds^{2}=-g_{\mu\nu}dx^{\mu}dx^{\nu},
\label{n49}
\end{equation}
where $ds$ is the invariant interval between two neighbouring space-time
events. 

In the following, we shall restrict ourselves to spherically symmetric systems
with motions, if any, only in the radial directions.
Under these assumptions,
the metric can be written in the form
\begin{equation}
ds^{2}=e^{\nu}c^2 dt^{2}-r^{2}(d\theta^{2}+\sin^{2}\theta
d\phi^{2})-e^{\lambda}dr^{2},
\label{n50}
\end{equation}
where $\nu$ and $\lambda$ are functions of $r$ and $t$ only. The
energy-momentum tensor is assumed to be that for a perfect fluid
\begin{equation}
T^{\mu\nu}=P g^{\mu\nu}+(P+\epsilon)u^{\mu}u^{\nu},
\label{n51}
\end{equation}
where $u^{\mu}=dx^{\mu}/ds$ is the fluid four-velocity, $P$ is the isotropic 
pressure and $\epsilon$ is the energy density including the rest mass.
The mass contained within a sphere of radius $r$ is
\begin{equation}
M(r)=\frac{1}{c^2}  \int_{0}^{r}\epsilon(r)  4\pi r^{2}\,
dr \quad \Rightarrow \quad \frac{dM}{dr}=\frac{\epsilon}{c^2}4\pi r^2.
\label{n52}
\end{equation}
The total mass is
\begin{equation}
\label{n53}
M=\frac{1}{c^2}\int_0^R \epsilon(r)4\pi r^2\, dr,
\end{equation}
where $R$ is the size of the system. The mass-energy is ${\cal E}=Mc^2$.
In the nonrelativistic limit
$c\rightarrow
+\infty$, using $\epsilon\sim \rho c^2$ where $\rho c^2$ is
the rest-mass energy (see below), Eqs. (\ref{n52}) and (\ref{n53}) return Eqs.
(\ref{n3}) and (\ref{n14}).

\subsubsection{TOV equations}
\label{sec_tov}

The equations of general relativity governing the hydrostatic equilibrium of a
spherical distribution of matter are well-known. They are given by (see, e.g,
\cite{weinberg}):
\begin{equation}
{d\over dr}(re^{-\lambda})=1-{8\pi G\over c^{4}}r^{2}\epsilon,
\label{n56}
\end{equation}
\begin{equation}
\frac{dP}{dr}=-\frac{1}{2}(\epsilon+P)\frac{d\nu}{dr},
\label{n57}
\end{equation}
\begin{equation}
{e^{-\lambda}\over r}{d\nu\over dr}={1\over r^{2}}(1-e^{-\lambda})+{8\pi G\over
c^{4}}P.
\label{n58}
\end{equation}
These equations can be deduced from the Einstein equations (\ref{n48}). However,
Eq. (\ref{n57}) can be obtained more directly from the local law of
energy-momentum conservation, $D_{\mu}T^{\mu\nu}=0$, which is also contained in
the Einstein equations. It can be interpreted as the condition
of hydrostatic equilibrium in general relativity.  This equation was first
derived and
emphasized by
Tolman \cite{tolman}. In
the nonrelativistic limit $c\rightarrow +\infty$ it reduces to Eq. (\ref{n1})
(see Sec. \ref{sec_rnp}).

Equations (\ref{n56})-(\ref{n58}) can be combined to give
\begin{equation}
\label{n58b}
\frac{{\rm
d}P}{dr}=-\frac{\epsilon+P}{c^2}\frac{\frac{GM(r)}{r^2}+\frac{4\pi
G}{c^2}Pr}{1-\frac{2GM(r)}{r c^2}},
\end{equation}
where $M(r)$ is given by Eq. (\ref{n52}). This equation was first derived by
Oppenheimer and Volkoff \cite{ov}. It
extends the
classical condition of hydrostatic equilibrium for a star to the context of
general relativity.  In the
nonrelativistic limit $c\rightarrow +\infty$, using
$\epsilon\sim \rho c^2\gg P$, Eq. (\ref{n58b}) reduces to Eq. (\ref{n5}).

\subsubsection{Metric functions}
\label{sec_mf}

Integrating Eq. (\ref{n56}) with the condition $\lambda(r)\rightarrow 0$ at
infinity, we obtain
\begin{equation}
e^{-\lambda(r)}=1-\frac{2GM(r)}{rc^{2}}.
\label{n59}
\end{equation}
Then,  Eq. (\ref{n58}) can be rewritten as
\begin{equation}
{d\nu\over dr}={1+4\pi P r^{3}/M(r)c^{2}\over r  ({rc^{2}/ 2 GM(r)}-1
)}.
\label{n60}
\end{equation}
These equations determine the metric functions $\lambda(r)$ and $\nu(r)$. Eq.
(\ref{n60}) can be interpreted as a generalization of Newton's law.  In
the nonrelativistic limit $c\rightarrow +\infty$ it reduces to Eq. (\ref{n4})
(see Sec. \ref{sec_rnp}).

In the empty space outside the star, $P=\epsilon=0$. Therefore, if $M=M(R)$
denotes the mass-energy of the star, Eqs. (\ref{n59}) and (\ref{n60})
become
for $r>R$:
\begin{equation}
e^{-\lambda(r)}=1-{2GM\over rc^{2}} \qquad {\rm and}\qquad {d\nu\over
dr}={1\over
r ({rc^{2}/ 2
GM}-1 )}.
\label{n62}
\end{equation}
The second equation is readily integrated into 
\begin{equation}
\nu(r)=\ln\left (1-\frac{2GM}{rc^{2}}\right ),
\label{n63}
\end{equation}
where we have taken the constant of integration to be zero by convention. In
this manner $\nu(r)\rightarrow 0$ when $r\rightarrow +\infty$. We note that
$\nu(r)=-\lambda(r)$. From
these equations, we get
\begin{equation}
e^{-\lambda(R)}=1-{2GM\over Rc^{2}}  \qquad {\rm and}\qquad
\nu(R)=\ln\left
(1-\frac{2GM}{Rc^{2}}\right ).
\label{n64}
\end{equation}
Substituting the foregoing expressions for $\lambda$ and $\nu$ into Eq.
(\ref{n50}), we obtain the well-known Schwarzschild's form of the metric in the
empty space outside
a star \cite{schwarzschild1}:
\begin{eqnarray}
ds^{2}=\biggl (1-{2GM\over rc^{2}}\biggr
)c^2dt^{2}-r^{2}(d\theta^{2}+\sin^{2}\theta d\phi^{2})
-{dr^{2}\over 1-{2GM/rc^{2}}}.
\label{n65}
\end{eqnarray}
This form of the metric remains valid even if
the star is unsteady as long as it remains spherically symmetric
(Jebsen-Birkhoff theorem \cite{jebsen,birkhoff,deser}). In this respect,
Newton's theorem according to which the gravitational field external to a
spherical distribution of matter depends only on its total mass [see Eq.
(\ref{n4c})] is equally true in general relativity. The metric (\ref{n65})  is
singular at
\begin{equation}
r={2GM\over c^{2}}\equiv R_{S},
\label{n66}
\end{equation}
where $R_{S}$ is the Schwarzschild radius appropriate to the mass $M$. This does
not mean that spacetime is singular at that radius but only that this particular
metric is. Indeed, the singularity can be removed by a judicious change of
coordinate system \cite{eddington24,finkelstein,fronsdal,kruskal}. When
$R_{S}>R$,
the star is a
black hole and no particle or even light can leave the region $R<r<R_{S}$.
However, for a gaseous sphere in hydrostatic equilibrium the
discussion does not arise because $R_{S}<R$. Indeed, it can be shown that the
radius
of the configuration is necessarily restricted by the Buchdahl 
\cite{buchdahl}  inequality\footnote{This inequality was previously derived by
Schwarzschild \cite{schwarzschild2} in the case of equilibrium configurations
with uniform energy density.}
\begin{equation}
R\ge {9\over 8}\ {2GM\over c^{2}}={9\over 8}R_{S}.
\label{n67}
\end{equation}
Therefore, the points exterior to the star always satisfy $r>R_{S}$.

\subsection{Local variables}
\label{sec_lgr}

We consider a gas of relativistic fermions described by the  distribution
function $f({\bf r},{\bf p})$ such that $f({\bf r},{\bf p}) d{\bf r}d{\bf p}$
gives the number density of fermions at position ${\bf r}$ with impulse ${\bf
p}$. 
The local particle number density is
\begin{equation}
\label{n68}
n=\int f\, d{\bf p}.
\end{equation}
The energy density is
\begin{equation}
\label{n69}
\epsilon=\int f E(p) \, d{\bf p},
\end{equation}
where $E$ is the total (kinetic $+$ rest mass) energy of a particle given by 
\begin{equation}
\label{n70}
E(p)=\sqrt{p^2c^2+m^2c^4}.
\end{equation}
In can be written as
\begin{equation}
\label{n71}
E(p)=mc^2+E_{\rm kin}(p),
\end{equation}
where
\begin{equation}
\label{n71b}
E_{\rm kin}(p)=mc^2\left\lbrace \sqrt{\frac{p^2}{m^2c^2}+1}-1\right\rbrace
\end{equation}
is the kinetic energy of the particle. In
the
nonrelativistic limit $c\rightarrow +\infty$, the kinetic energy of a particle
reduces to Eq. (\ref{n9}). The
energy
density $\epsilon$ can be written
as
\begin{equation}
\label{n72}
\epsilon=\rho c^2+\epsilon_{\rm kin},
\end{equation}
where $\rho=nm$ is the rest-mass density and  
\begin{equation}
\label{n73}
\epsilon_{\rm kin}=\int f E_{\rm kin}(p) \, d{\bf p}
\end{equation}
is the
kinetic
energy density. In the
nonrelativistic limit $c\rightarrow +\infty$, we have $\epsilon\sim \rho
c^2$. The local pressure is given by \cite{chandrabook}
\begin{equation}
\label{n74}
P=\frac{1}{3}\int f p\frac{dE}{dp}\, d{\bf p}=\frac{1}{3}\int f
\frac{p^2c^2}{E(p)}\, d{\bf p}.
\end{equation}
In the
nonrelativistic limit $c\rightarrow +\infty$, using $E(p)\simeq mc^2+p^2/2m$, we
recover Eq. (\ref{n10}). Finally, the
Fermi-Dirac entropy density is given by 
\begin{eqnarray}
\label{n75}
s=-k_B\frac{g}{h^3}\int\Biggl\lbrace \frac{f}{f_{\rm max}}\ln \frac{f}{f_{\rm
max}}
+\left (1-\frac{f}{f_{\rm max}}\right )\ln\left
(1-\frac{f}{f_{\rm max}}\right )\Biggr\rbrace\, d{\bf p},
\end{eqnarray}
as in Sec. \ref{sec_lnr}.

\subsection{Global variables}
\label{sec_ggr}

The entropy of the fermion gas  is
given by
\begin{equation}
\label{n76}
S=\int_0^R s(r)\left \lbrack 1-\frac{2GM(r)}{rc^2}\right \rbrack^{-1/2}4\pi
r^2\, dr,
\end{equation}
where the entropy density $s(r)$ has been multiplied by the
proper volume element $e^{\lambda/2} 4\pi r^2\,
dr$ obtained by
using the expression (\ref{n59}) of the metric coefficient 
$\lambda(r)$.
Similarly, the particle number is
given by\footnote{Equation (\ref{n59}) is valid at
equilibrium (for a steady state). Therefore, the expressions (\ref{n76}) and
(\ref{n77}) of $S$ and $N$ are justified only at equilibrium. However, it is
shown in \cite{chandra64} that Eq. (\ref{n59}) remains valid for small
perturbations about equilibrium up to second order in the motion. This
justifies using
Eqs. (\ref{n76}) and (\ref{n77}) when we make perturbations  about
the equilibrium state
as in the variational problem considered below.
}
\begin{equation}
\label{n77}
N=\int_0^R n(r)\left \lbrack 1-\frac{2GM(r)}{rc^2}\right \rbrack^{-1/2}4\pi
r^2\, dr.
\end{equation}
We introduce the (binding) energy\footnote{The binding
energy is usually defined by $E_b=(Nm-M)c^2$ so that $E=-E_b$. It is the
difference between the
rest mass energy $Nmc^2$ (the energy that the matter of the star would have
if dispersed to infinity) and the total mass-energy $Mc^2$. In
order to simplify the discussion, we shall define the binding
energy as $E_b=(M-Nm)c^2$ so that $E=E_b$.}
\begin{equation}
\label{n78}
E=(M-Nm)c^2,
\end{equation}
where $Mc^2$ is the mass-energy given by Eq. (\ref{n53}). We note
that $M$, unlike $S$ and $N$, involves the coordinate volume,  {\it not} the
proper volume. In the
nonrelativistic limit $c\rightarrow +\infty$, the binding energy $E$ reduces to
the
Newtonian energy $E=E_{\rm
kin}+W$
including the kinetic energy and the potential energy (see Sec. \ref{sec_mneb}).

As in the Newtonian case, a statistical equilibrium state exists
only if the system is confined within a box of radius $R$ otherwise it would
evaporate. In the microcanonical ensemble, the mass-energy ${\cal E}=Mc^2$
and the particle number $N$  are conserved.  The statistical equilibrium state
of the system is obtained by maximizing the Fermi-Dirac entropy $S$ at fixed
mass-energy ${\cal E}=Mc^2$ and particle number $N$:
\begin{eqnarray}
\max\ \lbrace {S}\, |\,  {\cal E}, N \,\, {\rm fixed} \rbrace.
\end{eqnarray}
This determines the ``most
probable'' state of an isolated system. To solve this maximization problem, we
proceed
in two steps as we did previously in the nonrelativistic case (the one-step
process is discussed in Appendix \ref{sec_osts}).

\subsection{Maximization of the entropy density at fixed energy density and
particle
number density}
\label{sec_step1gr}

\subsubsection{Local thermodynamic equilibrium}
\label{sec_fdgr}

We first maximize the Fermi-Dirac entropy density (\ref{n75}) at
fixed  energy density (\ref{n69}) and particle
number density (\ref{n68}). We write the
variational problem for the first
variations (extremization) under the form
\begin{equation}
\label{n80}
\frac{\delta s}{k_B}-\beta(r)\delta\epsilon+\alpha(r)\delta n=0,
\end{equation}
where $\beta(r)$ and $\alpha(r)$ are local Lagrange multipliers.
This leads to the Fermi-Dirac distribution function
\begin{equation}
\label{n81}
f({\bf r},{\bf p})=\frac{g}{h^3}\frac{1}{1+e^{\beta(r) E(p)-\alpha(r)}},
\end{equation}
which corresponds to the condition of local thermodynamic equilibrium.
Introducing the local temperature $T(r)$ and the
local
chemical potential $\mu(r)$ by the relations
\begin{equation}
\label{n82}
\beta(r)=\frac{1}{k_B T(r)}\qquad {\rm and}\qquad  \alpha(r)=\frac{\mu(r)}{k_B
T(r)},
\end{equation}
the
Fermi-Dirac
distribution (\ref{n81}) can be rewritten as
\begin{equation}
\label{n83}
f({\bf r},{\bf p})=\frac{g}{h^3}\frac{1}{1+e^{\lbrack E(p)-\mu(r)\rbrack/k_B
T(r)}}.
\end{equation}
On the other hand, the variational
principle (\ref{n80}) reduces to the first law of thermodynamics
\begin{equation}
\label{b84}
d s=\frac{d\epsilon}{T}-\frac{\mu}{T} dn.
\end{equation}
This law is valid for an arbitrary form of entropy (see Appendix
\ref{sec_osts}).

\subsubsection{Local variables}
\label{sec_elgr}

Substituting the Fermi-Dirac distribution (\ref{n83}) into Eqs.
(\ref{n68})-(\ref{n74}) we get
\begin{equation}
\label{b85}
n(r)=\frac{g}{h^3}\int \frac{1}{1+e^{\lbrack
E(p)-\mu(r)\rbrack/k_B T(r)}}  \, d{\bf p},
\end{equation}
\begin{equation}
\label{b86}
\epsilon(r)=\frac{g}{h^3}\int \frac{E(p)}{1+e^{\lbrack
E(p)-\mu(r)\rbrack/k_B T(r)}}  \, d{\bf p},
\end{equation}
\begin{equation}
\label{b87}
\epsilon_{\rm kin}(r)=\frac{g}{h^3}\int \frac{E_{\rm kin}(p)}{1+e^{\lbrack
E(p)-\mu(r)\rbrack/k_B T(r)}}  \, d{\bf p},
\end{equation}
\begin{equation}
\label{b88}
P(r)=\frac{g}{3h^3}\int \frac{1}{1+e^{\lbrack
E(p)-\mu(r)\rbrack/k_B T(r)}} p\frac{dE}{dp}  \, d{\bf
p}=\frac{g}{h^3}k_B T(r)\int \ln\left (1+e^{-\lbrack
E(p)-\mu(r)\rbrack/k_B T(r)}\right )  \, d{\bf p},
\end{equation}
where we used an integration by parts to obtain the last equality in Eq.
(\ref{b88}). 
These equations determine the Lagrange multipliers $T(r)$ and $\mu(r)$ in terms
of $\epsilon(r)$ and $n(r)$. They also determine the equation of
state $P=P\lbrack n(r),T(r)\rbrack$ in implicit form. On the other hand,
substituting the
Fermi-Dirac
distribution function (\ref{n83}) into Eq. (\ref{n75}), and using Eq. 
(\ref{b85})-(\ref{b88}), we obtain after some calculations the  integrated
Gibbs-Duhem relation
\begin{equation}
\label{b90}
s(r)=\frac{\epsilon(r)+P(r)-\mu(r) n(r)}{T(r)}.
\end{equation}
This relation is valid for an arbitrary form of entropy (see Appendix
\ref{sec_gd}). Finally, combining the
first law of
thermodynamics (\ref{b84}) and the integrated Gibbs-Duhem relation (\ref{b90})
we
obtain the identity
\begin{equation}
\label{b91}
d\left (\frac{P}{T}\right )=n\, d\left (\frac{\mu}{T}\right
)-\epsilon \, d\left (\frac{1}{T}\right ).
\end{equation}
We also have the identities
\begin{equation}
\label{dent}
d\left (\frac{\epsilon}{n}\right )=-Pd\left
(\frac{1}{n}\right )+Td\left (\frac{s}{n}\right )\qquad {\rm and}\qquad s
dT-dP+nd\mu=0,
\end{equation}
which correspond to the standard form of the first law
of thermodynamics and the local Gibbs-Duhem relation (see Appendix
\ref{sec_gd}). These results are valid for an arbitrary form of entropy.

\subsection{Maximization of the entropy at fixed mass-energy and 
particle
number}
\label{sec_step2gr}

\subsubsection{Variational principle}
\label{sec_vp}

If we introduce the Gibbs-Duhem relation (\ref{b90}) into the entropy
(\ref{n76}), we obtain
\begin{equation}
\label{b92}
S=\int_0^R \frac{\epsilon(r)+P(r)-\mu(r) n(r)}{T(r)}\left \lbrack
1-\frac{2GM(r)}{rc^2}\right \rbrack^{-1/2}4\pi
r^2\, dr.
\end{equation}
The functionals $S$, $M$ and $N$ depend
on  $\epsilon(r)$ and $n(r)$. We now maximize
the entropy $S$ at fixed mass-energy ${\cal E}=Mc^2$ and
particle number $N$. From that point, we follow Bilic and Viollier
\cite{bvrelat} (see
also \cite{tolman,cocke,hk0,kh,khk,kam,kh2,hk,ipser80,sorkin,sy,sy2,bvrelat,
aarelat1,aarelat2,gaobis,gaoE,roupas1,gsw,fg,roupas1E,
schiffrin,psw,fhj} for
alternative
derivations and generalizations).\footnote{We note that
Bilic and
Viollier \cite{bvrelat} work in the
canonical ensemble while we work in the
microcanonical ensemble. However, as we have
already
indicated (see Sec. \ref{sec_tsei}), the statistical  ensembles are equivalent
at the
level of the first order variations (extremization
problem) so they determine the same
equilibrium states. For systems with long-range interactions, like
self-gravitating systems,
ensembles inequivalence may occur at the level of the
second order variations of the thermodynamical potential, i.e., regarding the
{\it stability} of the equilibrium
states.} We
write the
variational problem for the first
variations (extremization) under the form
\begin{equation}
\label{b93}
\frac{\delta S}{k_B}-\beta_0 c^2\delta M+\alpha_0\delta N=0,
\end{equation}
where $\beta_0$ and $\alpha_0$ are global (uniform) Lagrange multipliers.
Taking the first variations of $M$, $N$ and $S$ from Eqs.
(\ref{n53}), (\ref{n76}), and (\ref{n77}), and using the
first law of thermodynamics (\ref{b84}) and the integrated Gibbs-Duhem relation
(\ref{b90}), we
obtain 
\begin{eqnarray}
\label{b94}
\int \frac{\delta \epsilon}{k_B T}\, \chi \, dV-\int \frac{\mu}{k_B T}\delta n\,
\chi \, dV+\int \frac{\epsilon+P-\mu n}{k_B T}\, \delta\chi
\, dV
-\beta_0 \int \delta\epsilon \, dV+\alpha_0 \int
\delta n \, \chi \, dV+\alpha_0 \int
n \, \delta\chi \, dV=0,
\end{eqnarray}
where we have introduced the short-hand notations
\begin{equation}
\label{b95}
\chi(r)=\left \lbrack 1-\frac{2GM(r)}{rc^2}\right
\rbrack^{-1/2}\qquad {\rm and}\qquad dV=4\pi r^2 dr.
\end{equation}
We note for future reference that
\begin{equation}
\label{b96}
\delta\chi=\frac{\partial\chi}{\partial M}\delta M(r)=\left \lbrack
1-\frac{2GM(r)}{rc^2}\right
\rbrack^{-3/2}\frac{G\delta M(r)}{r c^2},
\end{equation}
\begin{equation}
\label{b97}
\delta M(r)=\frac{1}{c^2}\int_0^r \delta\epsilon\, 4\pi r^2\, dr,\qquad
\frac{d\delta M(r)}{dr}=\frac{1}{c^2} \delta\epsilon \, 4\pi r^2.
\end{equation}
In Eq. (\ref{b94}) the variations on $\delta n$  and $\delta\epsilon$ must
vanish 
independently.

\subsubsection{Variations on $\delta n$}
\label{sec_varn}

The vanishing of Eq. (\ref{b94}) with repsect to
variations on $\delta n$ gives
\begin{equation}
\label{b98}
\alpha(r)=\frac{\mu(r)}{k_B T(r)}=\alpha_0.
\end{equation}
This relation shows that the ratio between the local chemical potential and the
local
temperature is a constant. In the sequel, we will denote this
constant by $\alpha$
instead of $\alpha_0$. Therefore, we write
\begin{equation}
\label{b99}
\alpha=\frac{\mu(r)}{k_B T(r)}.
\end{equation}
Since $\mu/T$ is constant, Eq. (\ref{b91})
reduces to
\begin{equation}
\label{b100}
d\left (\frac{P}{T}\right )=-\epsilon
\, d\left (\frac{1}{T}\right ),
\end{equation}
implying
\begin{equation}
\label{b101}
\frac{dP}{dr}=\frac{\epsilon+P}{T}\, \frac{dT}{dr}.
\end{equation}
This equation was first obtained by Tolman \cite{tolman}.\footnote{A similar
equation $dP/dT=(\epsilon+P)/T$ is used in cosmology  in order to relate the
temperature $T$ of a cosmic fluid described by an
equation of state $P=P(\epsilon)$ to its energy density $\epsilon$ (see, e.g.,
\cite{cosmopoly1}). In that
context, it is derived from thermodynamical arguments  \cite{weinberg} by
assuming that $\mu=0$
like in the case of the black-body radiation. By contrast, in
the present calculation, we have simply used the fact that $\mu/T$ is
constant, not that $\mu$ is necessarily equal to zero.} Using Eq. (\ref{b99}),
we also
have
\begin{equation}
\label{b101b}
\frac{dP}{dr}=\frac{\epsilon+P}{\mu}\, \frac{d\mu}{dr}.
\end{equation}

\subsubsection{Variations on $\delta \epsilon$}
\label{sec_vars}

Using Eq. (\ref{b98}) and focusing now on the variations on $\epsilon$,
Eq. (\ref{b94}) reduces to
\begin{eqnarray}
\label{b108}
\int \frac{\delta \epsilon}{k_B T}\, \chi \, dV+\int
\frac{\epsilon+P}{k_B T}\, \delta\chi
\, dV
-\beta_0 \int \delta\epsilon \, dV=0.
\end{eqnarray}
Using the identities (\ref{b96}) and (\ref{b97}), the foregoing equation can be
rewritten as
\begin{eqnarray}
\label{b109}
c^2\int \left (\frac{\chi}{k_B T}-\beta_0\right )\,\frac{d\delta M(r)}{{\rm
d}r} \, dr
+\int
\frac{\epsilon+P}{k_B T}\, \frac{\partial\chi}{\partial M}\delta M(r)\, 4\pi r^2
dr=0.
\end{eqnarray}
Using an integration by parts, we obtain
\begin{eqnarray}
\label{b110}
c^2\left\lbrack \frac{\chi(R)}{k_B T(R)}-\beta_0\right\rbrack \delta
M(R)
-\int
\left\lbrack c^2\frac{d}{dr}\left (\frac{\chi}{k_B T}\right
)-\frac{\epsilon+P}{k_B T}\, \frac{\partial\chi}{\partial M}\, 4\pi
r^2\right\rbrack\delta M(r)\, dr=0.
\end{eqnarray}
The two terms in brackets must vanish individually. The vanishing of the first
bracket yields
\begin{eqnarray}
\label{b111}
\beta_0=\frac{\chi(R)}{k_B T(R)}=\frac{1}{k_B T(R)\sqrt{1-\frac{2GM}{Rc^2}}}.
\end{eqnarray}
The vanishing of the second bracket yields
\begin{eqnarray}
\label{b113}
c^2\frac{d}{dr}\left (\frac{\chi}{T}\right )=\frac{\epsilon+P}{T}\,
\frac{\partial\chi}{\partial M}\, 4\pi
r^2,
\end{eqnarray}
leading to
\begin{equation}
\label{b115}
\frac{1}{T}\frac{{d}T}{dr}=-\frac{1}{c^2}\frac{\frac{GM(r)}{r^2}+\frac{4\pi
G}{c^2}Pr}{1-\frac{2GM(r)}{r c^2}}.
\end{equation}

\subsubsection{Condition of hydrostatic equilibrium}
\label{sec_che}

Combining Eqs. (\ref{b101}) and (\ref{b115}), we obtain the OV equation 
\begin{equation}
\label{b115ov}
\frac{{d}P}{dr}=-\frac{\epsilon+P}{c^2}\frac{\frac{GM(r)}{r^2}+\frac{4\pi
G}{c^2}Pr}{1-\frac{2GM(r)}{r c^2}},
\end{equation}
expressing the condition of hydrostatic equilibrium. From Eqs. (\ref{n60}) and
(\ref{b115ov}) we obtain the Tolman  equation 
\begin{equation}
\frac{dP}{dr}=-\frac{1}{2}(\epsilon+P)\frac{d\nu}{dr},
\label{n57b}
\end{equation}
which also expresses the condition of hydrostatic equilibrium (see
Sec. \ref{sec_tov}). Therefore, the
condition of statistical equilibrium, obtained by extremizing the
entropy at fixed mass-energy and particle number, implies the condition of
hydrostatic equilibrium. This condition was not assumed in the preceding
calculations. It results from the thermodynamical
variational problem (maximization of entropy at fixed mass-energy and particle
number). In this sense, the maximum entropy principle leads to the TOV
equations. As explained in Sec. \ref{sec_about} below, this should not cause
surprise. This was already the case in
Newtonian gravity (see Sec. \ref{sec_chd}). The intrinsic
reason of this result will be given in Sec. \ref{sec_verg}.

\subsubsection{Tolman and Klein relations}
\label{sec_vart}

Combining Eqs. 
(\ref{b101}) and (\ref{n57b}), we get
\begin{equation}
\label{b102}
\frac{1}{T}\frac{dT}{dr}=-\frac{1}{2}\frac{d\nu}{dr}.
\end{equation}
Integrating this equation with respect to $r$, we obtain the Tolman relation
between the local temperature and the metric coefficient
\begin{equation}
\label{b103}
T(r)=T_{\infty} e^{-\nu(r)/2},
\end{equation}
where $T_{\infty}$ is a constant of integration. Since $\nu(r)\rightarrow 0$
when $r\rightarrow +\infty$ according to Eq. (\ref{n63}), we see that
$T_{\infty}$ represents the temperature measured by an observer at infinity.
For brevity we will call it the Tolman temperature (or the global temperature).
Using Eq. (\ref{b99}), we
get the Klein relation
\begin{equation}
\label{b104}
\mu(r)=\mu_{\infty} e^{-\nu(r)/2},
\end{equation}
where $\mu_{\infty}\equiv\alpha k_B T_{\infty}$ is a constant representing the
chemical potential measured by an observer at infinity. For brevity we will call
it the Klein  chemical potential (or the global chemical potential). For future
reference, we note that
\begin{equation}
\label{b105}
\alpha=\frac{\mu(r)}{k_B T(r)}=\frac{\mu_{\infty}}{k_B T_{\infty}},
\end{equation}
where $T(r)$ and
$\mu(r)$ are the local
temperature
and the local chemical potential while $T_{\infty}$ and $\mu_{\infty}$ are the
global temperature
and the global chemical potential. 
 Applying the Tolman relation (\ref{b103}) at the edge of
the system, we get
\begin{equation}
\label{b106}
T_{\infty}=T(R) e^{\nu(R)/2}.
\end{equation}
Using the value of $\nu(R)$ from Eq. (\ref{n64}), we obtain
\begin{equation}
\label{b107}
T_{\infty}=T(R) \sqrt{1-\frac{2GM}{Rc^2}}.
\end{equation}
Comparing this relation with Eq. (\ref{b111}), we conclude
that
\begin{eqnarray}
\label{b112}
\beta_0=\frac{1}{k_B T_{\infty}}\equiv \beta_{\infty}.
\end{eqnarray}
Therefore, the Lagrange multiplier $\beta_0$ is equal to the
inverse of the Tolman temperature. From now on, we shall write  $\beta_{\infty}$
instead of $\beta_0$.

\subsubsection{Entropy}
\label{sec_vartent}

Using Eqs. (\ref{b105}), the entropy density 
(\ref{b90}) can be rewritten as
\begin{equation}
\label{b90n}
s(r)=\frac{\epsilon(r)+P(r)}{T(r)}-\frac{\mu_{\infty}}{T_{\infty}}
n(r).
\end{equation}
Integrating Eq. (\ref{b90n}) over the whole configuration, we find that the 
entropy  is given at statistical equilibrium by
\begin{equation}
\label{b118}
S=\int_0^R \frac{P(r)+\epsilon(r)}{T(r)}\left \lbrack
1-\frac{2GM(r)}{rc^2}\right
\rbrack^{-1/2}4\pi
r^2\, dr-\frac{\mu_{\infty}}{T_{\infty}}N.
\end{equation}
We emphasize that the results derived in this section are valid for an
arbitrary form
of entropy.

\subsubsection{About the derivation of the TOV equations from thermodynamics}
\label{sec_about}

It seems that we have obtained the OV equation (\ref{b115ov}) from a
thermodynamical
variational principle without using the Einstein field equations
(\ref{n56})-(\ref{n58}). On this account, it has sometimes been suggested in the
literature that the Einstein equations could be derived from thermodynamics
(see the conclusion). This is, however, not quite true in the present
context for the following reasons:

(i) In order to
write the total entropy (\ref{n76}) and the total particle number (\ref{n77}),
we need the expression of the metric coefficient $\lambda(r)$ that appears in
the proper volume element. This is
how gravitational effects ($G$) arise in the thermodynamical
variational principle. We have assumed that $\lambda(r)$ is given by Eq.
(\ref{n59}). Therefore, we have implicitly used the Einstein equation
(\ref{n56}).

(ii) Under the above
assumption, the maximum entropy principle yields the OV equation
(\ref{b115ov}). This equation is
actually all that we need to determine the
equilibrium state of a spherical system (see Secs. \ref{sec_hett} and
\ref{sec_hydphi}). In particular, we do not need the metric function $\nu(r)$ --
the equivalent of the gravitational potential $\Phi(r)$ in Newtonian gravity.
However,
the Einstein equations contain more information than just the OV equation.
For example, in order
to derive the Tolman equation (\ref{n57b}) from the
OV equation (\ref{b115ov}), we need to use Eq. (\ref{n60}) which arises from the
Einstein equations (\ref{n56}) and (\ref{n58}). This equation relates the metric
function $\nu(r)$ to the cumulated mass $M(r)$. This is the
generalization of Newton's law (\ref{n4}).
This important equation cannot be derived from the present
thermodynamical approach. Therefore, we have not derived the whole set of
Einstein equations (\ref{n56})-(\ref{n58}). Our point of view is that the
maximum entropy principle just yields the condition of hydrostatic equilibrium,
not the whole set of Einstein equations.

Similarly, in Newtonian gravity, if we use the expression (\ref{wnice}) of the
gravitational energy (which is deduced from the Newton equations for a
spherically symmetric system), the
maximum entropy principle yields the condition of hydrostatic equilibrium under
the form of Eq. (\ref{n5}) -- the Newtonian analogue of the OV equation
(\ref{b115ov}). This equation is actually all that we need to
determine the equilibrium state of a spherical system. In particular, we do not
need the gravitational potential $\Phi(r)$. However, if we use the Newton
law (\ref{n4}), which is equivalent to the Poisson equation (\ref{n2}), we get
the condition of hydrostatic equilibrium under the form of Eq. (\ref{n1}) -- the
Newtonian analogue of the Tolman equation (\ref{n57b}). The Poisson equation --
the
Newtonian analogue of the Einstein equations (\ref{n48}) --
cannot
be derived from the present thermodynamical approach. Again, our point of view
is that the maximum entropy principle just yields the condition of hydrostatic
equilibrium, not the whole set of Newton equations.

Actually, the fact that the maximum entropy principle yields the condition of
hydrostatic equilibrium is a very general result. Indeed, it can be shown that
thermodynamical stability implies dynamical stability (see Appendix
\ref{sec_dsv}). More precisely, an extremum of entropy at fixed energy and
particle number is a steady state of the Vlasov equation (furthermore
an entropy maximum is dynamically stable). Therefore, it satisfies the
condition of hydrostatic equilibrium. For spherically symmetric systems in
general relativity, the condition of hydrostatic equilibrium takes the form of
the TOV equations. In this sense, it is not surprising that the maximum entropy
principle yields the TOV equations. This is not, in our opinion, the
manifestation of a fundamental relationship between general relativity and
thermodynamics, although such a relationship could arise in some theories of
emergent gravity as discussed in the conclusion.

\subsection{Canonical ensemble: Minimization of the free energy at
fixed particle number}
\label{sec_fece}

In the previous sections, we worked in the microcanonical ensemble in which the
particle number and the energy are fixed. We now consider the canonical ensemble
where the system is in contact with a heat
bath fixing the temperature $T$. In that case, the relevant thermodynamical
potential is the free energy
\begin{equation}
\label{n79old}
F=Mc^2-T_{\infty} S,
\end{equation}
where $T_{\infty}=1/k_B\beta_{\infty}$ is the (constant) temperature of the
thermal bath.\footnote{We have seen in Sec. \ref{sec_step2gr} that the
inverse Tolman temperature $\beta_{\infty}$ is the conjugate
variable to the energy. Therefore, this is the quantity to keep constant in the
canonical ensemble.} In the canonical ensemble, the statistical equilibrium
state of the system
is obtained by minimizing the Fermi-Dirac free energy $F$ at fixed particle
number $N$:
\begin{eqnarray}
\min\ \lbrace F \, |\,  N \,\, {\rm fixed} \rbrace.
\end{eqnarray}
This determines the ``most probable'' state of a system in contact with a
thermal bath. 

Minimizing the free energy
$F=Mc^2-T_{\infty}S$ at fixed
$N$ is equivalent
to maximizing
the Massieu function $J=S/k_B-\beta_{\infty} Mc^2$ at fixed $N$ (the Massieu
function
is the Legendre transform of the entropy with respect to the mass-energy). To
solve this
maximization problem, we proceed in
two steps. We first
maximize the Massieu function $J$ at fixed 
energy density $\epsilon(r)$ and particle number density $n(r)$ under
variations of $f({\bf r},{\bf v})$. Since the  energy
density $\epsilon(r)$ determines the mass-energy $M$,
this is equivalent to maximizing the entropy $S$  at fixed $\epsilon(r)$
and $n(r)$. This returns the
results of Sec. \ref{sec_step1gr}. Using these results, we can express the
Massieu function $J=S/k_B-\beta_{\infty} Mc^2$ in terms of $\epsilon(r)$
and $n(r)$.  We now maximize
$J$ at fixed particle number $N$ under
variations of $\epsilon(r)$ and $n(r)$. The
first variations (extremization) can be written as 
\begin{equation}
\label{jnr}
\delta \left (\frac{S}{k_B}-\beta_{\infty} Mc^2\right )+\alpha_0\delta N=0.
\end{equation}
Since $\beta_{\infty}$ is a constant, this variational problem 
is equivalent to Eq. (\ref{b93}) so we get the same
results as in Sec. \ref{sec_step2gr} (for the first variations).

In order to correctly recover the nonrelativistic results in the limit
$c\rightarrow +\infty$ (see Sec. \ref{sec_efe}), it is better to define the free
energy by
\begin{equation}
\label{n79}
F=E-T_{\infty} S,
\end{equation}
where $E$ is the binding energy defined by Eq. (\ref{n78}). This is
possible since $N$ is fixed. Using
Eqs. (\ref{n78}) and (\ref{b118}), we find that the free energy (\ref{n79})
is given at statistical equilibrium by
\begin{eqnarray}
\label{b119}
F=(M-Nm)c^2+\mu_{\infty} N
-\int_0^R \frac{T_{\infty}}{T(r)}(P(r)+\epsilon(r))\left \lbrack
1-\frac{2GM(r)}{rc^2}\right
\rbrack^{-1/2}4\pi
r^2\, dr.
\end{eqnarray}

\subsection{Equations determining the statistical equilibrium state in terms of
$T(r)$}
\label{sec_hett}

In this section, we present the full
set of equations determining the statistical equilibrium state of
self-gravitating
fermions within the context of general relativity. We express the results in
terms of the local temperature $T(r)$.

\subsubsection{Local variables in terms of $T(r)$ and $\alpha$}
\label{sec_lta}

Using Eq. (\ref{b99}), we can rewrite the
distribution function (\ref{n83}) and the local variables 
(\ref{b85})-(\ref{b88}) as 
\begin{equation}
\label{b120}
f({\bf r},{\bf p})=\frac{g}{h^3}\frac{1}{1+e^{-\alpha}e^{E(p)/k_B T(r)}},
\end{equation}
\begin{equation}
\label{b121}
n(r)=\frac{g}{h^3}\int \frac{1}{1+e^{-\alpha}e^{E(p)/k_B T(r)}} \, d{\bf
p},
\end{equation}
\begin{equation}
\label{b122}
\epsilon(r)=\frac{g}{h^3}\int \frac{E(p)}{1+e^{-\alpha}e^{E(p)/k_B T(r)}} \,
{\rm
d}{\bf p},
\end{equation}
\begin{equation}
\label{b123}
\epsilon_{\rm kin}(r)=\frac{g}{h^3}\int \frac{E_{\rm
kin}(p)}{1+e^{-\alpha}e^{E(p)/k_B
T(r)}} \, {\rm
d}{\bf p},
\end{equation}
\begin{equation}
\label{b124}
P(r)=\frac{g}{3h^3}\int \frac{1}{1+e^{-\alpha}e^{E(p)/k_B T(r)}}  p\frac{{\rm
d}E}{dp}  \, d{\bf p}=\frac{g}{h^3}k_B T(r)\int \ln\left
\lbrack 1+e^{\alpha}e^{-E(p)/k_B T(r)}\right \rbrack
 \, d{\bf p},
\end{equation}
where we recall that $\alpha$ is a constant while the temperature $T(r)$ depends
on the position (Tolman's effect). From these equations, we have
$n(r)=n[\alpha,T(r)]$, $\epsilon(r)=\epsilon[\alpha,T(r)]$,
$\epsilon_{\rm kin}(r)=\epsilon_{\rm kin}[\alpha,T(r)]$ and
$P(r)=P[\alpha,T(r)]$ leading to a barotropic equation of
state of the form $P(r)=P[\alpha,\epsilon(r)]$. Taking the derivative of $P(r)$
with respect to $r$,
we
recover Eq. (\ref{b101}). We show in Appendix \ref{sec_hea} that this result
is valid for an arbitrary form of entropy.

\subsubsection{The TOV equations in terms of $T(r)$}
\label{sec_tovt}

The TOV
equations can be written in terms of $T(r)$ as
\begin{equation}
\label{b126}
\frac{dM}{dr}=\frac{\epsilon}{c^2}4\pi r^2,
\end{equation}
\begin{equation}
\label{b125}
\frac{1}{T}\frac{{
d}T}{dr}=-\frac{1}{c^2}\frac{\frac{GM(r)}{r^2}+\frac{4\pi
G}{c^2}Pr}{1-\frac{2GM(r)}{r c^2}},
\end{equation}
with the boundary conditions
\begin{equation}
\label{b128}
M(0)=0\qquad {\rm and}\qquad T(0)=T_0.
\end{equation}
For a given value of $\alpha$ and $T_0$ we can solve Eqs.
(\ref{b126}) and (\ref{b125})
between $r=0$ and $r=R$ with the local variables (\ref{b121})-(\ref{b124}). The
particle number
constraint
\begin{equation}
\label{b127}
N=\int_0^R n(r)\left \lbrack 1-\frac{2GM(r)}{rc^2}\right
\rbrack^{-1/2}4\pi
r^2\, dr
\end{equation}
can be used to determine $T_0$ as a function of  $\alpha$ (there may be several
solutions for the same value of $\alpha$). The mass $M$ and the
temperature measured by an observer at infinity $T_{\infty}$ are then obtained
from the relations
\begin{equation}
\label{b130b}
M=M(R)\qquad {\rm and}\qquad T_{\infty}=T(R) \sqrt{1-\frac{2GM}{Rc^2}}.
\end{equation}
In this manner, we get the binding energy $E=(M-Nm)c^2$ and the Tolman
temperature $T_{\infty}$ as a function of $\alpha$.
By
varying $\alpha$ between $-\infty$ and $+\infty$, we can obtain the full caloric
curve $T_{\infty}(E)$ for a given value of $N$ and $R$. Finally, the entropy
and the free energy are given by Eqs. (\ref{b118}) and (\ref{b119}). Phase
transitions and instabilities in the general relativistic Fermi gas have been
studied in \cite{bvr,acf,rc}. The complete canonical and
microcanonical phase
diagrams are given in \cite{acf}.

{\it Remark:} We can interpret Eq. (\ref{b125}) in different manners: (i) We
can consider that the temperature  is related to the pressure by Eq.
(\ref{b101}); then Eqs. (\ref{b115})-(\ref{b102})
describe the balance between the pressure -- or temperature -- gradient and the
gravitational ``force''. This is the correct interpretation in the case of the
black-body radiation (see Appendix \ref{sec_bbr}). (ii) We can consider that
the temperature is a measure of the gravitational potential in general
relativity [see Eqs. (\ref{b115}), (\ref{b102}) and (\ref{b103})]; then Eqs.
(\ref{b101}),
(\ref{b115ov}) and (\ref{n57b}) describe an equilibrium between the pressure
force and the
gravitational force. This interpretation is suggested by the post-Newtonian
approximation of Sec. \ref{sec_tkpn}. More generally, we can
consider that, at statistical equilibrium, a
temperature gradient forms to balance the weight of matter and heat.

\subsection{Equations determining the statistical equilibrium state in terms
of $\varphi(r)$}
\label{sec_hydphi}

In this section we reformulate the previous results in terms of the
gravitational potential $\varphi(r)$. This formulation will allow us, in
particular, to
correctly recover the completely degenerate limit ($T=0$) in Sec.
\ref{sec_zerorg}.

\subsubsection{Gravitational
potential  $\varphi(r)$}
\label{sec_gp}

As noted by Tolman \cite{tolman}, Eq. (\ref{n57b}) may be interpreted as the
condition of hydrostatic
equilibrium in general relativity (see Sec. \ref{sec_tov}). The left
hand side is the
pressure
gradient and the metric coefficient on the right hand side plays the role of
the 
gravitational potential in Newtonian gravity. Instead of working with
the
metric coefficient $\nu(r)$ it is convenient to introduce the general
relativistic
gravitational potential $\varphi(r)$  
defined
by
\begin{equation}
\label{b131}
e^{\nu(r)}=\left (\frac{\mu_{\infty}}{m c^2}\right
)^2\frac{1}{1+\frac{\varphi(r)}{c^2}}\qquad \Rightarrow \qquad \nu(r)=-\ln\left
(1+\frac{\varphi(r)}{c^2}\right )+2\ln\left
(\frac{|\mu_{\infty}|}{mc^2}\right ).
\end{equation}
It satisfies $\varphi(r)>-c^2$. Using Eq. (\ref{b105}), we have
equivalently
\begin{equation}
\label{b132}
e^{\nu(r)}=\left (\frac{\alpha k_B T_{\infty}}{m c^2}\right
)^2\frac{1}{1+\frac{\varphi(r)}{c^2}}\qquad \Rightarrow \qquad\nu(r)=-\ln\left
(1+\frac{\varphi(r)}{c^2}\right )+2\ln\left
(\frac{|\alpha|k_B T_{\infty}}{mc^2}\right ).
\end{equation}
We can also relate the gravitational potential  $\varphi(r)$ to the
temperature $T(r)$. Combining Eq. (\ref{b131}) with the Tolman-Klein relations
(\ref{b103}) and (\ref{b104}) we obtain
\begin{equation}
\label{b134}
k_B T(r)=\frac{mc^2}{|\alpha|}\sqrt{\frac{\varphi(r)}{c^2}+1}\qquad {\rm
and}\qquad |\mu(r)|=mc^2\sqrt{\frac{\varphi(r)}{c^2}+1}.
\end{equation}
Finally, using Eq. (\ref{n64}) or Eq. (\ref{b107}), we find that
$\varphi(R)$ is determined by the relation
\begin{equation}
\frac{k_B T_{\infty}}{mc^2}=\frac{1}{|\alpha|}\sqrt{\frac{
\varphi(R)}{c^2}+1}\left
(1-\frac{2GM}{Rc^2}\right )^{1/2}.
\end{equation}

\subsubsection{Local variables in terms of $\varphi(r)$}
\label{sec_locphi}

Using Eq. (\ref{b134}), we can rewrite the distribution function
(\ref{b120}) and the local variables
(\ref{b121})-(\ref{b124}) in terms of $\alpha$ and $\varphi(r)$ as
\begin{equation}
\label{b136}
f({\bf r},{\bf
p})=\frac{g}{h^3}\frac{1}{1+e^{-\alpha}e^{\frac{|\alpha|E(p)}{
mc^2\sqrt{
1+\frac{\varphi(r)}{c^2}}}}},
\end{equation}
\begin{equation}
\label{b137}
n(r)=\frac{g}{h^3}\int \frac{1}{1+e^{-\alpha}e^{\frac{|\alpha|E(p)}{
mc^2\sqrt{
1+\frac{\varphi(r)}{c^2}}}}}\, d{\bf p},
\end{equation}
\begin{equation}
\label{b138}
\epsilon(r)=\frac{g}{h^3}\int \frac{E(p)}{1+e^{-\alpha}e^{\frac{|\alpha|E(p)}{
mc^2\sqrt{
1+\frac{\varphi(r)}{c^2}}}}}\, d{\bf
p},
\end{equation}
\begin{equation}
\label{b139}
\epsilon_{\rm kin}(r)=\frac{g}{h^3}\int \frac{E_{\rm
kin}(p)}{1+e^{-\alpha}e^{\frac{|\alpha|E(p)}{
mc^2\sqrt{
1+\frac{\varphi(r)}{c^2}}}}}\, d{\bf p},
\end{equation}
\begin{equation}
\label{b140}
P(r)=\frac{g}{3h^3}\int \frac{1}{1+e^{-\alpha}e^{\frac{|\alpha|E(p)}{
mc^2\sqrt{
1+\frac{\varphi(r)}{c^2}}}}} p\frac{{\rm
d}E}{dp} \, d{\bf p}=\frac{g}{h^3}\frac{mc^2}{|\alpha|}\sqrt{
1+\frac{\varphi(r)}{c^2}}\int
\ln\left
\lbrack 1+e^{\alpha}e^{-\frac{|\alpha| E(p)}{mc^2\sqrt{
1+\frac{\varphi(r)}{c^2}}}}\right \rbrack
 \, d{\bf p}.
\end{equation}
From these equations, we have
$n(r)=n[\alpha,\varphi(r)]$, $\epsilon(r)=\epsilon[\alpha,\varphi(r)]$,
$\epsilon_{\rm kin}(r)=\epsilon_{\rm kin}[\alpha,\varphi(r)]$ and
$P(r)=P[\alpha,\varphi(r)]$ leading to an equation of
state of the form $P(r)=P[\alpha,\epsilon(r)]$.

\subsubsection{The TOV equations in terms of $\varphi(r)$}
\label{sec_tovphi}

Taking the derivative of Eq. (\ref{b132}) with respect to $r$ we obtain
\begin{equation}
\label{b142}
\frac{d\nu}{{\rm
d}r}=-\frac{1}{1+\frac{\varphi}{c^2}}\frac{1}{c^2}\frac{{\rm
d}\varphi}{dr}.
\end{equation} 
Substituting this relation into Eq. (\ref{n57b}) we
get\footnote{This equation can be directly obtained by
taking the gradient of Eq. (\ref{b140}).}
\begin{equation}
\label{b144}
\frac{dP}{{\rm
d}r}=\frac{1}{2c^2}(\epsilon+P)\frac{1}{1+\frac{\varphi}{c^2}}\frac{{\rm
d}\varphi}{dr}.
\end{equation}
Alternatively, substituting Eq. (\ref{b142}) into Eq.
(\ref{b102}) we obtain
\begin{equation}
\label{b143}
\frac{1}{T}\frac{dT}{{\rm
d}r}=\frac{1}{2c^2}\frac{1}{1+\frac{\varphi}{c^2}}\frac{{\rm
d}\varphi}{dr}.
\end{equation}
The TOV equations can be written in terms of
$\varphi(r)$ as
\begin{equation}
\label{b147}
\frac{dM}{dr}=\frac{\epsilon}{c^2}4\pi r^2,
\end{equation}
\begin{equation}
\label{b145}
\frac{{\rm
d}\varphi}{dr}=-2\left \lbrack 1+\frac{\varphi(r)}{c^2}\right
\rbrack\frac{\frac{GM(r)}{r^2}+\frac{4\pi
G}{c^2}Pr}{1-\frac{2GM(r)}{r c^2}},
\end{equation}
with the boundary conditions 
\begin{equation}
\label{b128b}
M(0)=0\qquad {\rm and}\qquad \varphi(0)=\varphi_0>-c^2.
\end{equation}
For a given value of $\alpha$ and $\varphi_0$ we can solve Eqs.
(\ref{b147}) and (\ref{b145}) between $r=0$ and $r=R$ with the local
variables (\ref{b137})-(\ref{b140}).
The particle number
constraint
\begin{equation}
\label{b127b}
N=\int_0^R n(r)\left \lbrack 1-\frac{2GM(r)}{rc^2}\right
\rbrack^{-1/2}4\pi
r^2\, dr
\end{equation}
can be used to determine $\varphi_0$  as a function of $\alpha$ (there may be
several solutions for the same value of $\alpha$). The mass $M$ and the
temperature measured by an observer at infinity $T_{\infty}$ are then obtained
from the
relations
\begin{equation}
\label{b130}
M=M(R)\qquad {\rm and}\qquad
\frac{k_B T_{\infty}}{mc^2}=\frac{1}{|\alpha|}\sqrt{\frac{
\varphi(R)}{c^2}+1}\left
(1-\frac{2GM}{Rc^2}\right )^{1/2}.
\end{equation}
In this manner, we get the binding energy $E=(M-Nm)c^2$ and the Tolman
temperature 
$T_{\infty}$ as a function of
$\alpha$. By
varying $\alpha$ between $-\infty$ and $+\infty$, we can obtain the full caloric
curve $T_{\infty}(E)$ for a given value of $N$ and $R$. Finally, the entropy
and the free energy are given by Eqs. (\ref{b118}) and
(\ref{b119}) where $T(r)$ is related to $\varphi(r)$ by Eq. (\ref{b134}).

\subsection{Thermodynamical stability and ensembles inequivalence}
\label{sec_tsrg}

As discussed in Sec. \ref{sec_tsei}, the statistical equilibrium states
in the microcanonical and 
canonical ensembles (extrema of entropy or free energy) are the same.
However, their thermodynamical stability may
be different in the microcanonical and canonical ensembles. This corresponds to
the concept
of ensembles inequivalence for systems with long-range interactions
\cite{cdr,campabook}.

The thermodynamical stability of an equilibrium state can be investigated by
studying the sign of the second order variations of the appropriate
thermodynamic potential (entropy or free energy) and reducing this study to an
eigenvalue
problem. We refer to \cite{cocke,sorkin,aarelat2,roupas1,roupas1E,gsw,fhj} for a
detailed discussion of this stability problem in general
relativity.

The thermodynamical stability of the system can also be directly
settled from the topology of the series of equilibria by using the Poincar\'e
criterion. The discussion is the same as in Sec. \ref{sec_tsei} provided
that $E$ is replaced by ${\cal E}=Mc^2$ or $E=(M-Nm)c^2$ and $\beta$ is replaced
by
$\beta_{\infty}$. In short, a change of microcanonical stability can take place
at a turning point of energy and a change of canonical
stability can take place at a turning point of temperature. We refer to 
\cite{hk,ipser80,bvr,aarelat2,roupas,acf,rc} for
some applications of the Poincar\'e criterion in general
relativity.

\subsection{Dynamical stability}

\subsubsection{Vlasov-Einstein equations}
\label{sec_verg}

The distribution function $f({\bf r},{\bf p})$ of a system of
self-gravitating fermions at statistical equilibrium is given by Eq.
(\ref{b120}). Using the Tolman relation (\ref{b103}), it is of the form
$f=f(E e^{\nu(r)/2})$ with $f'(E e^{\nu(r)/2})<0$. Therefore, at statistical
equilibrium, the
distribution function depends only on the energy at infinity
$E e^{\nu(r)/2}$ and is monotonically decreasing. It is shown in Appendices
\ref{sec_osts} and \ref{sec_hea} that these properties
remain valid for a general form of 
entropy. Since $f({\bf r},{\bf p})$ is a function of the energy at infinity
$E
e^{\nu(r)/2}$, which is a constant of the motion, it is
a particular steady state of the Vlasov-Einstein equations. This
is a special case of the  relativistic Jeans
theorem \cite{zp,fackerell,it,bkt,ipser80}.\footnote{According
to the relativistic Jeans theorem \cite{zp,fackerell,it,bkt,ipser80}, a
spherical stellar system in collisionless
equilibrium has a distribution function of the form $f=f(E e^{\nu(r)/2},L)$
where $E e^{\nu(r)/2}$ is the energy  at infinity and $L$ is the angular
momentum. We note that an
extremum of entropy $S$ at fixed mass-energy $Mc^2$ and particle number $N$
leads to a distribution function that
depends only on $E e^{\nu(r)/2}$, not on $L$. Therefore, an extremum of
entropy at fixed mass-energy and particle number is necessarily isotropic. We
also note that the relativistic Jeans theorem implies the  Tolman
relation (\ref{b103}) when $f$ is the Maxwell-Boltzmann distribution.}
Therefore, a statistical
equilibrium state (extremum of entropy at fixed energy and particle
number) is
a steady  state of the Vlasov-Einstein equations. Furthermore, it
can be shown that
thermodynamical stability implies dynamical stability with respect to the
Vlasov-Einstein equations \cite{ipser80}.
Therefore, a stable
thermodynamical equilibrium state (maximum of entropy at fixed energy and
particle
number) is always dynamically
stable.\footnote{This result is very general and stems from the fact
that the entropy (which is a particular Casimir), the mass-energy and the
particle
number are
conserved by the
Vlasov-Einstein equations (see \cite{cc} and Appendix
\ref{sec_dsv}).} In
general, the reciprocal is wrong.
However, in
general
relativity, Ipser \cite{ipser80} has shown that dynamical and thermodynamical
stability (in the microcanonical
ensemble) coincide (see Appendix \ref{sec_dsvgr}). As
a result, using the Poincar\'e
criterion (see Sec. \ref{sec_tsrg}), we generically  conclude that the series of
equilibria
before
the first turning point of energy is dynamically stable while it becomes
dynamically unstable afterwards.
This is in sharp constrast with the Newtonian
gravity case where all the statistical
equilibrium states are dynamically stable, even those that are
thermodynamically unstable (see Sec. \ref{sec_dsnv}). To solve this apparent
paradox, one
expects that the growth rate $\lambda$ of the dynamical instability decreases
as relativity effects decrease and that it tends to zero in the
nonrelativistic limit $c\rightarrow +\infty$.

\subsubsection{Euler-Einstein equations}
\label{sec_eerg}

We have seen that the statistical equilibrium state of a system of
self-gravitating fermions in general relativity  is described by a barotropic
equation of state
of the form $P(r)=P[\epsilon(r)]$ and that it satisfies the TOV
equation (\ref{b115ov}) expressing the condition of
hydrostatic equilibrium. It is
shown in
Appendices \ref{sec_osts} and \ref{sec_hea} that these properties remain
valid for an arbitrary form of entropy. As a result, the system is in a steady
state of the
Euler-Einstein equations. Furthermore, it can be shown
\cite{roupas1,roupas1E,gsw,fhj} that the
thermodynamical stability of a self-gravitating system  in the canonical
ensemble is equivalent to its dynamical stability  with
respect to the Euler-Einstein equations.  This generalizes the result obtained
by \cite{aaantonov} in Newtonian gravity (see Sec.
\ref{sec_dsne}).

\subsection{Particular limits}

The Fermi-Dirac distribution (\ref{b136}) can be written as
\begin{equation}
\label{z0}
f({\bf r},{\bf p})=\frac{g}{h^3}\frac{1}{1+e^{\lbrack
E(p)-\mu(r)\rbrack/k_B
T(r)}},
\end{equation}
where $E(p)=\sqrt{p^2c^2+m^2c^4}$ and $|\mu(r)|=mc^2\sqrt{\varphi(r)/c^2+1}$.
Let us consider
particular limits of this distribution function.

\subsubsection{The completely degenerate Fermi gas (ground state)}
\label{sec_zerorg}

The completely degenerate limit corresponds to  
$T(r)\rightarrow
0$, $\mu(r)>0$
finite, and $\alpha(r)=\mu(r)/k_B T(r)\rightarrow +\infty$. In
that case, the chemical potential is positive and large compared to the
temperature. This
yields the Fermi distribution (or Heaviside
function):
\begin{eqnarray}
\label{z2}
f({\bf r},{\bf
p})=\frac{g}{h^3}\quad {\rm if}\quad E(p)< E_{F}(r) \quad (p<
p_F(r)),
\end{eqnarray}
and
\begin{eqnarray}
\label{z3}
f({\bf r},{\bf
p})=0 \quad {\rm if}\quad E(p)> E_{F}(r) \quad (p> p_F(r)),
\end{eqnarray}
where
\begin{equation}
\label{z1}
E_F(r)\equiv \mu(r)=mc^2\sqrt{\frac{\varphi(r)}{c^2}+1},
\end{equation}
and
\begin{equation}
\label{z4}
p_F(r)=m\sqrt{\varphi(r)}
\end{equation}
are the Fermi energy and the Fermi impulse.    We
note that
Eq. (\ref{z4}) imposes the condition $\varphi\ge 0$.
Setting
\begin{equation}
\label{z5}
x=\frac{p_F}{mc}=\frac{\sqrt{\varphi(r)}}{c},
\end{equation}
and using the results of \cite{chandrabook} we
find that the equation of state of the
relativistic Fermi
gas at $T=0$ is given by
\begin{equation}
\label{z6}
n=\frac{4\pi g m^3c^3}{3h^3}x^3,
\end{equation}
\begin{equation}
\label{z7}
\epsilon=\frac{\pi g m^4c^5}{2h^3}\left\lbrack x(2x^2+1)(1+x^2)^{1/2}-{\rm
sinh}^{-1}(x)\right\rbrack,
\end{equation}
\begin{equation}
\label{z8}
P=\frac{\pi g m^4c^5}{6h^3}\left\lbrack x(2x^2-3)(1+x^2)^{1/2}+3{\rm
sinh}^{-1}(x)\right\rbrack.
\end{equation}
For a given value of $n_0=n(0)$, we can solve the TOV equations (\ref{n52}) and
(\ref{n58b}) with the equation of state (\ref{z6})-(\ref{z8}) until the point
where the density vanishes: $n(R)=0$. This determines the radius $R$ of the
configuration (when
$T=0$ we do not need a box to confine the system). We can then compute the
corresponding mass $M$ and
particle number $N$. By
varying $n_0$, we get the mass-radius relation $M(R)$. The
mass-radius relation and the corresponding density profiles of the
general relativistic Fermi gas at $T=0$ have been
obtained by Oppenheimer and Volkoff \cite{ov} and Chandrasekhar and Tooper
\cite{ct}.  These
results can also be obtained by solving Eqs. (\ref{b147}) and
(\ref{b145}) where $\epsilon$ and $P$ are expressed in terms of $\varphi$ by
using Eqs. (\ref{z7}) and (\ref{z8}) with Eq. (\ref{z5}) \cite{bvr,acf}.

{\it Remark:} For the general relativistic Fermi gas at $T=0$, the free energy
$F$ reduces to the binding energy $E=(M-Nm)c^2$. Therefore, a stable equilibrium
state at $T=0$ is a minimum of $E$ at fixed $N$ or, equivalently,
a minimum of $M$ at fixed $N$. We have seen that  the general relativistic Fermi
gas
at $T=0$ is described by a barotropic equation of state of the form
$P=P(\epsilon)$ determined in parametric form by Eqs. (\ref{z7})
and (\ref{z8}). Therefore, for the self-gravitating Fermi gas at
$T=0$, the condition of thermodynamical stability
coincides with the condition
of dynamical stability with respect to the Euler-Einstein equations (see
Appendix \ref{sec_dsegr}). Actually, this equivalence is  valid at arbitrary
temperature in the canonical ensemble (see Sec.
\ref{sec_eerg}).

\subsubsection{The nondegenerate Fermi gas (classical limit)}

The nondegenerate (classical) limit corresponds to 
\begin{equation}
\label{mj2}
\frac{E(p)-\mu(r)}{k_B T(r)}\gg 1 \quad {\it i.e.} \quad \beta(r) E(p)-\alpha\gg
1.
\end{equation}
This yields the Maxwell-Juttner distribution
\begin{equation}
\label{mj3}
f({\bf r},{\bf p})=\frac{g}{h^3} e^{-\lbrack
E(p)-\mu(r)\rbrack/k_B
T(r)}.
\end{equation}
The condition (\ref{mj2}) is always fulfilled when $\alpha\rightarrow -\infty$,
whatever the value of $\beta(r) E(p)$. Therefore, the
condition $\mu(r)\rightarrow -\infty$, $T(r)$ finite and $\alpha=\mu(r)/k_B
T(r)\rightarrow -\infty$ implies the nondegenerate
(classical) limit.
In that case, the chemical potential is negative and large 
(in absolute value) compared to the temperature.
However, this
is not the only case where the Boltzmann distribution is
valid. We can be in the
classical limit for arbitrary values of $\alpha$ (positive or negative) provided
that $\beta(r) E(p)-\alpha\gg 1$. The classical limit
is specifically studied in Paper II.

\section{The nonrelativistic limit}
\label{sec_nr}

\subsection{Relation between $\nu$ and $\Phi$}
\label{sec_rnp}

The condition of hydrostatic equilibrium in general
relativity is given by the Tolman equation 
(\ref{n57}) where the metric coefficient $\nu(r)$ plays the role of the
gravitational
potential $\Phi(r)$ in Newtonian gravity. In the
nonrelativistic limit $c\rightarrow +\infty$, 
using $\epsilon\simeq \rho c^2\gg P$, it
reduces to
\begin{equation}
\label{nl1}
\frac{dP}{dr}\simeq -\frac{1}{2}\rho c^2\frac{d\nu}{{\rm
d}r}.
\end{equation}
 Comparing this equation with the
condition of hydrostatic equilibrium in Newtonian gravity given by Eq.
(\ref{n1}), we find that
\begin{equation}
\label{nl2}
\nu(r)\simeq \frac{2\Phi(r)}{c^2}+{C}, 
\end{equation}
where $C$ is a constant.
On the other hand, according to Eq. (\ref{n64}) we have
\begin{equation}
\label{bn3}
\nu(R)\simeq -\frac{2GM}{Rc^2}.
\end{equation}
Comparing Eqs. (\ref{nl2}) and (\ref{bn3}) with Eq. (\ref{n4b}),
we see that the constant $C$ in Eq. (\ref{nl2}) is equal to zero. Therefore, we
get
\begin{equation}
\label{bn3b}
\nu(r)\simeq \frac{2\Phi(r)}{c^2}\quad {\rm or}\quad e^{\nu(r)}\simeq
1+\frac{2\Phi(r)}{c^2}.
\end{equation}
On the other hand, in the nonrelativistic limit
$c\rightarrow +\infty$, using $\epsilon\simeq \rho c^2\gg P$, the OV
equations (\ref{n53}) and (\ref{n58b}) reduce to
the Newtonian equations (\ref{n3}) and (\ref{n5}). Then, Eqs. 
(\ref{n1}) and (\ref{n5}) can be combined to recover the Newton law
(\ref{n4}). This equation is also directly obtained from Eq. (\ref{n60}) in the
limit $c\rightarrow +\infty$. Finally, Eqs. (\ref{n3}) and (\ref{n4}) return the
Poisson equation
(\ref{n2}).

{\it Remark:} The relation (\ref{bn3b}) and the
Poisson equation $\Delta\Phi=4\pi G\rho$ can be obtained directly from the 
Einstein equations in the nonrelativistic
limit $c\rightarrow +\infty$. The metric takes the form 
\begin{equation}
ds^{2}=(c^2+2\Phi) dt^{2}-r^{2}(d\theta^{2}+\sin^{2}\theta
d\phi^{2})-dr^{2},
\label{metricN}
\end{equation}
which is a  consequence of the correspondance principle \cite{llf,cp}.

\subsection{Tolman-Klein relations in the post-Newtonian approximation}
\label{sec_tkpn}

From Eqs. (\ref{b103}), (\ref{b104}) and (\ref{bn3b}) we find that the local
temperature and the local chemical potential are given in the post-Newtonian
approximation by
\begin{equation}
\label{bn23}
\frac{T(r)}{T_{\infty}} \simeq \frac{\mu(r)}{\mu_{\infty}}
\simeq 1-\frac{\Phi(r)}{c^2}+O(1/c^4).
\end{equation}
If ${\bf g}=-\nabla\Phi$ denotes the gravitational force by unit of mass
(acceleration) one has
\begin{equation}
\label{bn23qw}
\frac{\nabla T}{T}=\frac{\bf g}{c^2}.
\end{equation}
This relation was first given by Tolman \cite{tolman} as a
preamble of his general result (\ref{b103}).  In the nonrelativistic limit
$c\rightarrow +\infty$, using Eq.
(\ref{bn23}) and $\epsilon\simeq \rho c^2\gg P$, we find that Eq.
(\ref{b101}) reduces to Eq. (\ref{n1}). On the other hand, the
TOV
equations (\ref{b126}) and (\ref{b125}) reduce to Eq.
(\ref{comb}). The Tolman relation (\ref{bn23}) clearly shows that, in the
post-Newtonian approximation, the
gravitational potential $\Phi({r})$ is ``hidden'' in the inhomogeneous
temperature $T({r})$ (see also the Remark at the end of Sec. \ref{sec_tovt}).

\subsection{Distribution function and $\alpha$}

The distribution function of a gas of self-gravitating fermions at statistical
equilibrium  is given by Eq. (\ref{b120}) where the energy of a
particle is given by Eq. (\ref{n70}). In the nonrelativistic limit
$c\rightarrow +\infty$, we have
\begin{equation}
\label{bn31}
E(p)\simeq mc^2+\frac{p^2}{2m}+O(1/c^2).
\end{equation}
Using Eq. (\ref{bn23}), we get
\begin{equation}
\label{bn32}
\frac{E(p)}{k_B T(r)}\simeq \frac{1}{k_B T_{\infty}}\left
\lbrack mc^2+\frac{p^2}{2m}+m\Phi(r)\right \rbrack.
\end{equation}
Therefore, the distribution function (\ref{b120}) becomes
\begin{equation}
\label{bn33}
f({\bf r},{\bf p})\simeq \frac{g}{h^3}\frac{1}{1+e^{-\alpha}e^{mc^2/k_B
T_{\infty}}e^{p^2/2mk_B T_{\infty}}e^{m\Phi(r)/k_B T_{\infty}}}.
\end{equation}
In order to obtain an expression independent of $c$ and consistent with the
expression (\ref{n37}) obtained in Newtonian gravity, i.e.,
\begin{equation}
\label{bn35}
f({\bf r},{\bf
p})=\frac{g}{h^3}\frac{1}{1+e^{-\alpha_{0}^{\rm NR}}e^{(p^2/2m+m\Phi(r))/k_B
T_{\infty}}},
\end{equation}
we have to write\footnote{We note that $\alpha\sim mc^2/k_B
T\rightarrow
+\infty$ in the
nonrelativistic limit $c\rightarrow +\infty$ (i.e. $k_B
T\ll mc^2$), and is therefore positive,  while
$\alpha^{\rm NR}_0$ may be of any sign.} 
\begin{equation}
\label{bn6}
\alpha=\frac{mc^2}{k_B T_{\infty}}+\alpha^{\rm NR}_0+O(1/c^2).
\end{equation}
We can then
define
\begin{equation}
\alpha_{\rm NR}(r)=\alpha_{0}^{\rm
NR}-\beta_{\infty} m \Phi(r)
\end{equation}
in agreement with Eq. (\ref{n35}).

\subsection{Chemical potential}
\label{sec_cp}

From Eqs. (\ref{b105}) and (\ref{bn6}) we get
\begin{equation}
\label{bn26}
\mu_{\infty}=mc^2+\mu^{\rm NR}_{0}+O(1/c^2)
\end{equation}
with
\begin{equation}
\label{bn27}
\mu^{\rm NR}_{0}=\alpha^{\rm NR}_{0}k_B T_{\infty}.
\end{equation}
Using Eqs. (\ref{bn23}) and (\ref{bn26}), we obtain
\begin{equation}
\mu(r) =mc^2+\mu_{0}^{\rm
NR}-m\Phi(r)+O(1/c^2).
\end{equation}
Therefore, we can write\footnote{We note that
$\mu(r)\sim
mc^2\rightarrow +\infty$ in the
nonrelativistic limit $c\rightarrow +\infty$, and is therefore positive, while
$\mu_{\rm NR}(r)$ may be of any sign.}
\begin{equation}
\mu(r)=mc^2+\mu_{\rm NR}(r)+O(1/c^2)
\end{equation}
with 
\begin{equation}
\mu_{\rm NR}(r)=\mu_{0}^{\rm NR}-m\Phi(r)
\end{equation}
in agreement with Eq. (\ref{n35}). 

\subsection{Relation between $\varphi$ and $\Phi$}

Combining Eqs. (\ref{b132}) and (\ref{bn6}), we get
\begin{equation}
\label{bn6b}
\nu(r)\simeq -\frac{\varphi(r)}{c^2}+\frac{2\alpha^{\rm NR}_0k_B
T_{\infty}}{mc^2}.
\end{equation}
Comparing this equation with Eq. (\ref{bn3b}), we find that
\begin{equation}
\label{bn6c}
\varphi(r)\simeq -2\Phi(r)+\frac{2\alpha^{\rm NR}_0k_B
T_{\infty}}{m}.
\end{equation}
Using Eq. (\ref{bn6c}), we note that Eq. (\ref{b144}) reduces to
the Newtonian condition of hydrostatic equilibrium
(\ref{n1}) when $c\rightarrow +\infty$. On the other hand, the TOV
equations   (\ref{b147}) and
(\ref{b145}) reduce to Eq. (\ref{comb}).

\subsection{Mass, particle number and energy}
\label{sec_mneb}

Using Eq. (\ref{n72}), the mass-energy 
(\ref{n53}) can be
written as
\begin{eqnarray}
\label{bn40}
Mc^2=\int \rho c^2\, dV+\int \epsilon_{\rm kin}\, {\rm
d}V.
\end{eqnarray}
Here, no approximation has been made. 
On the other hand, for $c\rightarrow +\infty$, using the approximation
\begin{equation}
\label{bn37}
\left
\lbrack
1-\frac{2GM(r)}{rc^2}\right
\rbrack^{-1/2}\simeq 1+\frac{GM(r)}{rc^2}+O(1/c^4)
\end{equation}
the rest mass (\ref{n77}) can be written as
\begin{eqnarray}
\label{bn41}
Nmc^2=\int \rho c^2\, dV+\int \rho \frac{GM(r)}{r}\, {\rm
d}V+O(1/c^2).
\end{eqnarray}
Therefore, in the nonrelativistic limit $c\rightarrow +\infty$, the
binding energy (\ref{n78}) takes the form
\begin{eqnarray}
\label{bn42}
E=Mc^2-Nmc^2=\int \epsilon_{\rm kin}\, {\rm
d}V-\int \rho \frac{GM(r)}{r}\, {\rm
d}V=E_{\rm kin}+W\qquad (c\rightarrow +\infty),
\end{eqnarray}
where we have used the expression (\ref{wnice}) of the Newtonian gravitational
energy valid for a spherically symmetric distribution of matter. As a result, in
the
nonrelativistic limit $c\rightarrow +\infty$, the binding energy (\ref{n78})
reduces to the
Newtonian energy which is equal to the sum of the kinetic and potential
energies.

\subsection{Entropy and free energy}
\label{sec_efe}

In the nonrelativistic limit $c\rightarrow +\infty$, using Eqs. 
 (\ref{n72}), (\ref{bn23}),
(\ref{bn26}),  (\ref{bn37}) and
(\ref{wnice}), we find that the entropy (\ref{b118}) takes the form
\begin{eqnarray}
\label{bn39}
T_{\infty}S=-N\mu_{0}^{\rm NR}+2W+\int P\, dV+E_{\rm kin}\qquad
(c\rightarrow +\infty).
\end{eqnarray}
Using Eqs. (\ref{bn42}) and (\ref{bn39}), the free energy reduces
to
\begin{eqnarray}
\label{bn43}
F=E-T_{\infty}S=N\mu_{0}^{\rm NR}-W-\int P\, dV\qquad (c\rightarrow
+\infty).
\end{eqnarray}
Recalling Eq. (\ref{bx6}), we recover the expressions (\ref{n46}) and
(\ref{n47}) obtained in the Newtonian  approach.

\section{Conclusion}

In this paper, elaborating upon previous works on the subject
\cite{tolman,cocke,hk0,kh,khk,kam,kh2,hk,ipser80,sorkin,sy,sy2,bvrelat,
aarelat1,aarelat2,gaobis,gaoE,roupas1,gsw,fg,roupas1E,
schiffrin,psw,fhj}, we have developed a general formalism to
determine the statistical equilibrium state of a system of particles in
general relativity. Although we have considered a gas of fermions described by
the Fermi-Dirac entropy  for
illustration, our formalism is valid for an arbitrary form of
entropy. This shows that the notion of ``generalized
thermodynamics'' developed in recent years in statistical
mechanics \cite{tsallisbook} can be
extended to the context of general relativity.\footnote{This
also suggests that what we call ``generalized thermodynamics'' is just
``standard thermodynamics'' with a generalized form of entropy taking into
account microscopic constraints \cite{nfp,entropy}.} For spherically
symmetric
systems, the extremization  of the entropy $S$ at fixed mass-energy $Mc^2$ and
particle number $N$ yields the TOV equations (\ref{b115ov}) and (\ref{n57b})
expressing the condition of
hydrostatic equilibrium and the Tolman-Klein relations (\ref{b103}) and
(\ref{b104}). In Newtonian gravity,
the maximum entropy principle implies the uniformity of the temperature [see
Eq. (\ref{n33})] and of
the {\it total} chemical potential (Gibbs law) [see
Eq. (\ref{n35})], and the condition of hydrostatic
equilibrium (\ref{henq}).

Research in general relativity has
shown that there is a deep connection between 
gravitation and thermodynamics. It is reflected, e.g., in the thermodynamical
interpretation given by Bekenstein \cite{bekensteinbis} and Hawking
\cite{hawkingbis} of the four laws of
black hole mechanics \cite{bch} which were derived from the Einstein equations.
This led to the concept of Hawking radiation \cite{hawkingbis}. There have
been
also some attempts by Jacobson \cite{jacobson}, Padmanabhan \cite{paddyqg} and
Verlinde \cite{verlinde} to derive the Einstein equations from
thermodynamics. All these attempts are based on some deep underlying principles,
like the
holographic principle or the concept of emergent gravity, which could lay the
foundation for a theory of quantum
gravity. 
In this connection, the fact that the TOV equations
can be derived from the
maximum entropy principle (as discussed in Sec. \ref{sec_step2gr}) has sometimes
been
regarded as a strong evidence for the
fundamental relationship between general relativity and
thermodynamics \cite{gaobis}.
However, for systems of particles described by classical general gravity such
as the ones that we have considered in this paper, this result is not really
surprising and
does not reflect, we believe, a special connection between  gravity and
thermodynamics.

Indeed, it is well-known that a statistical equilibrium state
is always a steady state of the Vlasov equations and that thermodynamical
stability
implies
dynamical stability (see Appendix \ref{sec_dsv}). This is due to the fact that
the entropy $S$ (a particular Casimir functional), the energy $E$ and the
particle number $N$ which appear in the
maximum entropy principle are conserved
by the Vlasov equation. As a result, an extremum of  $S$ at fixed $E$ and $N$ 
is a steady state of the Vlasov equation and a maximum of  $S$ at fixed $E$ and
$N$ is dynamically stable (in addition of being thermodynamically
stable). Therefore, the maximum entropy principle implies the
condition of hydrostatic equilibrium. For spherically symmetric systems in
general relativity, this leads to  the TOV
equations. This result -- the fact that the maximum
entropy principle
implies the condition of hydrostatic equilibrium -- is very general and is
valid for all systems with long-range interactions \cite{cc}. In the context
of general relativity, it was first noted by Tolman \cite{tolman}
at the end of his paper (see Appendix \ref{sec_tolman}) and
rediscovered by many other
authors in the sequel. It does not bear a
deeper significance. In particular, we
emphasized in Sec. \ref{sec_about} that the maximum entropy principle implies 
the condition of hydrostatic equilibrium (yielding the TOV equations for
spherically symmetric systems) but  does not provide the
whole set of Einstein equations. Similarly, in Newtonian gravity,  the maximum
entropy principle implies the condition of hydrostatic equilibrium, but not the
whole set of Newton equations.

The equations derived in this paper have been used to construct the 
caloric curves of self-gravitating fermions and study phase transitions between
gaseous states and condensed states in Newtonian gravity
\cite{ht,bvn,csmnras,robert,pt,dark,ispolatov,rieutord,ptdimd,ijmpb} and general
relativity \cite{bvr,acf,rc}. A rather
complete understanding of these phase transitions has now been reached and
general phase diagrams have been obtained in \cite{ijmpb,acf}. In the case
of classical self-gravitating systems there is nothing to halt the collapse so
that an equilibrium state with a high density core (condensed phase) is never
reached. The caloric
curves of classical self-gravitating systems have been obtained both in
Newtonian gravity
\cite{antonov,lbw,katzpoincare1,katzking,lecarkatz,paddyapj,paddy,katzokamoto,
dvs1,dvs2,aaiso,crs,sc,grand,katzrevue,lifetime,cn,clm1,ijmpb} and in general
relativity \cite{roupas,acb}. In Paper II, we adapt the present
formalism to the case of classical particles described by the Boltzmann entropy
and give all the necessary equations to understand these studies. We also
investigate precisely the
nonrelativistic and ultrarelativistic limits of the classical self-gravitating
gas.

\appendix

\section{General relations between the pressure and the energy density}
\label{sec_b} 

In this Appendix, we provide general relations between the pressure and the
energy density in the nonrelativistic and ultrarelativistic
limits. They are valid for an arbitrary distribution function.

\subsection{Nonrelativistic limit}
\label{sec_lnra}

In the nonrelativistic limit, the kinetic energy of a particle is
\begin{equation}
\label{bx4}
E_{\rm kin}=\frac{p^2}{2m}.
\end{equation}
In that case, the kinetic energy density and the pressure are given by
\begin{equation}
\label{bx5}
\epsilon_{\rm kin}=\int f\frac{p^2}{2m}  \, d{\bf p}\qquad {\rm and}\qquad
P=\frac{1}{3}\int f \frac{p^2}{m}\, d{\bf p}.
\end{equation}
We have the general  relations
\begin{equation}
\label{bx6}
P=\frac{2}{3}\epsilon_{\rm kin},\qquad E_{\rm kin}=\frac{3}{2}\int P\, d{\bf r}.
\end{equation}

\subsection{Ultrarelativistic limit}
\label{sec_lur}

In the ultrarelativistic limit, the energy of a particle is
\begin{equation}
\label{bx7}
E=E_{\rm kin}=pc.
\end{equation}
In that case, the energy density and the pressure are given by
\begin{equation}
\label{bx8}
\epsilon=\epsilon_{\rm kin}=\int f pc  \, d{\bf p}\qquad {\rm and}\qquad
P=\frac{1}{3}\int f p c\, d{\bf p}.
\end{equation}
We have the general  relations 
\begin{equation}
\label{bx9}
P=\frac{1}{3}\epsilon=\frac{1}{3}\epsilon_{\rm kin},\qquad {\cal E}=E_{\rm
kin}=3\int P\, d{\bf r}.
\end{equation}

\section{Virial theorem for Newtonian systems}
\label{sec_v}

In this Appendix, we establish the general expression of the equilibrium
scalar virial theorem for
Newtonian systems. For
the sake of generality, we allow the particles to be relativistic in the sense
of special relativity.

It can be shown that the virial of the gravitational force is equal to the
gravitational energy (see, e.g., Appendix G of
\cite{ggp}):
\begin{equation}
-\int\rho{\bf r}\cdot \nabla\Phi\, d{\bf r}=W.
\label{v1}
\end{equation}
This equation is general, being valid for steady and unsteady
configurations. It does not depend whether the system is spherically symmetric
or not. If we now consider  a spherically symmetric system (still allowed to be
unsteady), using the Newton law
(\ref{n4}), we find from Eq. (\ref{v1}) that
\begin{eqnarray}
\label{wnice}
W=-\int \rho \frac{GM(r)}{r}\, {d}V.
\end{eqnarray}
This formula is useful to calculate the gravitational potential energy of
a spherically
symmetric distribution of matter. It can be directly obtained by approaching
from
infinity a succession of spherical shells of
mass $dM(r)=\rho(r) 4\pi r^{2}dr$ with potential energy
$-GM(r)dM(r)/r$ in the field of the mass $M(r)$ already
present, and integrating over $r$ (see, e.g., Ref. \cite{ll}).

We now consider a self-gravitating system at equilibrium. Substituting the
condition of
hydrostatic equilibrium
\begin{equation}
\nabla P+\rho \nabla\Phi={\bf 0}
\label{v2}
\end{equation}
into Eq. (\ref{v1}), and integrating by parts, we get
\begin{equation}
W=-3\int P\, d{\bf r}+\oint P{\bf r}\cdot d{\bf S}.
\label{v3}
\end{equation}
If the system is not submitted to an external pressure, the second term on the
right hand side vanishes. On the other hand, if the external pressure is
uniform on the boundary of the system, i.e. $P({\bf
r})=P_{b}={\rm cst}$, we have
\begin{equation}
\oint P{\bf r}\cdot d{\bf S}=P_{b}\oint {\bf r}\cdot d{\bf S}=P_{b}\int
\nabla \cdot {\bf r} \, d{\bf r}=3 P_{b}V.
\label{v4}
\end{equation}
More generally, this relation can be taken as a definition of $P_{b}$.
Combining the foregoing relations, we obtain the general form of the equilibrium
scalar virial
theorem 
\begin{equation}
3\int P\, d{\bf r}+W=3 P_{b}V.
\label{v5}
\end{equation}
For nonrelativistic particles, using Eq. (\ref{bx6}), the equilibrium
scalar virial theorem
becomes
\begin{equation}
2 E_{\rm kin}+ W=3 P_{b}V.
\label{v7}
\end{equation}
Using $E=E_{\rm kin}+W$, it can be rewritten as
\begin{equation}
E=-E_{\rm kin}+3 P_{b}V.
\label{v9}
\end{equation}
For ultrarelativistic particles, using Eq. (\ref{bx9}), the equilibrium
scalar virial theorem
becomes
\begin{equation}
E=E_{\rm kin}+ W=3 P_{b}V.
\label{v10}
\end{equation}

\section{Derivation of the statistical equilibrium state for a general form of
entropy}
\label{sec_osts}

In this Appendix, we derive the statistical equilibrium state of a
self-gravitating system  for a general form
of entropy
\begin{eqnarray}
\label{qgd9}
s=-k_B\int C(f)\,
d{\bf p},
\end{eqnarray}
where $C(f)$ is any convex function (i.e. $C''(f)>0$). This is
what we call a generalized entropy \cite{gen,nfp,entropy}. These
functionals appeared
in relation to the notion of ``generalized thermodynamics'' pioneered by Tsallis
\cite{tsallis} who introduced a
particular form of non-Boltzmannian entropy (of a power-law type) called the
Tsallis entropy. These
functionals (which are particular Casimir integrals)
are also useful to obtain sufficient conditions of nonlinear dynamical
stability \cite{ipser74,ih,ipser80,aaantonov,cc} with respect to the Vlasov
equation describing a collisionless evolution of the system (see Appendix
\ref{sec_dsv}). Below we show that the maximum entropy principle can be applied
to an arbitrary form of entropy. We consider both
Newtonian and general relativistic systems. We first present a two-steps
derivation as in the main text, then a one-step derivation. 

\subsection{Two-steps derivation}
\label{sec_ts}

To maximize the entropy $S$ at fixed energy $E$ and particle number $N$,
we proceed in two steps as in Secs. \ref{sec_smnrf} and \ref{sec_grf}. We
first maximize the
entropy density $s(r)$
at fixed energy density $\epsilon(r)$ and particle number density $n(r)$ with
respect to variations
on $f({\bf r},{\bf p})$ following the steps of Secs.
\ref{sec_lmnr} and \ref{sec_step1gr}.
The variational principle
(\ref{n80}) for the extremization problem yields
\begin{eqnarray}
\label{qgd10}
C'(f)=-\beta(r)E(p)+\alpha(r).
\end{eqnarray}
Since $C$ is convex, this relation can be inverted. It determines a distribution
function of the form
\begin{eqnarray}
\label{qgd11}
f({\bf r},{\bf p})=F\left\lbrack \beta(r)E(p)-\alpha(r)\right\rbrack,
\end{eqnarray}
where 
\begin{eqnarray}
\label{qgd11b}
F(x)=(C')^{-1}(-x).
\end{eqnarray}
Since 
\begin{eqnarray}
\label{qgd11bb}
\delta^2 s=-k_B\int C''(f)\frac{(\delta f)^2}{2}\, d{\bf p}<0,
\end{eqnarray}
this distribution function is the global maximum of the entropy density
at fixed energy density and particle number density. This
corresponds to the condition of local thermodynamical equilibrium. Using the
integrated Gibbs-Duhem relation (\ref{gd19}) which is valid for
a general form of entropy (see Appendix \ref{sec_gd}), we can express the
entropy $S$ as a functional of $n(r)$ and $\epsilon(r)$. We now maximize the
entropy
$S$ at fixed energy and particle number with respect to variations on  
$n(r)$ and $\epsilon(r)$.

\subsubsection{Newtonian gravity} 
\label{sec_tsng}

We first consider the Newtonian gravity regime but, for the sake of generality,
we allow the particles to be relativistic in the sense of special relativity.
In Newtonian gravity, Eq. (\ref{qgd11}) is replaced by
\begin{eqnarray}
\label{qgd12}
f({\bf r},{\bf p})=F\left\lbrack \beta(r)E_{\rm kin}(p)-\alpha(r)\right\rbrack,
\end{eqnarray} 
where $E_{\rm kin}(p)$ is given by Eq. (\ref{n71b}). Repeating the steps of
Sec. \ref{sec_gmnr}, we obtain Eqs. (\ref{n33}) and (\ref{n35}) expressing the
uniformity of the temperature and of the total chemical potential (Gibbs law).
We also obtain the condition of
hydrostatic equilibrium (\ref{henq}). As a result,
the equilibrium distribution function at statistical equilibrium is given by
\begin{eqnarray}
\label{qgd13}
f({\bf r},{\bf p})=F\left\lbrack \beta (E_{\rm
kin}(p)+m\Phi(r))-\alpha_0\right\rbrack.
\end{eqnarray} 
In the nonrelativistic regime where $E_{\rm kin}=p^2/2m$ this is a function of
the form
\begin{eqnarray}
\label{qgd14}
f({\bf r},{\bf v})=f[\epsilon({\bf r},{\bf v})] \quad {\rm with} \quad
f'(\epsilon)<0,
\end{eqnarray} 
where $\epsilon({\bf r},{\bf v})=v^2/2+\Phi(r)$ is the energy of a particle by
unit of mass and
we have introduced the velocity ${\bf v}={\bf p}/m$ instead of the impulse
${\bf p}$. We note that an extremum of
entropy at fixed energy and particle number is necessarily
isotropic. Repeating the arguments of
Sec. \ref{sec_gmnr}, we can show that the gas corresponding to the distribution
function (\ref{qgd13}) is described by a barotropic equation of state
$P(r)=P[\rho(r),T]$, where the function $P(\rho,T)$ is determined by the
function $C(f)$ characterizing the entropy.

\subsubsection{General relativity}

We now consider the general relativity case.  Repeating the steps of
Sec.  \ref{sec_step2gr}, we obtain Eq. (\ref{b98}).
We also obtain the TOV
equations (\ref{b115ov}) and (\ref{n57b}) expressing the condition of
hydrostatic equilibrium and
the Tolman-Klein relations (\ref{b103}) and (\ref{b104}). As a
result, the equilibrium distribution function at statistical equilibrium is
given by
\begin{eqnarray}
\label{qgd15}
f({\bf r},{\bf p})=F\left\lbrack \frac{E(p)}{k_B
T(r)}-\alpha\right\rbrack.
\end{eqnarray}
Using Eq.  (\ref{b103}), it can be written as
\begin{eqnarray}
\label{qgd16}
f({\bf r},{\bf p})=F\left\lbrack \beta_{\infty}
e^{\nu(r)/2} E(p)-\alpha\right\rbrack.
\end{eqnarray}
This is a function of the form
\begin{eqnarray}
\label{qgd17}
f({\bf r},{\bf p})=f\left\lbrack e^{\nu(r)/2} E(p)\right\rbrack \quad {\rm
with} \quad f'\left\lbrack e^{\nu(r)/2} E(p)\right\rbrack<0,
\end{eqnarray} 
where $E(p)$ is the energy of a particle. We note that an extremum of
entropy at fixed mass-energy and particle number is necessarily
isotropic. Repeating the arguments of
Sec. \ref{sec_lta}, we can show that the gas corresponding to the
distribution
function (\ref{qgd15}) is described by a barotropic equation of state
$P(r)=P[\alpha,\epsilon(r)]$, where the function $P(\epsilon,\alpha)$ is
determined by the
function $C(f)$ characterizing the entropy.

\subsection{One-step derivation}
\label{sec_os}

We now present a one-step derivation of the preceding results. We 
first consider the Newtonian gravity case. The generalized entropy is
\begin{eqnarray}
\label{qgd9b}
S=-k_B\int C(f)\, d{\bf r}d{\bf p}.
\end{eqnarray}
The particle number and the mass are
given by 
\begin{equation}
\label{ax2}
M=Nm=m\int f\, d{\bf r} d{\bf p}=\int m n\, d{\bf r}=\int \rho\, d{\bf r}.
\end{equation}
The energy is given by
\begin{equation}
\label{ax3}
E=E_{\rm kin}+W=\int f E_{\rm kin}(p) \,d{\bf r} d{\bf
p}+\frac{1}{2}\int
\rho \Phi\, d{\bf r},
\end{equation}
where $E_{\rm kin}$ is the kinetic
energy and $W$ is the potential (gravitational) energy.

In the microcanonical ensemble, the statistical equilibrium state is obtained by
maximizing the entropy $S$ at fixed energy $E$  and particle number $N$ with
respect to
variations on $f({\bf r},{\bf p})$. We write the
variational problem for the first
variations (extremization) as
\begin{equation}
\label{ax9}
\frac{\delta S}{k_B}-\beta\delta E+\alpha_0\delta N=0,
\end{equation}
where $\beta$ and $\alpha_0$ are global Lagrange multipliers. Taking the
variations
with respect to $f({\bf r},{\bf p})$, we obtain
\begin{eqnarray}
\label{qgd18}
C'(f)=-\beta (E_{\rm kin}(p)+m\Phi({\bf r}))+\alpha_0,
\end{eqnarray}
leading to
\begin{eqnarray}
\label{qgd19}
f({\bf r},{\bf p})=F\left\lbrack \beta (E_{\rm
kin}(p)+m\Phi({\bf r}))-\alpha_0\right\rbrack,
\end{eqnarray} 
where $F(x)$ is defined by Eq. (\ref{qgd11b}). This returns the result
from Eq. (\ref{qgd13}).
The temperature $T$ and the chemical
potential $\mu_0$ are related to $\beta$
and $\alpha_0$ by
\begin{equation}
\label{ax11}
\beta=\frac{1}{k_B T},\qquad \alpha_0=\frac{\mu_0}{k_B T}.
\end{equation}
We can then rewrite Eq. (\ref{qgd19}) as
\begin{equation}
\label{ax12}
f({\bf r},{\bf p})=F\left\lbrack \frac{1}{k_B T}\left (E_{\rm
kin}(p)+m\Phi({\bf r})-\mu_0\right )\right\rbrack.
\end{equation}

For the Fermi-Dirac  entropy  
\begin{eqnarray}
\label{ax7}
S=-k_B\frac{g}{h^3}\int\Biggl\lbrace \frac{f}{f_{\rm max}}\ln
\frac{f}{f_{\rm
max}}
+\left (1-\frac{f}{f_{\rm max}}\right )\ln\left
(1-\frac{f}{f_{\rm max}}\right )\Biggr\rbrace\,  d{\bf r} d{\bf p},
\end{eqnarray}
we obtain the mean field Fermi-Dirac distribution 
\begin{equation}
\label{ax10}
f({\bf r},{\bf p})=\frac{g}{h^3}\frac{1}{1+e^{-\alpha_0}e^{\beta (E_{\rm
kin}(p)+m\Phi({\bf r}))}}
\end{equation}
or, equivalently,
\begin{equation}
\label{ax13}
f({\bf r},{\bf p})=\frac{g}{h^3}\frac{1}{1+e^{(E_{\rm kin}(p)+m\Phi({\bf
r})-\mu_0)/k_B
T}}.
\end{equation}
For the Boltzmann entropy 
\begin{eqnarray}
\label{ax14}
S&=&-k_B\int f\left\lbrack \ln \left (\frac{f}{f_{\rm
max}}\right )-1\right\rbrack \,  d{\bf r} d{\bf p},
\end{eqnarray}
we obtain the mean field Maxwell-Boltzmann distribution 
\begin{equation}
\label{ax15}
f({\bf r},{\bf p})=\frac{g}{h^3}e^{\alpha_0}e^{-\beta (E_{\rm
kin}(p)+m\Phi({\bf
r}))}
\end{equation}
or, equivalently,
\begin{equation}
\label{ax12b}
f({\bf r},{\bf p})=\frac{g}{h^3}e^{-(E_{\rm kin}(p)+m\Phi({\bf
r})-\mu_0)/k_B
T}.
\end{equation}

This one-step derivation of the statistical equilibrium state, valid  for a
generalized entropy of the form (\ref{qgd9}), was given in
Newtonian gravity by Ipser
\cite{ipser74,ih}, Tremaine {\it et al.} \cite{thlb} and Chavanis \cite{gen},
directly leading to Eq. (\ref{qgd13}).
It was extended in general relativity by Ipser \cite{ipser80},  directly leading
to Eq. (\ref{qgd16}).

\section{Condition of hydrostatic equilibrium for a general form of entropy}
\label{sec_hea}

In this Appendix, we show by a direct calculation that the condition of
statistical equilibrium, obtained by extremizing the entropy at fixed
energy and particle number, implies
the condition of hydrostatic equilibrium. We consider a general form of entropy
given by Eq.
(\ref{qgd9}). 

\subsection{Newtonian gravity}
\label{sec_heang}

We first consider the Newtonian gravity case but, for the sake of generality, we
allow the particles to be relativistic in the sense of special relativity. The
extremization of the entropy $S$ at fixed particle number $N$ and
energy $E$ leads to a distribution function of the form (see
Appendix \ref{sec_osts})
\begin{eqnarray}
\label{hea1}
f({\bf r},{\bf p})=F\left\lbrack \beta \left
(E_{\rm kin}(p)+m\Phi({\bf r})\right )-\alpha_0\right\rbrack,
\end{eqnarray}
where $F$ is defined by Eq. (\ref{qgd11b}) and where $\beta$ and $\alpha_0$ are
constant. According to Eqs. (\ref{n74}) and (\ref{hea1}) the
pressure is given by
\begin{eqnarray}
\label{hea2}
P=\frac{1}{3}\int f p \frac{dE_{\rm kin}}{dp}\, d{\bf p}=\frac{1}{3}\int
F\left\lbrack \beta
\left
(E_{\rm kin}(p)+m\Phi({\bf r})\right )-\alpha_0\right\rbrack p \frac{dE_{\rm
kin}}{dp}\,
d{\bf p}.
\end{eqnarray}
Taking its gradient with respect to ${\bf r}$, we get
\begin{eqnarray}
\label{hea3}
\nabla P=\frac{1}{3}\beta m \nabla\Phi \int F'\left\lbrack
\beta
\left
(E_{\rm kin}(p)+m\Phi({\bf r})\right )-\alpha_0\right\rbrack p \frac{dE_{\rm
kin}}{dp}  \,
d{\bf p}.
\end{eqnarray}
This can also be written as
\begin{eqnarray}
\label{hea4}
\nabla P=\frac{1}{3} m \nabla\Phi \int {\bf p}\cdot \frac{\partial
F}{\partial {\bf p}}\left\lbrack
\beta
\left
(E_{\rm kin}(p)+m\Phi({\bf r})\right )-\alpha_0\right\rbrack \,
d{\bf p}.
\end{eqnarray}
Integrating by parts, we can rewrite the foregoing equation as
\begin{eqnarray}
\label{hea5}
\nabla P=- m \nabla\Phi \int 
F\left\lbrack
\beta
\left
(E_{\rm kin}(p)+m\Phi({\bf r})\right )-\alpha_0\right\rbrack \,
d{\bf p}.
\end{eqnarray}
Since the  density is given by
\begin{eqnarray}
\label{hea6}
\rho = m\int f\, d{\bf p}=m  \int 
F\left\lbrack
\beta
\left
(E_{\rm kin}(p)+m\Phi({\bf r})\right )-\alpha_0\right\rbrack \,
d{\bf p},
\end{eqnarray}
we finally obtain the condition of hydrostatic equilibrium
\begin{eqnarray}
\label{hea7}
\nabla P=-\rho\nabla\Phi.
\end{eqnarray}

\subsection{General relativity}
\label{sec_heagr}

We now consider the general relativity case. The
extremization of the entropy $S$ at fixed mass-energy ${\cal E}$ and particle
number
$N$ leads to a distribution function of the form (see Appendix \ref{sec_osts})
\begin{eqnarray}
\label{hrg1}
f({\bf r},{\bf p})=F\left\lbrack \frac{E(p)}{k_B
T(r)}-\alpha\right\rbrack,
\end{eqnarray}
where $F$ is defined by Eq. (\ref{qgd11b}) and where $\alpha$ is
constant. According to Eqs. (\ref{n74}) and (\ref{hrg1}) the
pressure is given by
\begin{eqnarray}
\label{hrg2}
P=\frac{1}{3}\int f p \frac{dE}{dp}\, d{\bf p}=\frac{1}{3}\int F\left\lbrack
\frac{E(p)}{k_B
T(r)}-\alpha\right\rbrack p
\frac{dE}{dp}\, d{\bf p}.
\end{eqnarray}
Taking its derivative with respect to $r$, we obtain
\begin{eqnarray}
\label{hrg3}
\frac{dP}{dr}=-\frac{1}{3} \frac{1}{k_B T(r)^2}\frac{dT}{dr} \int
F'\left\lbrack
\frac{E(p)}{k_B
T(r)}-\alpha\right\rbrack p
\frac{dE}{dp}E(p)\, d{\bf p}.
\end{eqnarray}
This can also be written as
\begin{eqnarray}
\label{hrg4}
\frac{dP}{dr}=-\frac{1}{3} \frac{1}{T(r)}\frac{dT}{dr} \int_{0}^{+\infty}
\frac{\partial F}{\partial p}\left\lbrack
\frac{E(p)}{k_B
T(r)}-\alpha\right\rbrack p
E(p) 4\pi p^2\, dp.
\end{eqnarray}
Integrating by parts, we can rewrite the foregoing equation as
\begin{eqnarray}
\label{hrg5}
\frac{dP}{dr}=\frac{1}{3} \frac{1}{T(r)}\frac{dT}{dr} \int_{0}^{+\infty}
F\left\lbrack
\frac{E(p)}{k_B
T(r)}-\alpha\right\rbrack \left (3
E(p) p^2+p^3\frac{dE}{dp}\right ) 4\pi \, dp.
\end{eqnarray}
Since the pressure is given by Eq. (\ref{hrg2}) and the energy density
by
\begin{eqnarray}
\label{hrg6}
\epsilon=\int f E(p) \, d{\bf p}=\int
F\left\lbrack
\frac{E(p)}{k_B
T(r)}-\alpha\right\rbrack E(p)\, d{\bf p},
\end{eqnarray}
we finally obtain the equation
\begin{eqnarray}
\label{hrg7}
\frac{dP}{dr}=\frac{\epsilon(r)+P(r)}{T(r)} \frac{dT}{dr}.
\end{eqnarray}
Combined with Eq. (\ref{b115}) it leads to the OV equation (\ref{b115ov}). From
Eqs. (\ref{n56}),
(\ref{n58}) and (\ref{b115ov}) we
then obtain the Tolman equation (\ref{n57b})  which expresses the condition
of hydrostatic equilibrium.

\section{Gibbs-Duhem relation}
\label{sec_gd}

In this Appendix, we derive the Gibbs-Duhem and 
integrated
Gibbs-Duhem relations. We first
recall the usual
derivation of these relations which
explicitly uses
the extensivity of the entropy. Then, we provide a direct derivation of
the 
integrated Gibbs-Duhem relation for an arbitrary form of
entropy without explicitly using the extensivity assumption. This shows that our
thermodynamical formalism is valid for an arbitrary form of entropy.

\subsection{Standard derivation}
\label{sec_gds}

The first law of thermodynamics can be written as
\begin{eqnarray}
\label{gd1}
dE=-PdV+TdS+\mu dN.
\end{eqnarray}
An extensive variable (energy, entropy,...) is proportional to the absolute size
of the
system. In other words, if one doubles all extensive variables, all other
extensive quantities also become twice as large. For example,
\begin{eqnarray}
\label{gd2}
E(\alpha S,\alpha V,\alpha N)=\alpha E(S,V,N),
\end{eqnarray}
where $\alpha$ is the enlargement factor. One calls functions which have this
property homogeneous functions of first order. All extensive variables are
homogeneous functions of first order of the other extensive variables. On the
other hand, the intensive variables (temperature, pressure...) are homogenous
functions of zeroth order
of the extensive variables, i.e., they do not change if we divide or duplicate
the system. For example,
\begin{eqnarray}
\label{gd3}
T(\alpha S,\alpha V,\alpha N)=T(S,V,N).
\end{eqnarray}
According to the Euler theorem, we have
\begin{eqnarray}
\label{gd4}
E=-PV+TS+\mu N.
\end{eqnarray}
Differentiating this expression and using the first law of thermodynamics
(\ref{gd1}), we get the Gibbs-Duhem relation
\begin{eqnarray}
\label{gd5}
SdT-VdP+Nd\mu=0.
\end{eqnarray}
We note that the energy does not appear in this expression. The Euler equation
(\ref{gd4}) for thermodynamic variables is also called the integrated
Gibbs-Duhem relation.

Defining $s=S/V$, $n=N/V$ and $\epsilon=E/V$, the local Gibbs-Duhem and the
integrated Gibbs-Duhem relations can be written
as
\begin{eqnarray}
\label{gd6}
sdT-dP+nd\mu=0
\end{eqnarray}
and
\begin{eqnarray}
\label{gd7}
\epsilon=-P+Ts+\mu n.
\end{eqnarray}
Introducing $S=sV$, $N=nV$ and $E=\epsilon V$ in the first law of
thermodynamics
(\ref{gd1}), developing the expression, and using the integrated
Gibbs-Duhem relation (\ref{gd7}), we obtain
\begin{eqnarray}
\label{gd8}
d\epsilon=Tds+\mu dn.
\end{eqnarray}
We note that the pressure does not explicitly appear in this expression. This is
the local form of the first law of thermodynamics [see Eq.
(\ref{b84})]. 

\subsection{Direct derivation of the integrated Gibbs-Duhem relation for a
general form of entropy}
\label{sec_gda}

The local condition of thermodynamical equilibrium, obtained by maximizing the
local
entropy at fixed energy density and particle number density, is given by Eq.
(\ref{qgd11}) with Eq. (\ref{qgd11b}). Substituting Eq. (\ref{qgd11}) into Eqs.
(\ref{n68}), (\ref{n69}) and (\ref{n74}) we find that the
particle number density, the energy density and the pressure are given by
\begin{eqnarray}
\label{gd12}
n(r)=\int_0^{+\infty} F\left\lbrack \beta(r)E(p)-\alpha(r)\right\rbrack 4\pi
p^2\, dp,
\end{eqnarray}
\begin{eqnarray}
\label{gd13}
\epsilon(r)=\int_0^{+\infty} F\left\lbrack \beta(r)E(p)-\alpha(r)\right\rbrack
E(p) 4\pi
p^2\, dp,
\end{eqnarray}
\begin{eqnarray}
\label{gd14}
P(r)=\frac{1}{3}\int_0^{+\infty} F\left\lbrack
\beta(r)E(p)-\alpha(r)\right\rbrack
p E'(p) 4\pi
p^2\, dp.
\end{eqnarray}
On the other hand, the entropy density [see Eq. (\ref{qgd9})]
is given by
\begin{eqnarray}
\label{gd15}
s(r)=-k_B \int_0^{+\infty} C\left\lbrace F\left\lbrack
\beta(r)E(p)-\alpha(r)\right\rbrack\right\rbrace  4\pi
p^2\, dp.
\end{eqnarray}
Integrating this equation by parts and using $C'\lbrack F(x)\rbrack=-x$, we
get
\begin{eqnarray}
\label{gd16}
s(r)=-k_B \int_0^{+\infty} \left\lbrack
\beta(r)E(p)-\alpha(r)\right\rbrack F'\left\lbrack
\beta(r)E(p)-\alpha(r)\right\rbrack \beta(r) E'(p)  \frac{4\pi}{3}
p^3\, dp.
\end{eqnarray}
Integrating by parts on more time, we obtain
\begin{equation}
\label{gd17}
s(r)=k_B \int_0^{+\infty} \beta(r) E'(p)  F\left\lbrack
\beta(r)E(p)-\alpha(r)\right\rbrack  \frac{4\pi}{3} p^3\, dp+k_B
\int_0^{+\infty} \left\lbrack
\beta(r)E(p)-\alpha(r)\right\rbrack F\left\lbrack
\beta(r)E(p)-\alpha(r)\right\rbrack 4\pi p^2\, dp.
\end{equation}
Comparing Eq. (\ref{gd17}) with Eqs. (\ref{gd12})-(\ref{gd14}), we
find that
\begin{eqnarray}
\label{gd18}
\frac{s(r)}{k_B}=\beta(r)P(r)+\beta(r)\epsilon(r)-\alpha(r)n(r).
\end{eqnarray}
Using Eq. (\ref{n82}), we finally obtain the integrated Gibbs-Duhem relation
\begin{eqnarray}
\label{gd19}
s(r)=\frac{
\epsilon(r)+P(r)-\mu(r)n(r)}{T(r)}.
\end{eqnarray}
This calculation emphasizes the fact that the  integrated
Gibbs-Duhem relation is valid for an arbitrary form of entropy and
for an arbitrary level of relativity.\footnote{This is an interesting result
because
there is a lot of polemic related to the notion of ``generalized
thermodynamics'' introduced by Tsallis 
\cite{tsallis}. The present calculation shows that standard
thermodynamics is actually valid for an arbitrary form of entropy (\ref{qgd9})
\cite{gen}.}

{\it Remark:} The calculations presented in this Appendix are equivalent to
those performed in Appendix B of \cite{aaantonov}, in Appendix C of
\cite{wddimd} and in Appendix D of \cite{nfp} although the connection with the
integrated Gibbs-Duhem relation was
not realized at that time.

\section{Entropy and free energy as functionals of the density for
Newtonian self-gravitating systems}
\label{sec_alt}

We consider a Newtonian self-gravitating system but, for the sake of
generality, we allow the particles to be relativistic in the sense of special
relativity. We also consider a general form of entropy given by Eq.
(\ref{qgd9b}). The statistical equilibrium state
is
obtained by maximizing the entropy at fixed energy and particle number in the
microcanonical ensemble, or
by minimizing the free energy at fixed particle number in the canonical
ensemble.  In Appendix
\ref{sec_os}, we have introduced entropy and free energy functionals of the
distribution
function $f({\bf
r},{\bf v})$. In Sec. \ref{sec_gmnr} and in Appendix \ref{sec_tsng}, we have
introduced entropy and free energy functionals of the local density $n(r)$ and
local kinetic energy
$\epsilon_{\rm
kin}(r)$. In this Appendix, we introduce entropy and free energy functionals
of the local density $n(r)$.

\subsection{Microcanonical ensemble}
\label{sec_altmce}

In the microcanonical ensemble,  the statistical equilibrium state is
obtained by maximizing the entropy $S[f]$ at fixed energy $E$ and particle
number $N$. To solve this
maximization problem, we proceed in two steps. We first maximize 
$S[f]$ at fixed $E$, $N$
{\it and} particle density $n(r)$. Since $n(r)$ determines the particle number 
$N[n]$ and the gravitational energy $W[n]$,
this is equivalent to maximizing $S[f]$  at fixed kinetic energy $E_{\rm kin}$
and
particle density $n(r)$. The variational problem
for the first variations (extremization) can be written as
\begin{equation}
\label{alt2}
\frac{\delta S}{k_B}-\beta\delta E_{\rm kin}+\int\alpha(r)\delta n \, d{\bf
r}=0,
\end{equation}
where $\beta$ is a global (uniform) Lagrange multiplier and $\alpha(r)$ is a
local (position dependent) Lagrange multiplier. This variational problem, which
is equivalent to
\begin{equation}
\label{alt2b}
\frac{\delta s}{k_B}-\beta\delta \epsilon_{\rm kin}+\alpha(r)\delta n=0,
\end{equation}
returns
the  results of Appendix  \ref{sec_ts}, except
that $\beta(r)$ is replaced by $\beta$. Therefore, it yields
\begin{eqnarray}
\label{alt3}
f({\bf r},{\bf p})=F\left\lbrack \beta E_{\rm kin}(p)-\alpha(r)\right\rbrack.
\end{eqnarray}
In this manner, we immediately find that $T$ is uniform at statistical
equilibrium. This results from the conservation of energy. As in Appendix
\ref{sec_ts}, we can show that the distribution (\ref{alt3}) is the global
maximum of
$S[f]$  at fixed $E_{\rm kin}$ and $n(r)$. Substituting Eq.
(\ref{alt3}) into Eqs. (\ref{n7}),
(\ref{n8}) and (\ref{n10}), we get
\begin{equation}
\label{alt4}
n(r)=\int  F\left\lbrack \beta E_{\rm kin}(p)-\alpha(r)\right\rbrack \, d{\bf
p},
\end{equation}
\begin{equation}
\label{alt5}
\epsilon_{\rm kin}(r)=\int  F\left\lbrack \beta E_{\rm
kin}(p)-\alpha(r)\right\rbrack E_{\rm
kin}(p)\, d{\bf
p},
\end{equation}
\begin{equation}
\label{alt6}
P(r)=\frac{1}{3}\int  F\left\lbrack \beta E_{\rm
kin}(p)-\alpha(r)\right\rbrack  p E_{\rm
kin}'(p)\, d{\bf
p}.
\end{equation}
The Lagrange
multiplier $\alpha(r)$ is determined by the density $n(r)$ according to Eq.
(\ref{alt4}).  On the other hand, the temperature $T$ is determined by the
kinetic energy $E_{\rm kin}[n(r),T]=E-W[n(r)]$ using Eq. (\ref{alt5})
integrated over the volume. In other words, the temperature  is determined by
the energy constraint
\begin{eqnarray}
\label{alt22}
E=E_{\rm kin}[n(r),T]+W[n(r)].
\end{eqnarray}
We note that $T$ is a functional of the density $n(r)$ but, for brevity, we
shall not write this dependence explicitly. 

Repeating the
steps of Appendix \ref{sec_gda},
we can derive
the integrated Gibbs-Duhem relation (\ref{gd19}), except
that $T(r)$ is replaced by $T$. Therefore, we get  
\begin{eqnarray}
\label{alt7}
s(r)=\frac{\epsilon_{\rm kin}(r)+P(r)-\mu(r)n(r)}{T}\quad {\rm with}\quad
\mu(r)=\alpha(r) k_B T.
\end{eqnarray}
Since  $T$ is uniform, Eq. (\ref{n29}) reduces to
\begin{equation}
\label{alt8}
d\mu=\frac{dP}{n}.
\end{equation}
On the other hand, eliminating formally $\alpha(r)$ between Eqs.
(\ref{alt4}) and (\ref{alt6}), we see that the equation of state is
barotropic: $P(r)=P[n(r),T]$ (we have explicitly written the temperature
$T$ because it is uniform but not constant when we consider variations of
$n(r)$ as explained above). Therefore, according to Eq. (\ref{alt8}) we have
$\mu(r)=\mu[n(r),T]$ with
\begin{equation}
\label{alt9}
\mu'(n,T)=\frac{P'(n,T)}{n}, \quad {\rm i.e.}
\quad \mu(n,T)=\int^{n}\frac{P'(n',T)}{n'}\, dn',
\end{equation}
where the derivative is with respect to $n$.\footnote{This relation determines
the chemical potential $\mu$ up to an additive constant that may depend on the
temperature $T$. The complete expression of the chemical potential can be
obtained from Eq. (\ref{alt4}).} 

We can now simplify the expression of the entropy.  Using the integrated
Gibbs-Duhem relation (\ref{alt7}), we have
\begin{eqnarray}
\label{alt10}
S=\frac{1}{T}\left (E_{\rm kin}+\int P(r)\, d{\bf r}-\int \mu(r)n(r)\, d{\bf
r}\right ).
\end{eqnarray}
The entropy can be written as a functional of
the density as
\begin{eqnarray}
\label{alt23}
S[n(r),T]=\frac{1}{T}\left (E_{\rm kin}[n(r),T]-U[n(r),T]\right
),
\end{eqnarray}
or, using Eq. (\ref{alt22}), as
\begin{eqnarray}
S[n(r),T]=\frac{1}{T}\left (E-W[n(r)]-U[n(r),T]\right
),
\end{eqnarray}
where $U[n(r),T]$ is the internal energy given by 
\begin{equation}
\label{alt14}
U[n(r),T]=\int V(n(r),T)\, d{\bf r}\quad {\rm with}\quad  V(n,T)=n
\mu(n,T)-P(n,T).
\end{equation}
Combining Eqs. (\ref{alt9}) and (\ref{alt14}), we get
\begin{equation}
\label{alt15}
V'(n,T)=\mu(n,T).
\end{equation}
Therefore, the pressure $P(n,T)$ is related to the density of internal energy
$V(n,T)$ by
\begin{equation}
\label{alt16}
P(n,T)=n\mu(n,T)-V(n,T)=n
V'(n,T)-V(n,T)=n^2\left\lbrack\frac{V(n,T)}{n}\right\rbrack'.
\end{equation}
Inversely, the  density of internal energy is determined by the equation of
state $P[n(r),T]$ according to the relation
\begin{equation}
\label{alt17}
V(n,T)=n\int^n \frac{P(n',T)}{{n'}^2}\, dn'.
\end{equation}
We note the identities
\begin{equation}
\label{alt18}
V'(n,T)=\int^{n} \frac{P'(n',T)}{n'}\, dn'\qquad {\rm and}\qquad 
V''(n,T)=\frac{P'(n,T)}{n}.
\end{equation}
The internal energy can be written explicitly as
\begin{equation}
\label{alt17b}
U[n(r),T]=\int n\int^n \frac{P(n',T)}{{n'}^2}\, dn' d{\bf r}.
\end{equation}

Finally, the statistical equilibrium state in the microcanonical ensemble is
obtained by maximizing the entropy $S[n]$ at fixed particle
number $N$, the
energy constraint being taken into account in the determination of the
temperature $T[n]$ through
the relation (\ref{alt22}).  The
variational problem
for the first variations (extremization) can be written as
\begin{equation}
\label{alt18b}
\frac{\delta S}{k_B}+\alpha_0\delta N=0.
\end{equation}
The conservation of energy implies [see Eq. (\ref{alt22})]:
\begin{eqnarray}
\label{alt22b}
0=\delta E_{\rm kin}+\int m\Phi\delta n\, d{\bf r}.
\end{eqnarray}
Using Eqs. (\ref{alt2}) and  (\ref{alt22b}), we get
\begin{equation}
\label{alt18c}
\frac{\delta S}{k_B}=-\beta \int m\Phi\delta n\, d{\bf r}-\int\alpha(r)\delta n
\, d{\bf
r}.
\end{equation}
As a result, the variational problem (\ref{alt18b}) yields
\begin{equation}
\label{alt18d}
\alpha(r)=\alpha_0-\beta m\Phi(r).
\end{equation}
We then recover all the results of Sec. \ref{sec_smnrf}. The interest of this
formulation it that it allows us to solve more easily the stability problem
related
to
the sign of the second variations of entropy. This problem  has
been studied in detail in \cite{paddyapj,paddy,sc} for the Boltzmann entropy and
in \cite{ts1,ts2,lang} for the Tsallis entropy. It has also been studied in
\cite{aarelat1} for the Boltzmann entropy within the framework of special
relativity.

\subsection{Canonical ensemble}
\label{sec_altce}

In the canonical ensemble,  the statistical equilibrium state is
obtained by minimizing the free energy $F[f]=E[f]-TS[f]$ at fixed particle
number $N$, or equivalently, by maximizing the Massieu function  
$J[f]=S[f]/k_B-\beta E[f]$ at fixed  particle number $N$. To solve this
maximization problem, we proceed in two steps. We first maximize 
$J[f]=S[f]/k_B-\beta E[f]$ at fixed $N$
{\it and} particle density $n(r)$. Since $n(r)$ determines the particle
number $N[n]$ and the gravitational energy $W[n]$,
this is equivalent to maximizing $S[f]/k_B-\beta E_{\rm kin}[f]$  at fixed
particle density $n(r)$. The variational problem
for the first variations (extremization) can be written as
\begin{equation}
\label{calt1}
\delta \left (\frac{S}{k_B}-\beta E_{\rm
kin}\right )+\int\alpha(r)\delta n \, d{\bf
r}=0,
\end{equation}
where $\alpha(r)$ is a local (position dependent)  Lagrange multiplier.
Since $\beta$ is constant in the canonical ensemble, this is equivalent to the
conditions (\ref{alt2}) and (\ref{alt2b}) yielding the distribution function
(\ref{alt3}). This distribution
is the global  maximum of
$S[f]/k_B-\beta E_{\rm kin}[f]$  at fixed $n(r)$. We then obtain the
same results as in Appendix \ref{sec_altmce}, except that  $T$ is fixed
while it was previously determined by the conservation of energy (\ref{alt22}).

We can now simplify the expression of the free energy. The
entropy is given by Eq. (\ref{alt23}) and the energy by Eq. (\ref{alt22}). 
Since $F=E-TS$, we obtain 
\begin{equation}
\label{alt13}
F[n(r),T]=U[n(r),T]+W[n(r)],
\end{equation}
where $U[n]$ is the internal energy given by Eq.
(\ref{alt14}). The statistical equilibrium state in the canonical ensemble is
obtained by minimizing the free energy $F[n]$ at fixed particle
number $N$. The variational problem
for the first variations (extremization) can be written as
\begin{equation}
\label{alt19b}
\delta J+\alpha_0\delta N=0.
\end{equation}
Decomposing the Massieu function as $J[f]=S[f]/k_B-\beta E_{\rm
kin}[f]-\beta W[n]$ and using Eq. (\ref{calt1}), we get
\begin{equation}
\label{alt18cb}
\delta J=-\int\alpha(r)\delta n
\, d{\bf
r}-\beta \int m\Phi\delta n\, d{\bf r}.
\end{equation}
As a result, the variational problem (\ref{alt19b}) yields
\begin{equation}
\label{calt18d}
\alpha(r)=\alpha_0-\beta m\Phi(r).
\end{equation}
We then recover all the results of Sec. \ref{sec_smnrf}. The interest of this
formulation is that it allows us to solve more easily the stability problem
related
to
the sign of the second variations of free energy. This
problem has
been studied in detail in \cite{aaiso,sc} for the Boltzmann free energy and
in \cite{aapoly,lang,ts2} for the Tsallis free energy.
It has also been studied
in
\cite{aarelat1} for the Boltzmann free energy within the framework of special
relativity. 

{\it Remark:} Using Eq. (\ref{alt17b}), we see that the free
energy
(\ref{alt13}) can be written as\footnote{A more direct
derivation of this result is given in Appendix \ref{sec_tins}.}
\begin{equation}
\label{alt19}
F[\rho]=\int \rho\int^{\rho} \frac{P(\rho')}{{\rho'}^2}\, d\rho'\, d{\bf
r}+\frac{1}{2}\int \rho\Phi\, d{\bf r}.
\end{equation}
We have not explicitly written the temperature $T$ since it is a constant in
the canonical ensemble. Up to
the kinetic term, Eq. (\ref{alt19}) coincides with the energy
functional (\ref{dseNG1}) associated with the Euler-Poisson equations
describing a
gas with a
barotropic equation of state $P=P(\rho)$ (see Appendix \ref{sec_dseNG}).  As a
result, {\it the thermodynamical
stability of a self-gravitating system in the canonical ensemble is equivalent
to the dynamical stability of the corresponding barotropic gas described by
the Euler-Poisson equations}.  This returns  the general result established in
\cite{aaantonov}. It is valid for an arbitrary form of
entropy. According to the Poincar\'e turning point criterion, the
series of
equilibria
becomes both thermodynamically unstable (in the canonical ensemble) and
dynamically
unstable with respect to the Euler-Poisson equations at the first turning point
of temperature (or, equivalently, at the first turning point of mass).

\subsection{Scaling of the equation of state in the nonrelativistic and
ultrarelativistic limits}
\label{sec_scaling}

We have seen that the equation of state implied by the distribution function
(\ref{alt3}) is of the form $P(r)=P[n(r),T]$. A simple scaling of this
equation of state can be obtained
in the nonrelativistic and ultrarelativistic limits.

In the nonrelativistic limit, using $E_{\rm kin}=p^2/2m$,  Eqs.
(\ref{alt4})-(\ref{alt6}) reduce to
\begin{equation}
\label{alt4nr}
n(r)=\int  F\left\lbrack  \frac{\beta p^2}{2m}-\alpha(r)\right\rbrack \, d{\bf
p},
\end{equation}
\begin{equation}
\label{alt5nr}
\epsilon_{\rm kin}(r)=\int  F\left\lbrack \frac{\beta
p^2}{2m}-\alpha(r)\right\rbrack \frac{p^2}{2m}\, d{\bf
p},
\end{equation}
\begin{equation}
\label{alt6nr}
P(r)=\frac{1}{3}\int  F\left\lbrack \frac{\beta
p^2}{2m}-\alpha(r)\right\rbrack  \frac{p^2}{m}\, d{\bf
p}.
\end{equation}
Making the change of variables ${\bf x}=(\beta/m)^{1/2}{\bf p}$, we obtain the
scaling
\begin{equation}
\label{scalnr}
P(n,T)=T^{5/2}{\Pi}_{\rm NR}\left (\frac{n}{T^{3/2}}\right ).
\end{equation}
Therefore, the internal energy (\ref{alt17b}) takes the form
\begin{equation}
\label{alt17bnr}
U[n(r),T]=T\int n\int^{n/T^{3/2}} \frac{\Pi(x)}{{x}^2}\, dx d{\bf r}.
\end{equation}
For the Boltzmann entropy in phase space $S_{B}[f]$, leading to the isothermal
equation of state $P=n k_B T$, the free energy is of the form $F[n]=W[n]-T
S_B[n]$ where $S_B[n]$ is the Boltzmann entropy in configuration space
(see \cite{sc} for details). For the Tsallis entropy in phase space $S_{q}[f]$,
leading to the polytropic equation of state $P=K(T) n^{\gamma}$, the free
energy is of the form $F[n]=W[n]-K(T) S_{\gamma}[n]$ where $S_{\gamma}[n]$
is the Tsallis entropy in configuration space (see
\cite{lang} for details). In general, we do {\it not} have $F[n]=W[n]-T
S[n]$ (except for Boltzmann) nor $F[n]=W[n]-\Theta(T)
S[n]$ (except for Tsallis).

In the ultrarelativistic limit, using $E_{\rm kin}=pc$,  Eqs.
(\ref{alt4})-(\ref{alt6}) reduce to
\begin{equation}
\label{alt4ur}
n(r)=\int  F\left\lbrack  \beta p c -\alpha(r)\right\rbrack \, d{\bf
p},
\end{equation}
\begin{equation}
\label{alt5ur}
\epsilon_{\rm kin}(r)=\int  F\left\lbrack \beta p c
-\alpha(r)\right\rbrack pc \, d{\bf
p},
\end{equation}
\begin{equation}
\label{alt6ur}
P(r)=\frac{1}{3}\int  F\left\lbrack \beta p c-\alpha(r)\right\rbrack 
p c \, d{\bf
p}.
\end{equation}
Making the change of variables ${\bf x}=\beta {\bf p} c$, we obtain the
scaling
\begin{equation}
\label{scalur}
P(n,T)=T^{4}{\Pi}_{\rm UR}\left (\frac{n}{T^{3}}\right ).
\end{equation}
Therefore, the internal energy (\ref{alt17b}) takes the form
\begin{equation}
\label{alt17bur}
U[n(r),T]=T\int n\int^{n/T^{3}} \frac{\Pi(x)}{{x}^2}\, dx d{\bf r}.
\end{equation}

\subsection{General relativity}
\label{sec_preliminaire}

Let us briefly consider the general relativity case. 
In the microcanonical ensemble,  the statistical equilibrium
state is
obtained by maximizing the entropy $S[f]$ at fixed mass-energy $Mc^2$ and
particle
number $N$. To solve this
maximization problem, we proceed in two steps. We first maximize 
$S[f]$ at fixed $Mc^2$, $N$
{\it and} energy density $\epsilon(r)$. Since $\epsilon(r)$ determines
$Mc^2$, this is equivalent to maximizing $S[f]$  at fixed $N$
and  $\epsilon(r)$. The variational problem
for the first variations (extremization) can be written as
\begin{equation}
\frac{\delta S}{k_B}-\int\tilde\beta(r)\delta \epsilon \, d{\bf
r}+\alpha\delta N=0,
\end{equation}
where $\tilde\beta(r)$ is a
local (position dependent) Lagrange multiplier and  $\alpha$ is a global
(uniform) Lagrange multiplier. Noting that $M(r)$ -- which appears in the
expressions of $S$ and $N$ -- is
fixed since it is determined by $\epsilon(r)$, this variational problem yields
\begin{eqnarray}
\label{alt3b}
C'(f)=-\beta(r)E(p)+\alpha
\end{eqnarray}
with $\beta(r)\equiv {\tilde\beta}(r)[1-2GM(r)/rc^2]^{1/2}$,
leading to Eq.
(\ref{qgd15}). In this manner, we immediately find that $\alpha$ is
uniform at statistical
equilibrium. This results from the conservation of $N$. Substituting Eq.
(\ref{alt3b}) into the expressions of $S$, $M$ and $N$ may help solving the
stability problem.

\section{Dynamical stability of a self-gravitating barotropic gas with respect
to the Euler equation}
\label{sec_dse}

\subsection{Newtonian gravity: Euler-Poisson equations}
\label{sec_dseNG}

We consider a Newtonian gaseous star with a barotropic equation of state
$P=P(\rho)$ described by the Euler-Poisson equations. These
equations conserve the mass $M$ [see Eq.
(\ref{n14})] and the energy
\begin{equation}
\label{dseNG1}
{\cal W}[\rho,{\bf u}]=\frac{1}{2}\int \rho {\bf u}^2\, d{\bf r}+\int
\rho\int^{\rho} \frac{P(\rho')}{{\rho'}^2}\, d\rho'\, d{\bf
r}+\frac{1}{2}\int \rho\Phi\, d{\bf r},
\end{equation}
which is the sum of the kinetic energy $\Theta_c$, the internal energy
$U$, and
the gravitational energy $W$ (see Appendix \ref{sec_tinbs}). It can be shown
that the minimization problem
\begin{eqnarray}
\min\ \lbrace {\cal W}\, |\,  M\,\, {\rm fixed} \rbrace
\label{dseNG2}
\end{eqnarray}
determines an equilibrium state of the Euler-Poisson equations that is
dynamically stable \cite{lbs,bt,aaantonov}. This
is a criterion of nonlinear
dynamical
stability resulting from the fact that ${\cal W}$ and $M$ are conserved by  the
Euler-Poisson equations \cite{holm}. It provides a necessary and
sufficient condition of dynamical stability since it takes into
account all the invariants of the Euler-Poisson equations.

The
variational principle for the first variations
(extremization) can be written as 
\begin{eqnarray}
\label{dseNG3}
\delta {\cal W}-\frac{\mu_0}{m}\delta M=0,
\end{eqnarray}
where $\mu_0$ is a Lagrange multiplier. This yields ${\bf u}={\bf 0}$
and 
\begin{equation}
\label{dseNG4}
\int^{\rho}\frac{P'(\rho')}{\rho'}\, d\rho'+\Phi(r)-\frac{\mu_0}{m}=0.
\end{equation}
Taking the gradient of this relation, we obtain the condition of hydrostatic
equilibrium
\begin{equation}
\label{dseNG5}
\nabla P+\rho\nabla\Phi={\bf 0}. 
\end{equation}
Therefore, an extremum of ${\cal W}$ at fixed $M$ is a steady state of the
Euler-Poisson equations. Then, considering the
second variations of ${\cal W}$, it can be shown that the star
is linearly stable with respect to the Euler-Poisson equations if, and only if,
it is a local minimum of ${\cal W}$ at fixed mass $M$. This is also equivalent
to
its spectral stability. Indeed, the complex pulsations
$\omega$ of the normal modes of the linearized  Euler-Poisson equations
\cite{eddington18,ledouxpekeris,chandraN63,chandraN64,bt,aaantonov} satisfy
$\omega^2>0$ for all modes if, and only
if, $\delta^2 {\cal W}>0$ for all
perturbations that conserve $M$. Using
the Poincar\'e criterion \cite{poincare}, we can generically conclude 
\cite{aaantonov} that the series of equilibria is dynamically stable before
the turning points of mass $M$ or energy ${\cal W}$ (they coincide) and that it
becomes
dynamically unstable
afterwards. Furthermore, the curve ${\cal
W}(M)$ displays spikes at its extremal points (since $\delta{\cal
W}=0\Leftrightarrow \delta M=0$). We refer to
\cite{lbs,bt,aaantonov} for the
derivation of these results.

{\it Remark:} In the case of isothermal and polytropic equations of states, the
marginal mode of instability has been explicitly determined
in \cite{aaiso,aapoly,grand}.

\subsection{General relativity: Euler-Einstein equations}
\label{sec_dsegr}

We consider a relativistic gaseous star with a barotropic equation of state
$P=P(\epsilon)$ described by the Euler-Einstein equations. We restrict
ourselves to spherically symmetric systems. The Euler-Einstein
equations conserve the mass-energy $Mc^2$ [see Eq.
(\ref{n53})] and the
particle number $N$ [see Eq. (\ref{n77})]. Here, the energy density is equal
to $\epsilon=\rho c^2+u$ where $u$ is the density of internal energy (see
Appendix \ref{sec_tigrg}). It can be shown that the
minimization  problem
\begin{eqnarray}
\min\ \lbrace M\, |\,  N\,\, {\rm fixed}  \rbrace
\label{dsegr1}
\end{eqnarray}
or, equivalently, the maximization
problem
\begin{eqnarray}
\max\ \lbrace N\, |\,  M\,\, {\rm fixed} \rbrace
\label{dsegr2}
\end{eqnarray}
determine an equilibrium state of the Euler-Einstein equations that is
dynamically stable \cite{zeldovichNM,htww65,cocke,weinberg}.\footnote{The
optimization problems (\ref{dsegr1})
and (\ref{dsegr2}) are equivalent to $\min\lbrace
E |  N\, {\rm fixed}  \rbrace$ and $\min\lbrace
E |  M\, {\rm fixed}  \rbrace$, where $E=(M-Nm)c^2$ is the binding energy. In
the nonrelativistic limit,
they reduce to the optimization problem (\ref{dseNG2}). 
Indeed, when $c\rightarrow +\infty$, repeating the steps of Sec. \ref{sec_mneb}
with
$u$
in place of $\epsilon_{\rm kin}$, we get $E\rightarrow U+W$.}
These are criteria of
nonlinear
dynamical stability
resulting from the fact that $M$ and $N$ are conserved by  the
Euler-Einstein equations \cite{holm}. They provide a necessary and
sufficient condition of dynamical stability since they take into
account all the invariants of the Euler-Einstein equations.

The variational
principle for the first
variations
(extremization) can be written as
\begin{eqnarray}
\label{dsegr3}
\delta M-\sigma\delta N=0 \qquad {\rm or}\qquad  \delta
N-\frac{1}{\sigma}\delta
M=0,
\end{eqnarray}
where $\sigma$ is a Lagrange multiplier. It leads to the TOV equations
(\ref{n52}) and (\ref{n58b}) which express the
condition of hydrostatic equilibrium. Then, considering the
second variations of $N$, it can be shown that the star
is linearly stable with respect to the Euler-Einstein equations if,
and only if,
it is a local maximum of $N$ at fixed
$M$. This is also equivalent to
its spectral stability.  Indeed, the complex pulsations
$\omega$ of the normal modes of the linearized  Euler-Einstein equations
\cite{chandra64,yabushitaR2} satisfy
$\omega^2>0$ for all modes if,
and only
if, $\delta^2 N<0$ for all
perturbations that conserve $M$.  Using
the Poincar\'e criterion \cite{poincare}, we can generically conclude that the
series of equilibria is dynamically stable before
the turning points of mass-energy $M$, particle number
$N$, or binding energy $E$ (they
all coincide) and that it becomes dynamically unstable
afterwards.\footnote{The  Poincar\'e turning point criterion \cite{poincare} 
is equivalent to the mass-radius theorem of Wheeler \cite{htww65} introduced in
the physics of compact objects like white dwarfs and neutron stars.}
Furthermore, the curve ${M}(N)$ displays spikes at its
extremal points (since $\delta M=0\Leftrightarrow \delta N=0$). We
refer to \cite{htww65,cocke,aarelat1,aarelat2} for the derivation of
these results.

{\it Remark:} In the case of a linear equation of state $P=q\epsilon$, the
marginal mode of instability has been explicitly determined
in \cite{aarelat1,aarelat2}.

\section{Dynamical stability of collisionless self-gravitating systems with
respect to the Vlasov equation}
\label{sec_dsv}

\subsection{Newtonian gravity: Vlasov-Poisson equations}
\label{sec_dsvNG}

We consider a Newtonian collisionless stellar system
described by the Vlasov-Poisson equations. These
equations conserve the energy $E$ [see Eq. (\ref{ax3})] and an infinite number
of Casimir integrals $I_h=\int h(f)\, d{\bf r}d{\bf v}$, where $h$ is an
arbitrary function, including
the particle number $N$
[see Eq. (\ref{ax2})]. The
minimization problem
\begin{eqnarray}
\min\ \lbrace {E}\, |\, I_h  \,\, {\rm fixed\, for \, all }\, h
\rbrace
\label{dsvNG1}
\end{eqnarray}
determines a steady state of the Vlasov-Poisson equations that is
dynamically stable \cite{lbs,lbhf,bartholomew,doremus71b,ih,kandrup91,cc}.
This is a
criterion of nonlinear dynamical
stability resulting from the fact that $E$ and $I_h$ are conserved by the
Vlasov-Poisson equations \cite{holm}.  It provides a necessary and
sufficient condition of dynamical stability since it takes into
account all the invariants of the Vlasov-Poisson equations.

It can be
shown that any steady state of the Vlasov-Poisson equations
extremizes the energy ($\delta E=0$) under phase-preserving, or symplectic,
perturbations (those that conserve all the Casimirs). Restricting
ourselves to steady states of the form $f=f(\epsilon)$ with
$f'(\epsilon)<0$ and considering the
second variations of $E$, it can be shown that a stellar system
is linearly stable with respect to the Vlasov-Poisson equations if, and only if,
it is a local minimum of $E$ under symplectic (phase-preserving) perturbations.
This is also equivalent to its spectral stability. Indeed, the complex
pulsations $\omega$ of the normal modes of the linearized Vlasov-Poisson
equations
\cite{antonov61} satisfy $\omega^2>0$ (one can show
that $\omega^2$ is real) for all modes if, and only
if, $\delta^2 E>0$ for all
perturbations that conserve the Casimirs at first order. We refer to
\cite{lbs,lbhf,bartholomew,doremus71b,ih,kandrup91,cc} for the derivation
of these results.

It can be shown furthermore that the maximization problem
(``microcanonical'' criterion)
\begin{eqnarray}
\max\ \lbrace {S}\, |\,  E, N \,\, {\rm fixed} \rbrace
\label{dsvNG2}
\end{eqnarray}
and the minimization problem (``canonical'' criterion)
\begin{eqnarray}
\min\ \lbrace F=E-TS \, |\,  N \,\, {\rm fixed} \rbrace
\label{dsvNG3}
\end{eqnarray}
for a generalized ``entropy'' of the form (\ref{qgd9}) provide {\it sufficient}
conditions of
dynamical stability with respect to the Vlasov-Poisson 
equations.\footnote{Here, the
analogy with thermodynamics is effective.} These are criteria of nonlinear
dynamical
stability resulting from the fact that $S$, $E$, $F$ and $N$ are conserved by 
the Vlasov-Poisson equations \cite{holm}. They provide just
sufficient conditions of dynamical stability because they take into
account the conservation of only certain invariants of the Vlasov-Poisson
equations, not all of them. It can be
shown that
``canonical stability'' implies ``microcanonical stability'' which implies
``dynamical stability''. We have
\begin{eqnarray}
(\ref{dsvNG3}) \Rightarrow (\ref{dsvNG2})
\Rightarrow (\ref{dsvNG1}). 
\label{implications}
\end{eqnarray}
This is similar to a situation of ensembles inequivalence in thermodynamics.
These results were established in
\cite{ih,aaantonov,cc}. Since
``microcanonical'' stability implies dynamical stability, using the Poincar\'e
criterion \cite{poincare}, we can generically conclude  that the
series of equilibria is
dynamically stable at least until the turning point of energy
\cite{ih,aaantonov}.\footnote{We can use this method to show graphically
(without calculation) that
all the
stellar polytropes are stable \cite{aaantonov}. This result was originally
proven by
Antonov \cite{antonovlaw} with rather complicated calculations.} This is
a general
result
valid for all systems
with
long-range interactions \cite{cc}. Now, in the case of Newtonian
self-gravitating systems, it can be shown
\cite{doremus71,doremus73,gillon76,sflp,ks,kandrup91} that {\it all} the
distribution functions of
the form $f=f(\epsilon)$ with $f'(\epsilon)<0$ are dynamically stable with
respect to the Vlasov-Poisson equations. Therefore, the whole series of
equilibria is dynamically stable,  even the equilibrium states that lie after
the first turning point of energy.

{\it Antonov first law:} Let us consider an isotropic stellar system with a
distribution
function of
the form $f=f(\epsilon)$ with $f'(\epsilon)<0$. Introducing the
density $\rho(r)=\int f(v^2/2+\Phi(r))\, d{\bf v}$ and the pressure
$P(r)=\frac{1}{3}\int f(v^2/2+\Phi(r)) v^2\, d{\bf v}$, and
eliminating formally $\Phi(r)$ between these two expressions, we find that the
corresponding gas is barotropic: $P(r)=P[\rho(r)]$. Then, proceeding as in
Appendix \ref{sec_hea}, we can show that it satisfies the condition of
hydrostatic equilibrium $\nabla P+\rho\nabla\Phi={\bf 0}$. Therefore, to any
stellar system described by a distribution function of the form $f=f(\epsilon)$
with $f'(\epsilon)<0$ we can
associate a corresponding barotropic star with an equation of state $P=P(\rho)$
that satisfies the condition of hydrostatic equilibrium.  Using the Schwarz
inequality, Antonov
 \cite{antonovlaw} and
Lynden-Bell and Sanitt \cite{lbs} have shown that a stellar system
with $f=f(\epsilon)$ and
$f'(\epsilon)<0$ is stable  with respect to the Vlasov-Poisson equations
whenever the corresponding barotropic star is stable with respect to the
Euler-Poisson equations. This is what Binney and Tremaine \cite{bt} have  called
the Antonov first
law. We can recover this result with a different method related to
the concept of ensembles
inequivalence. It can
be shown (see \cite{aaantonov} and Appendices  \ref{sec_altce} and
\ref{sec_dseNG})
that the ``canonical'' criterion of dynamical stability (\ref{dsvNG3}) for a
stellar system described
by the Vlasov-Poisson equations is
equivalent to the
criterion of dynamical stability (\ref{dseNG2}) for the corresponding
barotropic
star described by the Euler-Poisson equations. Since  ``canonical stability''
implies dynamical stability for collisionless stellar systems [see Eq.
(\ref{implications})], we conclude that
\begin{eqnarray}
(\ref{dseNG2}) \Leftrightarrow (\ref{dsvNG3})
\Rightarrow (\ref{dsvNG1}). 
\label{implications2}
\end{eqnarray}
Therefore, the dynamical stability with respect to the Euler-Poisson equations
implies the  dynamical stability with respect to the Vlasov-Poisson
equations. However, the converse is wrong.\footnote{There is
an exception. In the case of an infinite and homogeneous medium, collisionless
stellar systems (Vlasov) and self-gravitating fluids (Euler) behave in the same
way with respect to the Jeans instability in the sense that they lead to the
same criterion for instability \cite{nyquistjeans}.} This provides a new
derivation of
the Antonov first law
\cite{aaantonov} in terms of ensembles inequivalence. In particular, this
derivation is valid for nonlinear
dynamical stability while the original proof \cite{antonovlaw,lbs,bt} was
restricted to linear (spectral) dynamical stability.

\subsection{General relativity: Vlasov-Einstein equations}
\label{sec_dsvgr}

The preceding results (\ref{dsvNG1})-(\ref{implications}) can be extended to
the context of general
relativity \cite{it,ipser69,ipser69b,fackerell70,ipser80}. In particular, since
``microcanonical'' stability implies dynamical stability, using the Poincar\'e
criterion \cite{poincare}, we can generically conclude  that the
series of equilibria is
dynamically stable at least until the turning point of binding
energy \cite{ipser80}. Now, there is a conjecture by Ipser \cite{ipser80} that,
in general relativity,
the ``microcanonical'' criterion (\ref{dsvNG2}) is {\it equivalent} to the 
criterion of dynamical
stability (\ref{dsvNG1}), contrary to the Newtonian case. We have
\begin{eqnarray}
(\ref{dsvNG3}) \Rightarrow (\ref{dsvNG2})
\Leftrightarrow (\ref{dsvNG1}). 
\label{implications3}
\end{eqnarray} 
Accordingly, the
series of equilibria is dynamically stable
before the turning point of binding energy and becomes unstable afterwards. This
result has been established numerically for
heavily truncated isothermal distributions and stellar
polytropes \cite{ipser69b,st2}. The
conjecture consists in extending its validity to all distribution functions.

{\it Relativistic Antonov first law:} Let us consider a star cluster with an
isotropic  distribution
function of
the form $f=f(Ee^{\nu(r)/2})$ with $f'(Ee^{\nu(r)/2})<0$. Introducing the energy
density $\epsilon(r)=\int f(E(p)e^{\nu(r)/2})E(p)\, d{\bf p}$ and the pressure
$P(r)=(1/3)\int f(E(p)e^{\nu(r)/2})p E'(p)\, d{\bf p}$, and eliminating formally
$\nu(r)$ between these two expressions, we find that the
corresponding gas is barotropic: $P(r)=P[\epsilon(r)]$. Then, proceeding as in
Appendix \ref{sec_heagr}, we can show that it
satisfies the TOV equations expressing the condition
of hydrostatic equilibrium (\ref{n57}).\footnote{To make the correspondance
with Appendix \ref{sec_heagr} we just need to replace $k_B T(r)$ by
$e^{-\nu(r)/2}$. In that case, Eq. (\ref{hrg7}) reduces to Eq. (\ref{n57}).}
Using the Schwarz
inequality, Ipser
\cite{ipser69} has obtained a relativistic generalization of the linear
Antonov first law. On the other hand, it can
be shown \cite{roupas1,roupas1E,gsw,fhj} that the
``canonical''
criterion of dynamical stability (\ref{dsvNG3}) for a star
cluster described
by the Vlasov-Einstein equations is
equivalent to the 
criterion of dynamical stability (\ref{dsegr1})
for the corresponding
barotropic
star described by the Euler-Einstein equations. 
Since ``canonical
stability''
implies dynamical stability for collisionless star clusters [see Eq.
(\ref{implications3})], we conclude that 
\begin{eqnarray}
(\ref{dsegr1}) \Leftrightarrow (\ref{dsvNG3})
\Rightarrow (\ref{dsvNG1}). 
\label{implications2b}
\end{eqnarray}
Therefore, the dynamical stability with respect to the Euler-Einstein equations
implies the dynamical stability with respect to the Vlasov-Einstein
equations. This
provides a generalization of the nonlinear Antonov first law obtained in
Newtonian
gravity \cite{aaantonov}.

\section{Black-body radiation in general relativity}
\label{sec_bbr}

In this Appendix, we consider a gas of photons (black-body radiation) that is so
intense that general
relativity must be taken into account. This leads to the concept of ``photon
stars'' or self-gravitating black-body radiation. This problem has been studied
in detail in \cite{sorkin,aarelat2}. Below, we recall the basic equations
determining the statistical equilibrium state of a gas of photons in general
relativity and compare these results with those obtained  in Sec.
\ref{sec_grf} for material particles such as self-gravitating fermions.

\subsection{Thermodynamics of the black-body radiation}

The distribution function of a gas of photons is
\begin{eqnarray}
\label{bb1}
f({\bf p})=\frac{1}{h^3}\frac{1}{e^{\beta pc}-1}.
\end{eqnarray}
This corresponds to the Bose-Einstein statistics in the ultrarelativistic
limit ($E=pc$) and with a vanishing chemical potential ($\mu=0$). These
simplifications arise because the photons have no rest mass. Using Eq.
(\ref{n69}), we find that the energy density is related to the
temperature by the  Stefan-Boltzmann law
\begin{eqnarray}
\label{bb2}
\epsilon=\frac{24\pi}{h^3c^3}(k_B T)^4\frac{\pi^4}{90}.
\end{eqnarray}
The factor in front of $T^4$ is the Stefan-Boltzmann constant. Using Eq.
(\ref{n74}), we find that the pressure is given by
\begin{eqnarray}
\label{bb2b}
P=\frac{8\pi}{h^3c^3}(k_B T)^4\frac{\pi^4}{90}.
\end{eqnarray}
It is related to the
energy density by the linear equation of state
\begin{eqnarray}
\label{bb3}
P=\frac{1}{3}\epsilon.
\end{eqnarray}
This linear relationship, with a coefficient $1/3$, is valid for an arbitrary
ultrarelativistic gas (see Appendix \ref{sec_lur}). Using Eq.
(\ref{n68}), we find that the particle density is given by
\begin{eqnarray}
\label{bb4}
n=\frac{8\pi}{h^3c^3}(k_B T)^3\zeta(3),
\end{eqnarray}
where $\zeta(3)=1.202056...$ is the Ap\'ery constant (Riemann zeta function in
$x=3$). The pressure is related to the
particle density through the polytropic equation
of state
\begin{eqnarray}
\label{bb5}
P=Kn^{4/3}\quad {\rm with}\quad K=\frac{\pi^4}{90}\frac{hc}{\lbrack
8\pi\zeta(3)^4\rbrack^{1/3}}.
\end{eqnarray}
Finally, using the integrated Gibbs-Duhem relation (\ref{b90}) with
$\mu=0$, we find that the entropy
density is given by\footnote{It can also be obtained by
substituting Eq. (\ref{bb1}) into the 
the Bose-Einstein entropy $s=-k_B\frac{1}{h^3}\int\lbrace
\frac{f}{f_{*}}\ln \frac{f}{f_{*}}
- (1+\frac{f}{f_{*}} )\ln
(1+\frac{f}{f_{*}})\rbrace\, d{\bf p}$ where $f_*=1/h^3$.}
\begin{eqnarray}
\label{bb6}
s=k_B \frac{32\pi^5}{90 h^3c^3}(k_B T)^3.
\end{eqnarray}
We see that the entropy density is proportional to the particle density:
\begin{eqnarray}
\label{bb7}
s=\lambda  n k_B \quad {\rm with}\quad 
\lambda=\frac{4\pi^4}{90\zeta(3)}.
\end{eqnarray}
More details about these relations and their
consequences can be found in Ref. \cite{aarelat2}.

\subsection{Mechanical derivation of the Tolman relation}

Substituting the relation $\epsilon=3P$ from Eq. (\ref{bb3}) into Tolman's
equation of hydrostatic equilibrium (\ref{n57}), we get
\begin{eqnarray}
\label{bb8}
\frac{d\ln P}{dr}=-2\frac{d\nu}{dr}.
\end{eqnarray}
On the other hand, according to Eq. (\ref{bb2b}), we have
\begin{eqnarray}
\label{bb9}
\frac{d\ln P}{dr}=4\frac{d\ln T}{dr}.
\end{eqnarray}
These two equations directly imply the Tolman relation
\begin{equation}
\label{bb10}
\frac{d\ln T}{dr}=-\frac{1}{2}\frac{d\nu}{dr} \qquad\Rightarrow\qquad 
T(r) e^{\nu(r)/2}={\rm cst}.
\end{equation}
This derivation is valid only for the black-body radiation. It was given by
Tolman
\cite{tolman} as a particular example of his relation before
considering the general case of an
arbitrary perfect fluid.

{\it Remark:} This derivation presupposes the condition of hydrostatic
equilibrium (\ref{n57}). In the following section, we show that this equation
can be obtained from the maximization of the entropy $S$ at fixed mass-energy
$Mc^2$.

\subsection{Equivalence between dynamical and thermodynamical stability for
the self-gravitating black-body radiation}

According to Eq. (\ref{bb7}), the entropy of the black-body radiation is
proportional to
the particle number:
\begin{eqnarray}
\label{bb11}
S=\lambda N k_B\quad {\rm with}\quad 
\lambda=\frac{4\pi^4}{90\zeta(3)}.
\end{eqnarray}
The condition
of thermodynamical stability, corresponding to the maximization of the
entropy at fixed mass-energy:
\begin{eqnarray}
\label{bb12}
\max\, \lbrace S\, |\, {\cal E}=Mc^2\,\, {\rm fixed}\rbrace,
\end{eqnarray}
turns out to be equivalent to the maximization of the particle number at fixed
mass-energy:
\begin{eqnarray}
\label{bb13}
\max\, \lbrace N\, |\, M\,\, {\rm fixed}\rbrace,
\end{eqnarray}
which is itself equivalent to the minimization
of the mass-energy at
fixed particle number:
\begin{eqnarray}
\label{bb14}
\min\, \lbrace M\, |\, N\,\, {\rm fixed}\rbrace,
\end{eqnarray}
corresponding to the condition of dynamical stability for a
barotropic fluid in general relativity (see Appendix \ref{sec_dsegr}).
Therefore, in the case of the
self-gravitating black-body radiation, it
is straightforward to show the equivalence between dynamical and
thermodynamical stability. This is a particular case where
Ipser's conjecture \cite{ipser80} (see Appendix \ref{sec_dsvgr})
can be easily demonstrated.\footnote{However, it is important to
realize that, for the self-gravitating black-body
radiation, dynamical stability
refers to the Euler-Einstein equations while, for a collisionless star cluster,
it refers to the Vlasov-Einstein equations.
This is a  difference of fundamental importance.}

The maximization problem (\ref{bb12}) determining the
thermodynamical stability
of the self-gravitating black-body radiation in general
relativity was first studied by Tolman \cite{tolman},
and
later by Cocke \cite{cocke}, Sorkin {\it et al.} \cite{sorkin} and
Chavanis \cite{aarelat2}. The
variational principle for  the first variations
(extremization) can be written as
\begin{eqnarray}
\label{bb15}
\delta S-\frac{1}{T_{\infty}}\delta {\cal E}=0 \qquad \Rightarrow \qquad  \delta
N-\frac{1}{\lambda}\beta_{\infty}c^2\delta M=0,
\end{eqnarray}
where $1/T_{\infty}$ is a Lagrange multiplier. It leads to the TOV equations
(equivalent to the condition of hydrostatic equilibrium) and to
the Tolman
relation (the Lagrange multiplier
$T_{\infty}$
corresponds to the Tolman temperature). Then, considering the
second variations of $S$, it can be shown that the self-gravitating black-body
radiation
is linearly stable with respect to the Euler-Einstein equations if,
and only if, it is a local maximum of $S$ at fixed $M$. This is also equivalent
to its spectral stability. Indeed, the complex pulsations
$\omega$ of the normal modes of the linearized  Euler-Einstein
equations  \cite{chandra64,yabushitaR2} satisfy
$\omega^2>0$ for all modes if,
and only if, $\delta^2S<0$ for all perturbations that conserve $M$. 
Using the
Poincar\'e criterion \cite{poincare}, we can show  \cite{aarelat2} that the
series of equilibria is thermodynamically and dynamically stable before
the turning points of mass-energy $M$, particle number
$N$, binding energy $E$, or entropy $S$ (they
all coincide) and that it becomes thermodynamically and dynamically unstable
afterwards. Furthermore, the curve $S({\cal E})$ displays spikes
at its extremal points (since $\delta S=0\Leftrightarrow \delta {\cal E}=0$).
We
refer to \cite{cocke,sorkin,aarelat2} for the derivation of
these results.

{\it Remark:} In the case of material particles, the
statistical equilibrium state is obtained by maximizing the
entropy at fixed mass-energy and particle number. In the case of the
self-gravitating black-body radiation, the statistical equilibrium state is
obtained by maximizing the entropy, which is proportional to the
particle number, at fixed mass-energy. How can we understand this difference?
First, we have
to realize that, in the case of the black-body radiation, the particle number is
not fixed. What is fixed instead is the ratio between the chemical potential
and the temperature. Therefore, the correct manner to
treat the thermodynamics of the self-gravitating black-body radiation is to work
in the grand microcanonical ensemble \cite{lecarkatz,grand} where
$\alpha=\mu/k_B T$ and ${\cal
E}=Mc^2$ are fixed. In that ensemble, the thermodynamic potential is ${\cal
K}=S+\alpha k_B N$. The statistical equilibrium state is then obtained by
maximizing ${\cal K}$ at fixed mass-energy:
\begin{eqnarray}
\label{bb15b}
\max\, \lbrace {\cal K}\, |\, {\cal E}=Mc^2\,\, {\rm fixed}\rbrace.
\end{eqnarray}
The extremization problem (first variations) yields
\begin{eqnarray}
\delta {\cal K}-\frac{1}{T_{\infty}}\delta {\cal E}=0.
\end{eqnarray}
Now, for (massless) photons, the chemical potential vanishes: $\mu=0$. This
implies $\alpha=0$ and ${\cal K}=S$. In that case, the maximization
problem (\ref{bb15b}) reduces to  (\ref{bb12}).

\section{The Tolman-Klein relations}
\label{sec_sem}

In this Appendix, we review the main results given in the seminal papers of
Tolman \cite{tolman} and Klein \cite{klein}.

\subsection{Tolman's (1930) paper}
\label{sec_tolman}

In a paper published in 1930, Tolman \cite{tolman} investigated ``the
weight of heat and thermal
equilibrium in general relativity''. His main finding is that, even at
thermodynamic equilibrium, the temperature is inhomogeneous in the presence of
gravitation. He discovered a definite relation connecting the distribution of
temperature $T(r)$ throughout the system to the gravitational potential
(or metric coefficient) $\nu(r)$. Tolman's relation  [see Eq. (\ref{b103})]
between
equilibrium
temperature and gravitational potential was something essentially new in
thermodynamics since, until his work, uniform temperature throughout any
system which
has come to thermal equilibrium had hitherto been taken as an inescapable part
of thermodynamic theory.

Tolman first considered the case of a weak gravitational field described by
Newtonian gravitation. By maximizing the entropy for an
isolated system he obtained an approximate relation between the temperature
distribution and the Newtonian gravitational potential [see Eq. (\ref{bn23})].
This
can be viewed as a post-Newtonian relation since the temperature gradient is
inversely proportional to the square of the velocity of light.

He then considered the case of the black-body radiation. By performing a
purely mechanical treatment of temperature distribution based on the Einstein
equations (using Eq. (\ref{n57}) representing the general relativistic extension
of the condition of hydrostatic equilibrium) he obtained an exact general
relativistic relation between the proper temperature and the metric coefficient
$\nu$ [see Eq. (\ref{bb10})].

He then recovered this result by maximizing the entropy of the
self-gravitating black-body radiation by using the
formalism of relativistic thermodynamics that he had developed a
few years earlier. Therefore, in the simple case of the black-body radiation
where a
mechanical treatment can be given, the thermodynamical and mechanical
treatments of temperature distribution under the action of gravity lead to the
same result.

Finally, he generalized his  thermodynamical approach (maximum entropy
principle) to the case of any perfect fluid and obtained the Tolman relation
(\ref{b103})
in a general setting.

He noted at the end of his paper that the maximum entropy
principle implies the  general relativistic  condition of hydrostatic
equilibrium [see Eq. (\ref{n57})] contained in the Einstein equations.
 He
noted: ``It may seem strange that this purely mechanical equation holding
within the interior of the system should be derivable from the application of
thermodynamics to the system as a whole. The result,
however, is the
relativity analogue to the equation for change in pressure with height obtained
by Gibbs (``Scientific Papers,'' Longmans, Green 1906, equation 230, p. 145) in
his thermodynamic treatment of the conditions of equilibrium
under the influence of gravity. Indeed the whole treatment of this article may
be regarded as the relativistic extension of this part of Gibbs' work.''

\subsection{Klein's (1949) paper}

In a paper entitled ``On the thermodynamical equilibrium of fluids in
gravitational fields'' published in 1949, Klein \cite{klein} managed to derive
the Tolman relation
(\ref{b103}), together with a similar relation between the chemical potential
$\mu(r)$
and the metric coefficient $\nu(r)$ [see Eq. (\ref{b104})] with almost no
calculation,\footnote{The calculations of Tolman \cite{tolman} based on the
maximum entropy principle are comparatively much more complicated.} by
using essentially the Gibbs-Duhem relation and the first
principle of thermodynamics. We give below a summary of Klein's
calculations.

Klein started from the first principle of thermodynamics
\begin{equation}
\label{k1}
dE=-PdV+TdS+\mu dN.
\end{equation}
Since $E$ is a homogeneous function of the first degree in the three variables
$V$, $S$ and $N$, the Euler theorem implies that
\begin{equation}
\label{k2}
E=-PV+TS+\mu N,
\end{equation}
which is the Gibbs-Duhem relation (see
Appendix \ref{sec_gd}). From Eqs. (\ref{k1}) and (\ref{k2}),
we get
\begin{equation}
\label{k2b}
d\left (\frac{P}{T}\right )=\frac{N}{V}d\left (\frac{\mu}{T}\right
)-\frac{E}{V}d\left (\frac{1}{T}\right ).
\end{equation}
Written under a
local form, with the variables $s=S/V$, $n=N/V$ and $\epsilon=E/V$, Eqs.
(\ref{k1})-(\ref{k2b}) return Eq. (\ref{b84}) and Eqs. (\ref{b90})-(\ref{dent}).
  In turn, Eq.
(\ref{b91}) can be written as
\begin{equation}
\label{k3}
\frac{dP}{dr}=\frac{\epsilon+P}{T}\frac{dT}{dr}+nT\frac{d}{dr}\left
(\frac{\mu}{T}\right ).
\end{equation}
Combined with the condition of hydrostatic equilibrium from Eq. (\ref{n57}), we
get
\begin{equation}
\label{k4}
\frac{1}{2}\frac{d\nu}{dr}+\frac{1}{T}\frac{dT}{dr}=-\frac{nT}{\epsilon+P}\frac
{ d } { dr } \left
(\frac{\mu}{T}\right ).
\end{equation}
At that point, Klein considered several independent substances
present in the same gravitational field and argued that an equation of the
type (\ref{k4}) holds for each of them separately with the same values of
$\nu$ and $T$. As one such substance, we always have the radiation for
which $\mu=0$. Thus, we get
\begin{equation}
\label{k5}
\frac{1}{2}\frac{d\nu}{dr}+\frac{1}{T}\frac{dT}{dr}=0\qquad\Rightarrow\qquad 
T(r) e^{\nu(r)/2}={\rm cst},
\end{equation}
which is Tolman's relation. Then, for all other substances
\begin{equation}
\label{k6}
\frac{\mu(r)}{k_B T(r)}={\rm cst} \qquad\Rightarrow\qquad 
\mu(r) e^{\nu(r)/2}={\rm cst},
\end{equation}
which is Klein's relation. As emphasized by Klein \cite{klein}, this relation
consitutes the relativistic
generalization of the well-known Gibbs \cite{gibbsklein} condition for the
equilibrium in a gravitational field.

\section{Thermodynamic identities}
\label{sec_ti}

In this Appendix, we regroup useful thermodynamic identities valid for
Newtonian and general relativistic barotropic gases.

\subsection{Newtonian isentropic or cold barotropic gases}
\label{sec_tinbs}

The first principle of
thermodynamics writes
\begin{equation}
\label{tinbs1}
d\left (\frac{u}{n}\right )=-Pd\left
(\frac{1}{n}\right )+Td\left (\frac{s}{n}\right ),
\end{equation}
where $u$ is the density of internal energy.  We assume
that $Td(s/n)=0$. This corresponds
to cold ($T=0$) or isentropic ($s/n=\lambda={\rm cst}$) gases. In that case, Eq.
(\ref{tinbs1}) reduces to
\begin{equation}
\label{tinbs2}
d\left (\frac{u}{n}\right )=-Pd\left
(\frac{1}{n}\right )=\frac{P}{n^2}dn.
\end{equation}
For a barotropic equation of state $P=P(n)$, Eq.
(\ref{tinbs2}) can be integrated into
\begin{equation}
\label{tinbs3}
u(n)=n\int^n \frac{P(n')}{{n'}^2}\, dn'.
\end{equation}
The internal energy is
\begin{equation}
\label{tinbs4}
U=\int n\int^n \frac{P(n')}{{n'}^2}\, dn'\, d{\bf r}.
\end{equation}
We also have
\begin{equation}
\label{tinbs5}
du=\frac{P+u}{n} \, dn=h\, dn
\end{equation}
and
\begin{equation}
dh=\frac{dP}{n},
\end{equation}
where 
\begin{equation}
\label{tinbs6}
h=\frac{P+u}{n}
\end{equation}
is the enthalpy. 
We note the identities
\begin{equation}
\label{tinbs7}
u'(n)=h(n), \qquad u''(n)=h'(n)
\end{equation}
and
\begin{equation}
\label{tinbs8}
P(n)=n h(n)-u(n)=n u'(n)-u(n).
\end{equation}
The energy of a Newtonian isentropic or cold barotropic self-gravitating gas is
${\cal W}=U+W$ where $U$ is the internal
energy and $W$ is the gravitational energy. A stable equilibrium state of the
Euler-Poisson equations is a
minimum of energy ${\cal W}$ at fixed particle number $N$ (see
Appendix \ref{sec_dseNG}).\footnote{In the isentropic case $s/n=\lambda$
we have $S=\lambda N$. Therefore a minimum of ${\cal W}$ at fixed $N$ is
also a maximum
of $S$ at fixed ${\cal W}$.} If the pressure can be written as
$P(n)=T\, \Pi(n)$, we get ${\cal W}=W-TS_{\rm eff}$ where $S_{\rm
eff}=-\int n\int^n \frac{\Pi(n')}{{n'}^2}\, dn'\, d{\bf r}$ is a generalized
entropy of the density $n$ \cite{nfp,entropy}.

{\it Remark:} For an ideal gas at $T=0$, the thermodynamic identities of Sec.
\ref{sec_smnrf} reduce to $d\epsilon_{\rm kin}=\mu\, dn$, $\epsilon_{\rm
kin}+P-\mu n=0$ and $dP=n \, d\mu$. We can check that they coincide with Eqs.
(\ref{tinbs2})-(\ref{tinbs8}) with $u=\epsilon_{\rm kin}$ and $h=\mu$.

\subsection{General relativistic isentropic or cold barotropic gases}
\label{sec_tigrg}

In general relativity, the first principle of
thermodynamics writes
\begin{equation}
\label{tinbs1rg}
d\left (\frac{\epsilon}{n}\right )=-Pd\left
(\frac{1}{n}\right )+Td\left (\frac{s}{n}\right ),
\end{equation}
where $\epsilon=\rho c^2+u$ is the mass-energy density and $\rho=n
m$ is the rest-mass density. The relations of Appendix \ref{sec_tinbs} remain
valid with $u$ or with $\epsilon$. When  $Td(s/n)=0$, Eq. (\ref{tinbs1rg})
reduces to
\begin{equation}
\label{tigrg1}
d\epsilon=\frac{P+\epsilon}{n} \, dn. 
\end{equation}
For a barotropic equation of state
of the form $P=P(n)$, we obtain
\begin{equation}
\label{tigrg2}
\epsilon(n)=nmc^2+n\int^n \frac{P(n')}{{n'}^2}\, dn'.
\end{equation}
Since $\epsilon$ is a function of $n$, the pressure is a function 
$P=P(\epsilon)$ of the energy density. Therefore, Eq. (\ref{tigrg1}) can be
integrated into
\begin{equation}
\label{tigrg3}
n(\epsilon)=e^{\int^{\epsilon} \frac{d\epsilon'}{P(\epsilon')+\epsilon'}}.
\end{equation}
The binding energy of a general relativistic
isentropic or cold barotropic gas
is $E=(M-Nm)c^2$ where $M$ is the mass and $N$ is the particle number. A
stable equilibrium state of the Euler-Einstein equations is a
minimum of energy $E$ at fixed particle number $N$ (see
Appendix \ref{sec_dsegr}).\footnote{In the isentropic case $s/n=\lambda$
we have $S=\lambda N$. Therefore a minimum of $E$ at fixed $N$ is also a maximum
of $S$ at fixed $E$ (or ${\cal E}$).} In the Newtonian limit, $E\rightarrow
U+W={\cal W}$  (see Sec. \ref{sec_mneb} with $u$
in place of $\epsilon_{\rm kin}$) and
we recover the results of Appendix \ref{sec_tinbs}.

{\it Remark:} For a linear equation of state
\begin{equation}
\label{tigrg4}
P=q\epsilon \qquad {\rm with} \qquad q=\gamma-1,
\end{equation}
and for $Td(s/n)=0$ we obtain 
\begin{equation}
\label{tigrg5}
P=Kn^{\gamma}\qquad {\rm and}\qquad \epsilon=\frac{K}{q} n^{\gamma},
\end{equation}
where $K$ is a constant of integration.
When $\mu=0$ the integrated
Gibbs-Duhem relation (\ref{b90}) reduces to 
\begin{equation}
\label{tigrg6}
s=\frac{\epsilon+P}{T}.
\end{equation}
When $T=0$ we obtain $P=-\epsilon$. Therefore, the equation of state of dark
energy corresponds to a relativistic gas at $T=0$ with $\mu=0$. When $s=\lambda
n$ we obtain
\begin{equation}
\label{tigrg7}
T=\frac{q+1}{q} \frac{K}{\lambda} n^{\gamma-1}.
\end{equation}
The case $\mu=0$ applies to the black-body radiation for which
$q=1/3$. In that case, we recover the relations of Appendix
\ref{sec_bbr} but the constant $K$ is not determined by the present method.

\subsection{Newtonian self-gravitating gases at statistical equilibrium}
\label{sec_tins}

We consider a Newtonian self-gravitating gas at statistical equilibrium (see
Sec. \ref{sec_smnrf}). The
first
principle of thermodynamics writes
\begin{equation}
\label{tins1}
d\left (\frac{\epsilon_{\rm kin}}{n}\right )=-Pd\left
(\frac{1}{n}\right )+Td\left (\frac{s}{n}\right ).
\end{equation}
Since  the temperature $T$ is uniform at statistical equilibrium, we have
\begin{equation}
\label{tins2}
d\left (\frac{\epsilon_{\rm kin}-Ts}{n}\right )=-Pd\left
(\frac{1}{n}\right )=\frac{P}{n^2}dn.
\end{equation}
On the other hand, we have seen in Sec. \ref{sec_smnrf} that the  gas has a
barotropic equation of
state $P=P(n)$. Therefore, the
foregoing equation can be integrated into
\begin{equation}
\label{tins3}
\epsilon_{\rm kin}(n)-Ts(n)=n\int^n \frac{P(n')}{{n'}^2}\, dn'.
\end{equation}
Introducing the internal energy defined by Eq. (\ref{tinbs3}) we obtain the
important relation
\begin{equation}
\label{tins4}
\epsilon_{\rm kin}(n)-Ts(n)=u(n).
\end{equation}
Integrating this relation over the whole configuration, we find that 
the entropy is given by
\begin{equation}
\label{tins5}
S=\frac{E_{\rm kin}-U}{T},
\end{equation}
which returns Eq. (\ref{alt23}). On the other hand, the total energy is given
by $E=E_{\rm kin}+W$. In the microcanonical ensemble, a stable equilibrium state
is a maximum of entropy $S$ at fixed energy $E$ and particle number $N$. On
the other hand, in the canonical ensemble, a stable equilibrium
state
is a minimum  of free energy $F$ at fixed particle number $N$. Using Eq.
(\ref{tins5}), we find that the free energy is given by 
\begin{equation}
\label{tins6}
F=E-TS=E_{\rm kin}+W-TS=U+W,
\end{equation}
which returns Eq. (\ref{alt13}). We see that
\begin{equation}
\label{tins7}
F={\cal W}.
\end{equation}
Therefore, the criterion of thermodynamical stability in the canonical ensemble
(minimum of $F$ at fixed $N$) coincides with
the criterion of dynamical stability with respect to the Euler-Poisson
equations (minimum of ${\cal W}$ at fixed $N$). The calculations of this
Appendix  provide a direct proof of the nonlinear Antonov first law
obtained in \cite{aaantonov}.

{\it Remark:} At $T=0$ we see from Eq. (\ref{tins4}) that $\epsilon_{\rm
kin}(n)=u(n)$. This implies that $E_{\rm kin}=U$ so that $E=E_{\rm
kin}+W=U+W={\cal W}$. This is a particular
case of the general relation (\ref{tins7}). At $T=0$, the equilibrium
state is obtained
either by minimizing $E=E_{\rm
kin}+W$ at fixed particle number $N$ or by minimizing ${\cal
W}=U+W$ at fixed particle number
$N$.

\end{document}